\documentclass[useAMS,usenatbib]{mn2e}  % <---- correct one

\usepackage{graphicx}
\usepackage{epsfig}
\usepackage{subfigure}

\usepackage{microtype}
\usepackage[]{hyperref}  % PDF hyperlinks, with coloured links
\def\link_col{blue}
\hypersetup{colorlinks,citecolor=\link_col,filecolor=\link_col,linkcolor=\link_col,urlcolor=\link_col}

\usepackage{morefloats}

\newcommand{\be}{\begin{equation}}
\newcommand{\ee}{\end{equation}}

\def \aaps{AAPS}
\def \aap{AAP}
\def \apjl{ApJ}
\def \apj{ApJ}
\def \apjs{ApJS}
\def \araa{AAP}

\def \mnras{MNRAS.}

\def \nat{Nature}
\def \pasj{PASJ}

\def \aj{AJ}

\def \apss{APSS}
\def\msun{{\,M_\odot}}

\def\lesssim{\lower.5ex\hbox{$\; \buildrel < \over \sim \;$}}
\def\gtrsim{\lower.5ex\hbox{$\; \buildrel > \over \sim \;$}}

\newcommand{\nn}{\nonumber}
\newcommand{\bea}{\begin{eqnarray}}
\newcommand{\eea}{\end{eqnarray}}

\title[GC Mass and Energy Flows] {
Non-thermal Insights on 
Mass and Energy Flows Through the Galactic Centre and into the Fermi Bubbles
}
\author[Roland M. Crocker et al.]
{R.~M.~Crocker$^{1}$\thanks{E-mail: Roland.Crocker@mpi-hd.mpg.de}%\thanks{IIF Marie Curie Fellow} 
\\
$^{1}$Max-Planck-Institut f{\" u}r Kernphsik, P.O. Box 103980 Heidelberg, Germany}
\begin{document}

%%%%%%%%%%%%%%%%%%%%%%%%%%%%%%%%%%%%%%%%%%%%%%%%%%%%%%%%%%%%%%%
%%%%%%%%%%%%%%%%%%%%%%%%%%%%%%%%%%%%%%%%%%%%%%%%%%%%%%%%%%%%%%%

\date{Accepted XXX. Received XXX; in original form XXX}

\pagerange{\pageref{firstpage}--\pageref{lastpage}} \pubyear{2011}

\maketitle

\label{firstpage}

\begin{abstract}

We construct a simple model of the star-formation- (and resultant supernova-) driven mass and energy flows through the inner $\sim$200 pc (in diameter) of the Galaxy.
Our modelling is constrained, in particular, by the non-thermal radio continuum and $\gamma$-ray signals detected from the region.
The modelling points to a current star-formation rate of $0.04-0.12 \msun$/year at $2\sigma$ confidence within the region with best-fit value in the range $0.08-0.12 \msun$/year which -- if sustained over 10 Gyr -- would fill out the $\sim 10^9 \msun$ stellar population of the nuclear bulge.
Mass is being accreted on to the Galactic centre (GC) region  at a rate $\dot{M}_{IN} \sim 0.3 \msun/$year.
The region's star-formation activity drives an outflow of plasma, cosmic rays, and entrained, cooler gas.
Neither the plasma nor the entrained gas reaches the gravitational escape speed, however, and all this material fountains back on to the inner Galaxy.
The system we model  can naturally account for the recently-observed $\gtrsim 10^6 \msun$ `halo' of molecular gas surrounding the Central Molecular Zone out to 100-200 pc heights.
The injection of cooler, high-metallicity material into the Galactic halo above the GC may catalyse the subsequent cooling and condensation of hot plasma out of this region and explain
the presence of relatively pristine, nuclear-unprocessed gas in the GC. 
This process may also be an important ingredient in understanding the long-term stability of the GC star-formation rate.
The plasma outflow from the GC reaches a height of a few kpc and is compellingly related to the recently-discovered Fermi Bubbles by a number of pieces of evidence.
These include that the outflow advects precisely i) the power in cosmic rays required to sustain the Bubbles' $\gamma$-ray luminosity in saturation; ii) the hot gas required to compensate for gas cooling and drop-out from the Bubbles; and iii) the magnetic field required to stabilise the walls of these structures.
Our modelling demonstrates that $\sim 10^9 \msun$ of hot gas is processed through the GC over 10 Gyr. 
We speculate that the continual star-formation in the GC over the age of the Milky Way has kept the SMBH in a quiescent state thus preventing it from significantly heating the coronal gas, allowing for the continual accretion of  gas on to the disk and the sustenance of star formation on much wider scales in the Galaxy.
In general, our investigations  explicitly reveal  the GC's important role in the Milky Way's wider  stellar ecology. 

\end{abstract}


%\maketitle

\renewcommand{\thefootnote}{\arabic{footnote}}

\section{Introduction}

The inner 200 pc (in diameter) of the Milky Way 
features a spectacular confluence of unusual and energetic astrophysical phenomena.
Within this region of the Galaxy -- circumscribed by the Inner Lindblad Resonance associated with the non-axisymmetric gravitational potential of the Galactic bar -- 
the distribution of stars cusps sharply into the distinct population of the so-called nuclear bulge \citep{Serabyn1996}.
Correspondingly, over the same region the current, inferred areal density of star formation, $\dot{\Sigma}_*$, sharply peaks to $\sim 200 \msun/$kpc$^2$/yr. 
This is approximately three orders of magnitude higher than the mean value in the Galactic disk. 
With such a high $\dot{\Sigma}_*$,  observations of the nuclei of external, star-forming galaxies tell us to expect a star-formation-driven outflow; there is much empirical and theoretical evidence that such an outflow exists in the GC as we have explored in a number of recent papers \citep{Crocker2010b,Crocker2011a,Crocker2011b}.
This paper adds significantly to that evidence.

The very high star formation rate (SFR) density likewise sustains a very high energy density in all phases of the GC ISM.
Most directly, the optical and UV output of the many young, hot stars in the region is reprocessed by thick, ambient dust into a dominantly infrared photon background of $\sim$100 eV cm$^{-3}$.
Radio continuum and $\gamma$-ray observations allow one to place a robust lower limit of $\sim$50 $\mu$G on the typical magnetic field throughout the entire inner $\sim$800 pc (in diameter) of the Galaxy \citep{Crocker2010a}; modelling \citep{Crocker2010b,Crocker2011b} points to a magnetic field in the inner 200 pc that is at least 100 $\mu$G.
In association with and, in fact, as a necessary precondition to,  the high SFR, observations reveal an enormous agglomeration of hot, dense, and highly-turbulent molecular gas of mass $3 \times 10^7 \msun$ \citep{Dahmen1998,Molinari2011}, 5-10\% of the Milky Way's entire $H_2$ allocation.
This gas forms an asymmetric distribution extended along the plane referred to as the Central Molecular Zone \citep[CMZ:][]{Serabyn1996}.
Recent infrared observations of the CMZ by $Herschel$ \citep{Molinari2011} place much of this $H_2$ on a $\sim 100 $ pc-radius ring, apparently akin to the star-forming rings observed in the nuclei of many face-on galaxies.  

The CMZ $H_2$ is constantly bombarded by an extended, hard-spectrum cosmic ray ion population which results in
a diffuse glow of hard-spectrum, $\sim$TeV $\gamma$-rays coextensive with the gas \citep{Aharonian2006}.
GC X-ray observations \citep{Koyama1989} {\it apparently} reveal the existence of a very hot ($\sim$ 7 keV), extended thermal plasma which would have an energy density similar to the lightfield, turbulent molecular gas, and magnetic field \citep{Spergel1992}.

It must also be remarked that the GC hosts  the Milky Way's resident supermassive ($\sim4.3 \times 10^6 \msun$: \citealt{Gillessen2009}) black hole (SMBH).
Though currently in a state of apparently unusual quiescence, this must certainly have been much more active at various times in the past \citep[e.g.][]{Ponti2010}.
On the other hand, we have found from our recent work that 
the mechanical power delivered by supernovae 
 -- occurring at a rate consistent with that pointed to by the region's current star-formation as traced by FIR emission -- 
is completely sufficient to sustain the currently-observed non-thermal emission ($\sim$GHz radio continuum and $\sim$TeV $\gamma$-ray)
from the $\sim$200 pc scales of interest here.
Thus, from the point of view proffered by the non-thermal data, it is not {\it necessary} 
that the SMBH have any significant role beyond the inner few pcs; our investigations below confirm this in general.

Finally, one of the most interesting recent discoveries in high energy astrophysics is of the `Fermi Bubbles', so-called because these 
structures were revealed \citep{Su2010} in $\sim$ GeV $\gamma$-ray data collected by the $Fermi$-LAT.
The Bubbles are north-south symmetric about the Galactic plane and centred on the Galactic nucleus.
Given this morphology they are compellingly associated with some sort of activity in the GC.
Given, then, the Bubbles' large angular extent (they rise to $\pm 50^\circ$ in $b$) they are enormous structures extending  10 kpc from the plane.
The Bubbles' $\gamma$-ray emission might be due to inverse Compton emission from a rather young (given short energy loss time) population of cosmic ray electrons.
Alternatively, the emission might be due to hadronic collisions experienced by a hard-spectrum cosmic ray proton and heavier ion population \citep{Crocker2011a,Crocker2011c,Zubovas2011}.
We set out the evidence connecting the Fermi Bubbles with multi-Gyr-scale GC-star-formation-driven injection of cosmic ray {\it protons} into the Galactic halo below.

\subsection{Motivating Questions}

The preceding tour of GC and inner Galaxy phenomenology
helps motivate a number of questions which we seek to address in this paper:
\begin{enumerate}

\item{\bf Gas accretion:} What is the rate at which the GC region typically accretes gas through the Galactic plane? 
How do we understand the presence \citep{Lubowich2000,Riquelme2010} of relatively pristine gas (i.e., that has undergone relatively little nuclear processing) in the GC?
 
\item {\bf Star formation:} What is the efficiency with which the GC converts gas into stars?
Given the unusual conditions in the GC environment, is GC star-formation biased towards the production of massive stars?

\item {\bf GC ISM conditions:} Is the very hot plasma putatively revealed by X-ray observations real or not?
What contribution do the non-thermal ISM phases, in particular the cosmic rays, make to the overall energy density in the region?
What is the dynamical importance, if any, of the cosmic rays?

\item {\bf Outflows:}  There is multi-wavelength evidence (reviewed below) for outflows from the GC over size scales from pcs to 10 kpcs.
Are these outflows all different aspects of the same overarching phenomenon and how are they driven?
What is the wider importance of the GC outflow(s) to the Galactic ecology?
Is the material expelled from the nucleus lost to extra-galactic space or does it fountain back on to the Galactic disk?
How do the recently-discovered Fermi Bubbles \citep{Su2010} relate to activity in the GC?

\end{enumerate}
More broadly, we aim in this paper to produce a first draft of a coherent explanation of all the disparate phenomena listed above that is itself physically plausible and motivated.
Overall, we shall see that it is  star-formation (driven by secular accretion processes over long timescales) -- rather than processes associated directly with the SMBH -- that seems to control the overall dynamics of the GC.

\subsection{Conventions and Assumptions}

We assume a distance to the Galactic centre of 8 kpc in this work. 
We  use {\tiny MATHEMATICA} notation: $f[x]$ denotes $f$ in a function of parameter $x$.
Formally,  the region we are investigating and attempting to model is that centred on $(l,b) = (0,0)$ and extending to $\pm 0.8^\circ$ in Galactic longitude and $\pm 0.3^\circ$ in Galactic latitude; this is the region for which the HESS telescope reported \citep{Aharonian2006} a diffuse flux of $\sim$TeV $\gamma$-rays.

\section{Background}

%\subsubsection{Ejected Mass}

\subsection{Star Formation in the GC}

\subsubsection{Star formation rate in the GC}
\label{sctn_sfrGC}

Using standard prescriptions and assuming a `normal' initial mass function (IMF), $\psi \equiv dN/dM$, the current  SFR over the central few degrees of the Galaxy 
can be estimated from the region's inferred Lyman continuum photon output of $\sim 10^{52}$ photons s$^{-1}$ \citep{Cox1989}
to be $0.3-0.6 \msun$/year \citep{Gusten1989}.
\citet{Yusef-Zadeh2009} have determined  a SFR over the central 400 pc  that has ranged between 0.14 and $\sim0.007 \msun$/yr over the last $\sim10^7$ years on the basis of mid-infrared and other data\footnote{The 0.007 $\msun$/yr reported by \citet{Yusef-Zadeh2009} is obtained from their estimate of the 1.4 GHz radio continuum flux; such an estimate suffers from the problem  that cosmic ray electrons are removed from this system before they can radiate much of their energy: see below and also \citet{Crocker2010b,Crocker2011b}. Put another way, were this advective removal of cosmic rays electrons not taking place one would derive a RC estimate of the SFR much closer to  determinations relying on other data.}
These authors also estimate an upper limit on the SFR in the region over the last 10 Gyr of $0.15 \msun/$yr and claim a probable average value in the range 0.04-0.08 $ \msun/$yr.
On the basis of 5-38 $\mu$m observations with the Spitzer
Infrared Spectrograph of young stellar objects in the CMZ (and assuming a \citet{Kroupa2001} IMF)
\citet{Immer2011} have also recently determined a  SFR in the region of $\sim 0.08 \msun/$year over the last Myr.
\citet{Figer2004} obtain a lower limit of $0.01 \msun$/yr on the recent GC SFR by the simple expedient of dividing the mass in known, recently-formed stars by the duration of the star formation episodes that formed those stars.
Accounting for the fact that recent studies of Paschen-$\alpha$ emission \citep{Mauerhan2010} have shown that approximately half the region's massive stars are located outside known clusters, a SFR of $\sim 0.02  \msun$/yr (for the central few tens of pcs) is suggested.
This SFR is close to the value suggested by the stellar luminosity function analysis of \citet{Figer2004}.
The detailed modelling of \citet{Kim2011} suggests that the SFR within the X2 orbits ranges between 0.04 and 0.09 $\msun$/year.
Somewhat in tension with the above determinations, \citet{Yasuda2008} have claimed a low level of current GC star-formation activity on the basis of a low ratio of the [C{\sc ii}]
fine-structure emission line (due to  photo-dissociation and H{\sc ii} regions) to the total 
FIR emission, leading them to the conclusion that the region's radiative output is actually dominated by old stars.

\subsubsection{GC IMF}
\label{sctn_IMF}

Note that though the different SFR determinations listed above apply to regions of somewhat different sizes, given the GC's stellar population is highly centrally peaked \citep{Serabyn1996,Launhardt2002}, there is a definite hint that these measures are discrepant: the region's inferred UV radiation output seems too high.
Given that massive stars tend to completely dominate the production of such radiation,  this discrepancy may, in fact, be an indication \citep{Figer2004} that the region's star formation is biased towards the production of massive stars \citep{Morris1993,Maness2007}, consistent with independent indicators \citep{Figer1999} that the region's initial mass function is significantly flattened.
That star-formation in the GC be biased towards the formation of more massive stars is a rather natural prediction: given the environment, any or all of the
region's strong tidal forces, high gas pressures, and magnetic fields might be expected \citep{Morris1993,Lis2001} to 
significantly alter the dynamics of the collapse of molecular gas into
stars.
Whether this expectation is supported observationally remains, however, a topic of hot debate.
Indeed, the GC, as the site of some of the most active massive star-formation in the Galaxy, has been a natural battle ground in the debate over whether the IMF is truly universal 
\citep[e.g.,][]{Bastian2010,Lockmann2010} or is
flattened (or has a higher low mass cut-off to the formed stellar population)  in the GC \citep[][]{Morris1993, Figer1999, Figer2004, Maness2007} and other star-burst-like environments.

\subsubsection{Continuous and steady star formation in the GC }
\label{sectn_SS}

Importantly for our purposes, 
the  luminosity function   analysis of \citet{Figer2004} also favours a SFR that has been sustained at more-or-less the current value for a 
timescale approaching 10 Gyr.
In fact, 
there seems to be accumulating evidence from different directions  \citep{Maness2007,Lockmann2010,Kim2011}   
that the GC has been continuously forming stars over this sort of timeframe.
Indeed,  we \citep{Crocker2011a} have recently suggested on the basis of our modelling of the non-thermal emission Fermi Bubbles (see below) that 
the currently-observed star formation rate in the GC is typical of the system's time-averaged value over the last $\sim 8$ Gyr.

A corollary of this sort of picture is that the drama associated with most of the accumulation of the mass of the SMBH at Sgr A* is pushed back to highish redshifts, probably accompanied by the formation of 
 most of the long-lived stellar population of the bulge.
In this context, the fact \citep{Gilmore2002} that the Milky Way underwent its last major merger activity $\gtrsim$11 Gyr ago and has subsequently experienced rather quiescent \citep{Yin2009} evolution  is significant.
On the other hand, recent modelling \citep{Purcell2011}
shows that at least some of the continuing gas feeding to the inner Galaxy could ultimately be driven by the on-going,
minor merger activity the Galaxy experiences

Other independent evidence that the GC SFR has been steady over long timescales comes from the observation (discussed further below: \S \ref{sectn_UV})
of separate parcels of highly-ionized high-velocity gas (in UV absorption spectroscopy along the sight lines to distant quasars) at low Galactic longitude but varying Galactic height (both north and south of the plane), both emerging from and, apparently, falling back on to the GC \citep{Keeney2006}.
Significantly, these gas parcels apparently form part of a Galactic fountain and can be inferred to have reached (or will reach) the {\it same maximum height} from the plane of $12 \pm 1$ kpc but must have been launched at different times in the past ranging from 20-50 Myr to more than 800 Myr, suggesting the operation of a common launching mechanism over at least the latter timeframe.
The most recent UV measurements \citep[towards a post-AGB in the inner Galaxy;][]{Zech2008} suggest, moreover, that at least some of this material is of super-solar metallicity.
This suggest both an inner-Galaxy origin and may point to a star-formation -- rather than AGN -- origin to the outflow(s) \citep[cf.][]{Su2010}.

\subsection{X-ray observations}
\label{sctn_Plasma}

In its plasma phase, X-ray continuum and Fe line observations apparently reveal a two-temperature plasma containing `hot' ($\sim$ 1 keV) and (mysteriously) `very hot' (6-9 keV) components \citep{Koyama1989,Yamauchi1990,Kaneda1997,Muno2004,Belanger2004}. 
The X-ray emission from the putative very hot component is strongly concentrated within the inner $\sim$150 pc (in diameter) of the Galaxy \citep{Yamauchi1990,Belmont2005}. 
As first observed by \citet{Spergel1992}, 
there may be pressure equilibrium \citep[at $(3-6) \times 10^6$ K cm$^{-3}$:][]{Koyama1996,Muno2004} between the kinetic pressure of the putative
very hot plasma phase  and the virial pressure implied by
the  
turbulent motions of the molecular gas.

{\it Prima facie}, the very-hot plasma presents a severe energetics problem, however: (assuming it is a hydrogen plasma) its sound speed at $\sim$1500 km/s would be considerably in excess of the local escape velocity of $\sim$ 900 km/s \citep{Muno2004} suggesting it should escape \citep{Yamauchi1990} on a short timescale. 
This suggests a steady-state situation would require a power considerably in excess of $10^{40}$ erg/s to sustain the outflow.
A second difficulty is that  there is no widely-accepted mechanism to heat the plasma to more than a few keV; Galactic disk supernova remnants, in particular, do not seem to heat plasma beyond $\sim 3$ keV after a couple of centuries \citep[hotter temperatures at earlier times are possible but the smooth distribution and overall energy of the GC's putative hot plasma cannot be reconciled with such a young explosion:][]{Muno2004,Belmont2006}.

There is no universally-accepted resolution to these anomalies. One interesting suggestion is that the 8 keV emission is due to a very hot {\it helium} plasma which would be gravitationally bound \citep{Belmont2005}.
Another suggestion is that the `plasma' is illusory, the emission actually being attributable to unresolved point sources\citep{Wang2002}.
Recent, deep Chandra observations around $l$ = 0.08, $b$ =1.42  (taken to be typical of the so-called X-ray Ridge) support this sort of picture \citep{Revnivtsev2009}.
On the other hand, the situation within the inner $\sim$150 pc -- where the 6.7 keV Fe line emission strongly peaks \citep{Yamauchi1990} --
may be quite different to that pertaining elsewhere in the Galaxy \citep{Dogiel2009_II}.
A deep observation \citep{Muno2004} of the inner 17$'$ with Chandra could only explain  $\lesssim 40 \%$ of the X-ray flux as due to dim point sources.
Moreover, recent results obtained with the SUZAKU X-ray telescope continue to clearly suggest \citep{Koyama2007,Dogiel2010,Koyama2011} the existence of a hot plasma covering at least the central 20$'$; this issue, therefore, has remained unresolved.
Below  the modelling we  present shows how the GC's star-formation activity {\it might} be able to sustain such a plasma in steady-state.

\subsection{Evidence for a Galactic Centre Outflow}

Observationally, there is evidence on multiple scales and at many different wavelengths   for an outflow or outflows from the GC, some of which we review briefly below.
The idea of an outflow has also received  theoretical support from our recent work \citep{Crocker2010b,Crocker2011a,Crocker2011b} which we also briefly review.

\subsubsection{GC Lobe and CMZ molecular halo}

\begin{figure}
 %\vspace{200pt}
 \epsfig{file=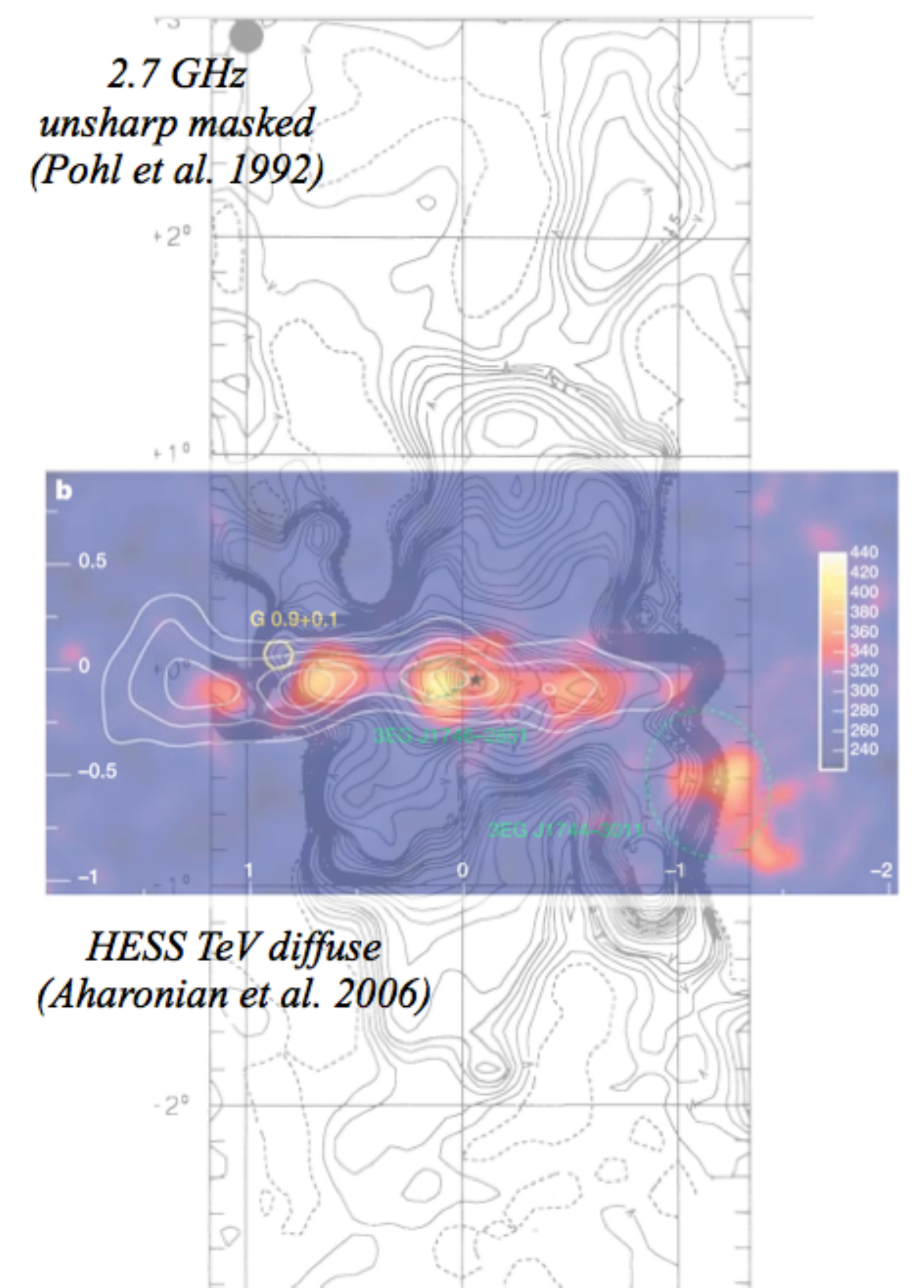,width= \columnwidth}
%{\includegraphics[width=\columnwidth]{plotRqrdWindSpeed.pdf}
\caption{Colours: $\sim$ TeV map of the GC region \citep{Aharonian2006} 
after removal of point-like sources coincident with Sgr A$^*$ and the SNR G0.9+0.1 and (white contours)
CS(1$\to$0) molecular line emission (a tracer of the total $H_2$ column);
black contours: 2.7 GHz unsharp-masked radio continuum image with 9.4' resolution \citep{Pohl1992}.
The radio continuum spurs above $l \simeq 0^\circ.1$ (the Radio Arc) and $\simeq -0^\circ.7$ (Sgr C) define the non-thermal GC Lobe structure.
Note that the footprints of the radio continuum spurs are coincident with regions of  high molecular column and $\sim$TeV $\gamma$-ray emission and star-formation.
HESS data figure is adapted by permission from Macmillan Publishers Ltd: [Nature, 439, 695], copyright (2006).
For 2.7 GHz radio contours credit:  \citet{Pohl1992}, A\&A,  262,  441, 1992, reproduced with permission \copyright ESO.
}
\label{HESSRadioCorres}
%} 
\end{figure}

Radio and optical recombination line observation observations by \citet{Law2009} reveal $2 \times 10^5 \msun$ of warm ($\sim4000$ K), ionized gas extending up to $\sim1^\circ$ north from the plane above the GC.
This gas is nested within a shell formed by the so-called Galactic Centre Lobe \citep[GCL;][]{Sofue1984} detected as a non-thermal radio continuum source at $\sim$GHz frequencies \citep[and references therein]{Law2010} and visible up to at least 10.5 GHz 
(\citealt{Sofue1996,Crocker2010a}; refer to the radio continuum contours in fig.~\ref{HESSRadioCorres} from \citealt{Pohl1992}: the GCL rises between $l \simeq 0^\circ.2$ and  $l \simeq -0^\circ.7$, i.e., roughly above the Radio Arc and Sagittarius C).
An outermost shell around this structure of entrained dust and PAHs  is detected at MIR wavelengths \citep{Bland-Hawthorn2003}, some emission revealing  helical topologies presumably tracing a complex magnetic field structure \citep{Morris2006}.
Finally, CO and CS  line emission from the region  reveals  molecular gas extended along spurs north  \citep{Uchida1994} and south of the plane and more-or-less coincident with (actually slightly inside-of)  the radio continuum features and apparently rotating \citep{Sofue1996}.
The mass of such molecular gas in the GCL has been estimated to be at least $\sim 3 \times 10^5 \msun$.

The number density of the warm gas from the radio recombination line observations 
can be estimated to be $\sim10^3$ cm$^{-3}$ and its pressure $P/k_B \sim 7  \times 10^6$ K cm$^{-3}$  
would put it in or close to pressure equilibrium with the other GC ISM phases \citep[including the very hot plasma were it real;][]{Law2009}.
The gas is also of high metallicity and  parts of the lobe's RC emission suffer from H{\sc i} absorption \citep{Law2009}; both these factors clearly point to the structure's location in the GC.
The warm gas has, however, a low filling factor, $f \sim 10^{-4}$, much smaller than typical for this phase in the Galactic disk but interestingly comparable to that inferred for other Galactic outflows \citep[][and references therein]{Law2009}.

Radio continuum observations also lend support to the notion that the GCL represents an outflow \citep{Law2010}: the non-thermal spectrum of the GCL steepens as a function of increasing Galactic latitude, a clear sign of a synchrotron-emitting electron population that is ageing as it is transported from the plane \citep[cf. the recent work by][ on the star-burst system NGC 253]{Heesen2009}. 

We can use the estimated $\sim 2 \times 10^5 \msun$ of warm, ionized gas filling the GCL \citep{Law2010} to a height $h\sim$140 pc north of the GC
to arrive at a lower limit to the mass flux, $\dot{M}$ in an outflow: with $M_{GCL} \simeq \dot{M} h/v_{wind} > M_{obs} = 2 \times 10^5 \msun$, we find $\dot{M} \gtrsim 0.3 \msun$/year $\times v_{wind}/100$ km/s (assuming a similar, but so-far unobserved, distribution of ionized gas south of the plane from the GC -- though see \S \ref{sctn_worms})\footnote{Scaling the results of \citet[figure 6]{Martin2005} according to the GC's estimated SFR areal density, we find that the expectation afforded by observations of external galaxies is that the GC should drive an outflow with a speed of $\sim$400 km/s.}.

The above is likely a conservative estimate as it neglects  mass in other gas phases.
In fact recent CO($2\to1$) and CO($1\to0$) line observations with the  Nanten-II telescope point to a halo of molecular gas around the entire CMZ with height $\sim1^\circ$ (or 100-200 pc) and total mass few $\times 10^6 \msun$ (Yasuo Fukui, private communication).
Spitzer IRS spectra also reveal high-latitude 17 and 28 $\mu$m $H_2$ emission lines around the CMZ, consistent with such a molecular gas halo (Mark Morris, private communication).
Finally, older OH absorption line observations of the region \citep{Boyce1994}  also point to the existence of a $\sim 200$ pc molecular halo around the GC and reveal individual molecular spur features coincident with radio continuum features (including the GCL).

\subsubsection{GC `worms'}
\label{sctn_worms}

\citet{Koo1992}  made an inventory of Galactic `worms': isolated structures appearing in both H{\sc i} and far infrared (60 and 100 $\mu$m) and generally identified as the walls around superbubbles that have broken through the thin gaseous disk of the Galaxy (to both north and south).
Many of these structures are coincident with H{\sc ii} regions and 408 MHz RC emission.
\citet{Koo1992} list four worms coincident with the GC region under consideration in this paper, with three (GW 357.4 + 5.4, 358.0 - 5.3, 0.5 - 5.9) appearing around $v_{LSR} = 0$ km/s (``LSR" denotes {\it local standard of rest}).
%, consistent with these worms' being located in the GC.
%
The inferred mass of GW 0.5 - 5.9 (which projects south of the Galactic plane) is $2 \times 10^5 \msun$, interestingly close to the mass of warm ionised gas observed in the GC lobe (which projects north).

\subsubsection{Extended molecular gas and dust distribution}
\label{sctn_extndH2}

There is multi-wavelength evidence for extended -- and probably outflowing -- cold gas around the GC.
Surveys of molecular lines and dust continuum emission at mm, sub-mm, and mid-infrared wavelengths \citep{Oka1998,Tsuboi1998,Pierce-Price2000,Bland-Hawthorn2003,Stolovy2006,Oka2010} reveal filaments, arcs, and shells, indicative of local, turbulent sources \citep{Tanaka2007} and explosive events.
A large-scale CO J=3-2 survey of the CMZ  with the Atacama Submillimeter-wave Telescope Experiment \citep{Oka2010}
detects an unusual population of high-velocity ($\Delta v \geq 50$ km/s) compact clouds (HVCCs) in the GC.
24 micron observations have revealed  dust emission from a fascinating `double helix' structure \citep{Morris2006} $\sim 0.^\circ7$ degrees north of the GC which  has recently been shown\footnote{Rei Enokiya, talk presented at `The emerging, multi-wavelength view of the Galactic Centre Environment', Heidelberg, Germany, October 17-20, 2011.}  to exhibit kinematically-related molecular emission and linearly-polarized radio continuum emission consistent with a highly-ordered magnetic field structure\citep{Tsuboi2010};
$^{12}$CO ($1\to0$) molecular line emission observations demonstrate the existence of giant molecular loops (GMLs) with large velocity dispersions in the region, 
argued to originate in the phenomenon of magnetic flotation controlled by the \citet{Parker1966} instability \citep{Fukui2006}.
Importantly for our purposes here, the most recent $^{12}$CO ($1\to0$) and ($2\to1$) molecular line observations by the Nanten-II group reveal an extended halo of molecular gas around the entire
CMZ with total mass of few $\times 10^6 \msun$ of $H_2$, with more in individually-identifiable outflows\footnote{Yasuo Fukui, private communication, and  Kazufumi Torii, talk presented at `The emerging, multi-wavelength view of the Galactic Centre Environment', Heidelberg, Germany, October 17-20, 2011.}.

\subsubsection{Evidence for disk-halo connection in the GC}
\label{sctn_diskHalo1}

Another interesting piece of the puzzle of GC gas dynamics emerges from studies indicating the presence of relatively pristine (i.e., relatively nuclear-unprocessed) or even primaeval gas in the GC.
A first piece of evidence suggesting accretion of such gas into the region comes from observation of the $J = 1\to0$ and $2\to1$ lines of  DCN in the `50 km s$^{-1}$' molecular cloud located $\sim$10 pc from the true GC \citep{Lubowich2000}.
Deuterium is destroyed in stellar interiors and -- unless it is continuously replenished in a heavily astrated region like the GC -- its abundance relative to H should be at the few parts per trillion level; \citet{Lubowich2000} determined levels five orders of magnitude higher than this.

This inference of the infall of fresh gas into the GC has recently found interesting confirmation in determinations of high $^{12}$C/$^{13}$C isotopic ratios (through measurement of the $1 \to 0$ lines of HCO$^+$, HCN, HNC and their $^{13}$C isotopologues) along a number of GC sightlines.
Overall the  $^{12}$C/$^{13}$C ratio is known to exhibit a gradient from high (80-90) to low (20-25) values going from the galactocentric distance of the solar system to the GC \citep{Wilson1999}.
This is consistent with the general picture that gas in the outer Galaxy should have experienced less nuclear processing than that in the inner: $^{12}$C is formed in  first-generation, metal-poor stars over shortish timescales whereas $^{13}$C is formed via CNO processing by low or intermediate mass stars of $^{12}$C seeded by earlier stellar generations \citep[and references therein]{Riquelme2010}.
\citet{Riquelme2010} have determined lower limits to $^{12}$C/$^{13}$C isotopic ratios 
for a number of places in the CMZ that are close to the values measured in the local ISM and inconsistent with the general gradient identified above.
The implication of this -- as for the deuterium observations -- is that some relatively less-processed gas is, somehow, finding its way directly to the GC.

\citet{Riquelme2010} have found evidence for such unprocessed gas, in particular, in the footprints of a number of GMLs identified by the Fukui group and, in general,
in gas whose phase space position places it in the $X1$ orbit family  or in transit from   $X1$ orbits to $X2$ \citep[recent higher resolution data from][seems to confirm this latter finding; see their fig.~18]{JonesBurton2011}.
Indeed, the measured $^{12}$C/$^{13}$C isotopic ratios may be more consistent with the idea \citep{Morris2006b,Torii2010,Riquelme2010} that the 
shocks induced in the rising portions of the loops sweep up and compress rarefied atomic gas in the halo above the GC, 
leading to rapid cooling and condensation into molecular gas.
Whatever the mechanism, the presence of relatively pristine gas in the GC, represents compelling evidence that some gas is being rather directly accreted out of the halo on to the GC.

\subsubsection{H{\sc i} `outflow(s)' from GC}

Making sense of 21 cm line data towards the GC is difficult.
The interpretation of features in such data as indicating some sort of outflow or multiple ejections from the nucleus has a venerable history \citep[e.g.,][]{Sanders1972,Mirabel1976}. 
Much of the apparently anomalous kinematics of the individual atomic hydrogen features within 
$|b| \lesssim 2^\circ$ over the $|l|  \lesssim 1^\circ$ longitude range of relevance here
has subsequently been reinterpreted as motion governed by the inner Galaxy's tilted H{\sc i} disk  identified by \citet{Burton1978}.
Even this latter work, however, found tentative evidence for streams of material moving out along the polar axis of this  disk (tilted by $12^\circ$ with respect to the vertical in the model) away from the nucleus at $\sim 200$ km/s and representing a total mass $\gtrsim 10^6 \msun$.
We remark in passing that these opposing streams -- moving into the north east and south west 
quadrants\footnote{The south-west feature is tracked by \citet{Mirabel1976} out to $b = -22^\circ$.}
 -- are roughly aligned to the corresponding, inner edges of the Fermi Bubbles (see below, \S\ref{sctn_FB}) and the biconical X-ray feature identified in ROSAT data by \citet{Sofue2000} (see below, \S\ref{sctn_ROSAT}); \citet{Sanders1972} identified an H{\sc i}  feature on similar size scales that seems to be similarly coincident with the north-$west$  edge of the northern Fermi Bubble and the same X-ray structure.

On much larger scales, individual, high and intermediate velocity H{\sc i} clouds or cloud complexes are seen all over the sky \citep[e.g.,][and references therein]{Winkel2011}; the distance to these features -- and their inferred masses and sizes -- is notoriously difficult to pin down.
A massive literature  describing these data exists; it is sufficient for current purposes to mention that a small fraction of such clouds may be associated with material ejected by a GC fountain.
The recent analysis of \citet{Winkel2011} of the high-velocity cloud complex Galactic center negative (covering the south-east complex out $\sim 80^\circ$ from the GC) 
finds a distinct sub-population of clouds which, while separated from the GC by up to 80$^\circ$, seems to be kinematically related to high-velocity dispersion gas in the central few degrees of the Galaxy; this might be tentatively ascribed to a large-scale GC ejection event or outflow, though analysis of this structure is on-going.

\subsubsection{UV absorption spectroscopy: evidence for a GC fountain}
\label{sectn_UV}

As prefaced above, UV spectroscopic data suggests both ejection of highly-ionized material from the GC \citep{Bland-Hawthorn2003} and the subsequent fountaining-back of some of this material \citep{Keeney2006}.
At least some of this material seems to have super-solar metaliicity \citep{Zech2008}.
Consistent with this evidence, we will see below that our modelling naturally predicts an outflow speed less than the gravitational escape speed.

\subsubsection{GC spur}
\label{sctn_GCS}

On the basis of an analysis of unsharp-masked 408 and 1408 MHz radio continuum data, \citet{Sofue1989} identified a large,  non-thermal (but hard spectrum) radio feature extending north of the GC up to latitudes of $\sim 20^\circ$ that they labelled the Galactic Centre Spur (GCS).
\citet{Jones2011} have recently shown that this same feature is visible in polarised emission in WMAP microwave data (at 23 and 33 GHz).
There is no obvious, corresponding feature seen to Galactic south in radio or microwave frequencies.
Rather interestingly in light of the discovery discussed in the next sub-section, by eye the GCS appears also partially coincident with a $\gamma$-ray feature extending north from the GC that was  claimed previously on the basis of EGRET data \citep[ though note the statistical significance of this feature could not be established]{Hartmann1997}.

Also notable is that -- despite its length and curvature -- the GCS remains well-collimated over its length having an almost constant width of $1.5-2^\circ$ \citep{Sofue1989}.
This essentially matches the width of the GC star-forming region we model in this paper.
Exactly how the structure remains collimated over distances of $\gtrsim$ 4 kpc is mysterious; regardless we make the point that it apparently represents a channel for the delivery of non-thermal particles to large distances from the plane with minimal adiabatic energy losses.

At 1.4 GHz the GCS divides into two strong radio spurs which very plausibly -- but not definitely -- join on to the radio spurs seen \citep{Pohl1992} above Sgr B and Sgr C at 2.7 GHz (see fig.~\ref{HESSRadioCorres}).
In addition to the above, a number of other pieces of evidence mark the GCS as a unique feature and suggests its GC location:
\begin{enumerate}
\item the feature terminates in the Galactic plane;
\item it is the brightest radio continuum spur after the North Polar Spur \citep{Sofue1989},  likely a local ISM feature;
\item while polarised emission from this structure is detectable at microwave frequencies  \citep{Jones2011}, polarised emission from the GCS disappears by 1.4 GHz: such behaviour is consistent with the `magnetic horizon' effect, i.e., the Faraday depolarisation due to the turbulent ISM which renders polarised emission at $\sim$GHz undetectable beyond a few kpc through the GC plane (thus the GCS is likely at at least this distance). 
\end{enumerate}

Finally, the GCS exhibits noticeable curvature to Galactic west; this curvature can be coherently explained within the general idea that the feature is due to a rather slowly moving, star-formation-driven outflow ($\sim$ 1000 km/s). 
In particular, if the feature is associated with an individual star-formation `event' occurring in the central $\sim$100 pc star-forming gas ring, rotating at $\sim 100$ km/s, \citep{Molinari2011}, then differential rotation would indicate a formation timescale $\sim \pi \ 100$ pc/100 km/s $\simeq 3 \times 10^6$ year (note that the inferred outflowing corkscrew has not yet executed one full turn) and an outflow speed $\sim 900$ km/s \citep[cf.][on the helical magnetic field structure of the nuclear outflow from NGC 253]{Heesen2011}.
These inferred parameters seem eminently reasonable; further study might be able associate the feature to the formation of a particular, GC super stellar cluster.
Note that while the inferred GCS outflow speed is somewhat faster than for the general outflow we identify (see below), the speed is rather well matched to that required to generate 
the sort of high latitude, high metallicity, and highly-ionized gas features identified by UV spectroscopy \citep{Keeney2006,Zech2008}.

\subsubsection{X-ray evidence of a giant, GC-centred biconical structure}
\label{sctn_ROSAT}

\citet{Sofue2000} identified a biconical structure extending north and south from the GC in 1.5 keV ROSAT data that extends out to $|b| \sim 20^\circ$ above and below the plane.
The dynamics of this structure were extensively investigated by both \citet{Sofue2000} and later \citet{Bland-Hawthorn2003}.
\citet{Sofue2000} suggested an association with very-large-angular scale structures in the 408 MHz sky, in particular, the North Polar Spur and suggested these multi-wavelength features were related to a nuclear star-burst that released $\sim 10^{56}$ erg energy, filling out a giant hyper shell centred on the GC.
\citet{Bland-Hawthorn2003} similarly drew a connection between the biconical X-ray feature and 
multi-wavelength data on different scales including the MIR features mentioned above and again suggested a 
likely star-burst origin for the apparent outflow, though with somewhat more modest energetics ($\sim 10^{55}$ erg).

\subsubsection{Fermi Bubbles}
\label{sctn_FB}
 
One of the most interesting recent discoveries  in high-energy astronomy is of the `Fermi Bubbles' \citep{Su2010,Dobler2010} introduced earlier.
The Bubbles are characterised by a rather uniform intensity and an unusually-hard spectrum, $dF_\gamma/dE_\gamma \propto E_\gamma^{-2.1}$ and have a total luminosity $4 \times 10^{37}$ erg/s.

Many researchers have focused on the general idea that the $\gamma$-ray emission from the Fermi Bubbles arises from the inverse-Compton (IC) emission from a (mysterious) population of cosmic ray electrons.
Given that the spectrum of the Bubbles displays no obvious variation with Galactic latitude, however, it is necessary that the photon background being up-scattered by this putative electron population is the CMB.
This, in turn, implies that the electrons have an energy scale $\sim$TeV and consequently short IC loss times, $\sim 10^6$ years \citep{Crocker2011a}.
Given the vast extension of the Bubbles ($\sim$ 10 kpc from the plane), 
these electrons either have to be delivered very quickly -- presumably on an AGN-type outflow originating at Sgr A* \citep{Guo2011} --
or accelerated in-situ by first \citep{Cheng2011} or second-order \citep{Mertsch2011} Fermi acceleration processes.

We \citep{Crocker2011a} have recently considered the alternative explanation that the Bubbles' $\gamma$-ray emission arises from the hadronic collisions of a population of cosmic ray {\it protons} (and heavier ions) populating their interiors.
Because of the long proton loss times on the low-density plasma of the Bubbles this escapes the timing difficulties facing any leptonic mechanism.
Our explanation requires i) (given adiabatic and ionisation energy losses) a total cosmic ray hadron power $\sim 10^{39}$ erg/s that ii) (essentially because of the same long loss time referred to above and the consequently long time required to reach steady state) has been injected quasi-continuously into the Bubbles for a timescale of $\gtrsim 8$ Gyr.
These requirements are precisely matched by the GC CR outflow that we \citep{Crocker2011b}
identified  {\it on the basis of completely independent considerations} to do with observations at radio continuum and TeV $\gamma$-ray wavebands of the inner $\sim 200$ pc of the Galaxy.
This putative solution fits nicely from a number of other perspectives: 
\begin{enumerate}

\item The hard-spectrum of the emission is also explained: by construction, the cosmic rays injected into the Bubbles are trapped so there is no energy-dependent escape process acting to modify the in-situ, steady state distribution away from the injection spectrum and the daughter $\gamma$-rays will trace this hard, parent proton distribution.

\item On the other hand, $\pi^0$-decay kinematics enforces a down-turn below $\sim$GeV on a spectral energy distribution plot of the emitted $\gamma$-radiation; such a downturn is robustly detected, at least qualitatively, in the Bubbles' spectra \citep{Su2010}.

\item The total enthalpy -- the energy required to supply the final internal energy of the Bubble and do the $p\ dV$ work of inflation against the pressure of the external medium -- of a slowly inflated Bubble is \citep[e.g.,][]{Hinton2007} given by $H \ = \ \gamma/(\gamma -1) \ p \ V \simeq (2-4) \times 10^{56}$ erg[\footnote{Note that \citealt{Guo2011} find from their modelling that typical total AGN energy release of $\sim 10^{57}$ erg is required in a fast-inflation scenario with  AGN jets.}] \citep[where $\gamma = 4/3-5/3$ and we adopt $4 \times 10^4$ K cm$^{-3}$ as the pressure in the Galactic disk towards the inner Galaxy:][and the total volume of the Bubbles is $\sim 2 \times 10^{67}$ cm$^{-3}$]{Kasparova2008}.
Setting $\dot{E}_{out} \ t_{inf} \equiv H$ where $\dot{E}_{out}$ is the rate at which the outflow does $P \ dV$ work and $t_{inf}$ is the inflation timescale, we find $t_{inf} \sim $10 Gyr $(\dot{E}_{out}/10^{39}$ erg/s $)^{-1}$;
where we normalise to the typical power we find that the GC delivers into freshly-accelerated cosmic rays (see below).
In reality,  the cosmic rays, advected, frozen-in magnetic field, and injected plasma will all contribute to inflating the Bubbles but the important point is that {\it this inflation will happen over the same multi-Gyr timescales  independently suggested by considerations around reproducing the $\gamma$-ray phenomenology of the Bubbles given the long $pp$ timescale}.
(While the outflow is likely to have a total power $> 10^{40}$ erg/s, both cosmic rays and the plasma injected into the outflow will experience significant radiative losses over these long timescales so adiabaticity is not satisfied: see below.)
Note that the alternative postulate that the  Bubbles are inflated by much shorter duration and much higher mechanical luminosity output from a recent AGN phase of the SMBH comes close to violating a  limit on such activity claimed \citep{Lubowich2000} on the basis of the DCN line observations referred to above.

\item The total plasma mass of the Bubbles is $\lesssim 10^8 \msun$ \citep{Su2010} -- this mass can also be explained given the rate of mass flux in the GC outflow and assuming the same long timescales (as we confirm in detail below).
Also note that the total  power fed into the base of the Bubbles by the outflow can sustain the thermal (X-ray) radiation from the Bubbles \citep{Crocker2011a}.

\item Dynamically, the Bubbles end up being slightly over-pressured but slightly under-dense with respect to the surrounding halo plasma, with internal energy density supplied approximately equally by cosmic rays and their interior hot plasma. They can, therefore, be expected to rise slowly under buoyancy. 

\item The hadronic scenario naturally predicts concomitant secondary electron production within the Bubbles; these secondaries would synchrotron-radiate on the Bubble's magnetic field, thereby explaining the coincident (at lower Galactic latitude) `WMAP haze' detected \citep{Finkbeiner2004,Dobler2008} at microwave frequencies. 

\end{enumerate}

Of course, all this requires that the Fermi Bubbles are very old structures -- almost as old as the Galaxy -- and that they can trap TeV cosmic rays for multi-Gyr timescales.
In fact,  our scenario implies that they would be calorimeters for GC activity over the history of the Milky Way.
This is an interesting prospect indeed.

We finally remark on a very recent development: Finkbeiner and co-workers\footnote{Douglas Finkbeiner, talk delivered at {\it The Emerging, Multi-Wavelength View of the Galactic Centre Environment}, Heidelberg, Germany, October 2011.} have recently claimed the discovery of $\gamma$-ray substructure within the Bubbles, in particular two counter-propagating
jet-like features intersecting the GC and extending into both Galactic hemispheres (slightly west of north and east of south, respectively) with even harder spectra than the overall spectrum of the Bubbles.
By eye, these features seem at least partially coincident with the GCS feature identified in radio continuum and polarised microwave emission and discussed above;
in contrast to the widescale Bubble emission, they may originate in IC emission from young, primary electrons carried out of the GC system.

\subsubsection{Non-thermal evidence for an outflow}

As briefly reviewed above, the GC displays extended, diffuse TeV emission \citep{Aharonian2006} spatially correlated with the column of molecular gas over the central $\sim 1.5^\circ$ in Galactic longitude.
On even wider scales than for the TeV emission \citep[$\sim 6^\circ$ in $l$, $\sim 2^\circ$ in $b$:][]{LaRosa2005,Crocker2010a},  radio continuum observations show that the GC is a distinct source of diffuse, $\sim$GHz, non-thermal emission (of which the GC lobe forms a part).
Such emission must be due to the synchrotron losses experienced by a wide-spread population of cosmic ray electrons inhabiting the GC.

Despite the fact of this wide-spread non-thermal emission, the GC is actually significantly {\it under}luminous in both radio continuum and $\sim$TeV (and $\sim$GeV) $\gamma$-ray wavebands given the amount of star-formation currently going on there -- as we now explain.
Firstly, placing the GC on a plot of  its 60 $\mu$m vs. 1.4 GHz luminosity, one determines that radio continuum emission from this system falls one order of magnitude  (i.e., $\sim4 \sigma$) short with respect to the expectation afforded by the FIR-radio continuum correlation \citep[e.g.,][]{Condon1992}.
Equally, confronted with the theoretical expectation for the numerical scaling \citep{Thompson2007} between a star-forming galaxy or region's (non-thermal) 
$\gamma$-ray  and its FIR emission (expected were the region calorimetric to the accelerated cosmic rays), the GC is significantly in deficit.
In fact, its TeV luminosity is at the level of $\sim 1\%$ expectation (the system's GeV emission, as measured by the $Fermi$ satellite \citep{Chernyakova2010}, is at about 10\% of expectation, but substantially polluted by point sources in the field).

As we have discussed at length previously \citep{Crocker2010b,Crocker2011b}, the explanation for these non-thermal deficits is that the GC is {\it not} a calorimeter for either the cosmic ray protons or electron populations it accelerates: some sort of transport process is acting to remove the non-thermal particles quickly enough that they do not have the opportunity to lose their energy radiatively {\it in situ}.
This explanation seems reasonable -- and, indeed, given the small scale of the GC region under consideration, we are probing length scales where the FIR-RC correlation is known to break down \citep{Hughes2006} -- but the `usual' explanation that cosmic ray {\it diffusion} is removing the particles is not available to us.
This follows for the simple reason that -- given the hardness of the detected non-thermal radiations -- there is
no evidence for diffusion steepening of the emitting non-thermal particle populations (cf. the situation in the Galactic plane). 
Indeed, the steady state particle populations appear to have a spectrum -- $dN/dE \propto E^{-2.2}$ or so -- completely consistent with the expectation for that at {\it injection} (following first-order Fermi acceleration at astrophysical shocks).
Thus, if some process is acting to transport particles away -- as apparently required on the basis of the evidence described above -- this process must act without prejudice as to particle energy.
This requirement is naturally met by a large-scale outflow or wind of a few hundred km/s.
Note also that the implication of this reasoning is that the GC is losing $\sim 10^{39}$ erg/s in non-thermal particles to the outflow: this is precisely the power required to sustain the $\gamma$-ray emission from the Fermi Bubbles in steady state as identified above (and also to inflate them over the same $\sim 10^{10}$ year timescale required for the $pp$-in-saturation explanation of the $\gamma$-rays).

\section{Modelling}

Given all the forgoing, we seek to understand the non-thermal $\sim$GHz radio continuum and $\sim$TeV $\gamma$-ray emission detected from the GC region.
To this end we have developed a single-zone model of the injection, cooling, and escape of relativistic particles from the  region.
Cooling and escape processes -- controlled by the environmental conditions
as described by our parameter space -- form the steady-state, non-thermal particle populations we model; we also make 
 the theoretically and empirically-motivated assumption that cosmic ray populations are injected into the ISM by their accelerators
as power-laws in momentum.
Relevant cooling processes are hadronic ($pp$) collisions and ionizing/Coulomb collisions for high- and low-energy protons, respectively, and ionizing/Coulomb collisions, bremsstrahlung, synchrotron,  inverse Compton (IC) emission for electrons, and adiabatic losses for all particle types.
Our code also accounts for advective particle loss.
We track production of electrons and positrons (`secondary electrons') through charged meson decay following $pp$ collisions and the radiation they produce.
Relevant radiative processes are, at radio continuum (and microwave) wavelengths, synchrotron emission by electrons and, at $\gamma$-ray wavelengths, bremsstrahlung and IC, by electrons and  neutral  meson decay following $pp$ collisions for protons.

Self-consistently, the radiation from the modelled steady-state non-thermal particle populations
should reproduce the emission we detect from the GC region\footnote{See \citet{Crocker2011b} for modelled broad-band spectra of the region.}; we search over our parameter space 
with a $\chi^2$ minimisation procedure (instantiated within {\tiny MATHEMATICA}), trying to reproduce the particle populations and environmental conditions that do this.
Our modelling of the non-thermal particles -- in principle, cosmic ray protons and heavier ions and electrons, but here taken to be simply protons and electrons for simplicity -- their radiation, and the secondary particles they produce, is largely as described in \citep{Crocker2010b,Crocker2011b} with 
some extensions and modifications as described below.

\subsection{Assumptions}

One difference with our previous modelling is that -- in order to cut-down the size of the parameter space we need to search -- we  assume that the ratio of electrons to protons at injection, $\kappa_{ep}$, follows the theoretical expectation \citep{Bell1978} for momentum power laws with identical spectral index $\gamma_e = \gamma_p \equiv \gamma$ and assuming equal overall numbers of  electrons and protons accelerated into non-thermal populations (e.g., for a typical best-fit spectral index value $\gamma = 2.2$ this expectation is $\kappa_{ep}[\gamma=2.2] \simeq 0.01$ at TeV).
This simplification is justified given that we found in our previous modelling \citep{Crocker2011b} that the best-fit values of a floating $\kappa_{ep}$ correspond exactly to this theoretical expectation.

Other important assumptions of our modelling of the non-thermal particle populations are that the system is in quasi-steady-state;
this condition was previously shown by us to be empirically reasonable \citep[refer fig. 4 of][]{Crocker2011b}.
Essentially this is guaranteed if the time between injection events is the smallest relevant timescale, shorter, in particular, than particle cooling and escape timescales.
This condition is met over the parameter space of relevance with possible exceptions for the case of very high energy electrons ($>$ TeV) in strongish magnetic fields ($>100$ $\mu$G) and where, in addition, rare hypernovae ($E_{SN} \geq 10^{52}$ erg as defined below) make a significant contribution to the total, time-averaged energy budget of the region.
Practically, however, this case is not a concern given that we find $\gamma$-radiation (IC emission) from these high-energy electrons is subdominant to hadronic $\gamma$-ray emission 
over best-fit regions of the parameter space.
Note that, even if the steady state condition is not formally met, the condition is  too stringent 
as it neglects the finite timescale for which each hypernova explosion is an effective accelerator.
Indeed, it is possible \citep{Crocker2011b} -- some have even argued likely \citep{Melia2011,Amano2011} -- that the unusually turbulent ISM conditions in the ISM mean that particle acceleration occurs not (or not exclusively) in association with individual SNRs {\it per se} but rather via second order (stochastic) acceleration on diffuse ISM turbulence.
If this is the case, the region should better be regarded as a single giant and continuous accelerator (even if ultimately {\it powered} by supernova explosions);
the fact that \citet{Wommer2008} found they needed $\gtrsim 50$ point, TeV $\gamma$-ray sources to reproduce the overall rather smooth distribution of $\gamma$-ray intensity over the GC region is consistent with such a picture.

Over longer timescales, we have already listed above (\S \ref{sectn_SS}) the evidence that the star-formation/supernova process itself is in steady state in the GC, certainly over the $\sim$30 Myr we estimate \citep[following the work of][]{Mo2010} necessary  to ensure a steady supernova rate is reached for the lowest mass ($M_{ZAMS} = 8 \msun$) SN progenitors and probably for much longer.

\subsection{Extensions to previous modelling}

We have significantly extended our treatment of the non-thermal phenomenology of the GC region to properly incorporate other data covering the region
and to introduce various physically-motivated constraints.
Most importantly we explicitly require energy and mass conservation.
Other constraints/data are explained below.
One other extension to our previous modelling is that we add an extra term into our $\chi^2$ function related to the diffuse thermal X-ray luminosity of the region (see \S\ref{sctn_Xraychi2}).

\subsection{Mass flows}

As described above, the SFR in the GC is constant when averaged over sufficiently long periods of time. 
Mass conservation then implies
\begin{equation}
\dot{M}_{IN} = \dot{M}_{OUT} + SFR + \dot{M}_{CMZ} 
\end{equation}
where, given the relationship between SFR and the total surface mass density of gas \citep{Kennicutt1998}, the total gas mass of the system is invariant on average,  $\dot{M}_{CMZ} \equiv 0$.
$\dot{M}_{IN}$ represents {\it all sources} of mass falling on to the inner $\sim$200 pc (in diameter) of the Galaxy.
This 
receives a substantial contribution from  mass accreted through the plane but other sources of mass -- including  gas falling in (or back) from the halo above the GC and
donated by the  bulge's old stellar population -- may also be important.
In steady state,
the difference between $\dot{M}_{IN}$ and the SFR is the mass that must be expelled from the system on an outflow,
$\dot{M}_{OUT}$.

Note that it is the $rate$ at which stars are formed -- not the total $number$ of stars -- that is in steady-state.
Nevertheless, given that stars have finite lifetimes, the numbers of 
stars with $M_{ZAMS}$ sufficiently large that their lifetimes are less than the age of the system will be in steady state.
Given the indications adduced above that the GC system has been forming stars at an approximately constant rate over a multi-Gyr timescale,
the number of stars sufficiently massive to generate a supernova at death $M_{ZAMS} > 8 \msun (\tau\left[8 \msun \right] \simeq$ 30 Myr; e.g., \citealt{Mo2010})
is  in approximate steady state.
In fact, the number of stars down to $M_{ZAMS} \simeq 1 \msun$  is approximately constant ($\tau\left[1 \msun \right] \simeq$ 10 Gyr: \citealt{Mo2010}) given the system has been operating as long as we think it has.
Of course, the number of lower mass stellar and sub-stellar objects (if they are formed) continues to grow with time as does the mass in compact stellar remnants.

\subsection{Power}

We assume that the mechanical power delivered by core collapse supernovae
\begin{eqnarray}
\dot{E}_{OUT} &\equiv& \dot{M}_{SF} \times \left(\int^{M_{up}}_{M_{frag}} M \ \psi[M] \ dM\right)^{-1} \nn \\
&&
\times \int^{M_{up}}_{M_{down}} E_{SN}[M] \ \psi[M] \ dM
%\frac{d N}{d M} [M]
\end{eqnarray}
(where $\psi \equiv dN/dM$ is the IMF)
drives the entire system with a (generally) subdominant contribution from stellar winds \citep[of $\sim 10^{39}$ erg/s; see Appendix B4 of][]{Crocker2011b} at the $\lesssim$10\% level.
Here the integrals are over, respectively the zero-age main sequence mass of progenitors, $M_{ZAMS}$
from the minimum mass object (quite possibly sub-stellar) into which cooling gas fragments, $M_{frag}$,  and from the minimum mass (at zero age) necessary for 
a star to explode as a core collapse  supernova, $M_{down}$, to the largest stellar mass arising from the star-formation process.
Here we will assume  that i) the fragmentation mass satisfies $0.07 \leq M_{frag}/\msun \leq 1.2$ where the upper limit to $M_{frag}$ is derived from the determination \citep{Figer2004} that stars down to at least this $M_{ZAMS}$ exist in the GC;
ii)  the 
lower mass limit for a star to explode as a core collapse SN is $\sim 8 \msun$ \citep[e.g.,][]{Smartt2009}; 
and iii) that the limiting upper stellar mass is 150 $\msun$ as derived by \citet{Figer2005}.
Note that our calculations are not very sensitive to the precise value of $M_{up}$.
 
For simplicity in our modelling we ignore the sub-dominant power input from thermonuclear supernovae.
There are other potential sources of power injection into the system \citep[see appendix B6 of][and references therein]{Crocker2011b} which, however, are not particularly well constrained; amongst these we count processes associated with the supermassive black hole.
As previously stated, one of the purposes of this paper is to demonstrate how the GC system can be kept ticking over without appealing to such processes.
Note also that we expect for the GC environment that any radiation driving of the global outflow is negligible \citep[cf.][]{Thompson2009}.

\subsection{Supernova energetics}
\label{sctn_ESN}

\begin{figure}
 %\vspace{200pt}
 \epsfig{file=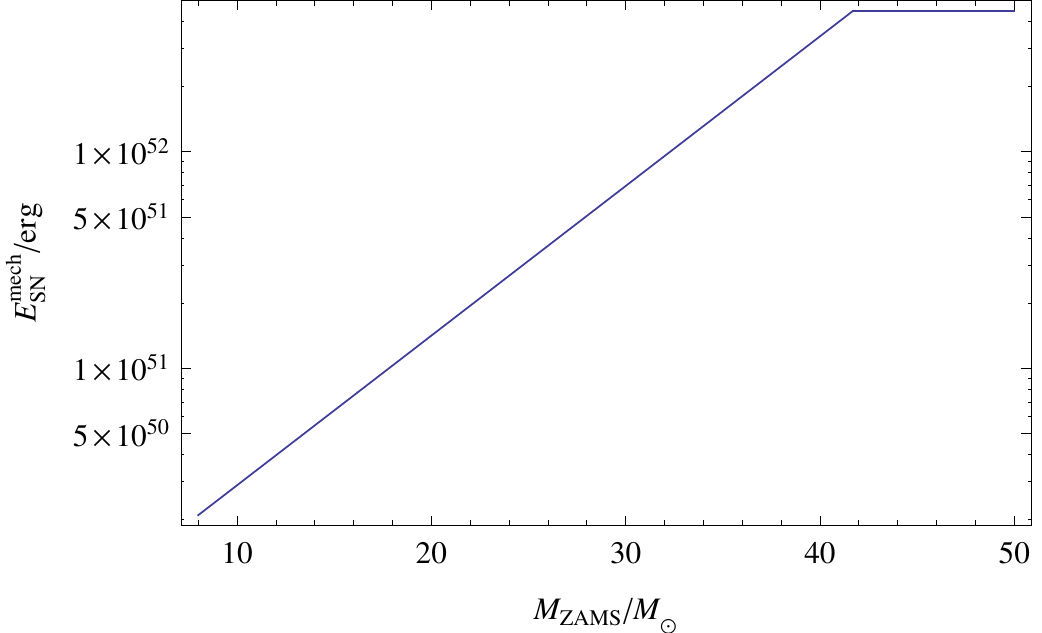,width=\columnwidth}
%{\includegraphics[width=\columnwidth]{plotRqrdWindSpeed.pdf}
\caption{A plot of the parameterization of the mechanical power released by a supernova as a function of the zero-age main sequence mass of the progenitor, $E_{SN}[M_{ZAMS}]$.
The function is a rough parameterization of the data presented by \citet{Nomoto2010}, disregarding  the `faint SN branch data points, and \citet{Utrobin2011} on SNIIP events.
Note that parameterization of SN mechanical energy tops out at $\sim 6 \times 10^{52}$ erg.
Also notice that, under the influence of the \citet{Utrobin2011} data, at low $M_{ZAMS}$ values the parameterized mechanical energy  is considerably {\it smaller} than the canonical $10^{51}$ erg.
As explained in the text, the parameterization is effectively an upper limit to the real (population-averaged) dependence of $E_{SN}$ on $M_{ZAMS}$.
}
\label{plotESN}
%} 
\end{figure}

We need a prescription in our modelling for how the mechanical energy delivered 
by each supernova, $E_{SN}$, evolves as a function of the zero-age mass of ($M_{ZAMS}$) of the progenitor:
here we will explore two limiting cases:
\begin{enumerate}
\item The `standard' assumption 
 that supernovae deliver a mechanical energy of $10^{51}$ erg invariant with respect to the zero-age mass of ($M_{ZAMS}$) of the progenitor.
\item On the other hand, given there does seem to be some evidence for growth of $E_{SN}$ with $M_{ZAMS}$ \citep[e.g.,][]{Nomoto2006,Nomoto2010,Utrobin2011}
we will also explore a parameterization of this apparent growth (shown in  fig.\ref{plotESN}).
We assume that the mechanical energy asymptotes to $6 \times 10^{52}$ erg; this is simply set by the  most energetic event (SN 2003lw) listed  by \citet{Nomoto2010} rather than being, necessarily, some fundamental physical limit.
Here (in the spirit of exploring the upper limit to the energy evolution) we ignore the possibility that the supernovae of some high-mass progenitors fizzle because of fall-back 
on to a newly-formed black hole \citep[e.g.,][]{Nomoto2006} and we also ignore the fact that the mechanical energies arrived at by \citet{Nomoto2006}
have been inferred under the assumption of spherical symmetry; the difference between the isotropic and real energies of any real SN is expected, however, to be less than a factor $\sim$2 \citep[see, e.g., supplementary material for][]{Maeda2008}.
With the energy evolution parameterization we have assumed a hypernova ($E_{SN} \geq 10^{52}$ erg) requires a progenitor with $M_{ZAMS} > 26.5 \msun$.
\end{enumerate}

\subsection{Power leaving system}

We require that the power being lost from the system, $\dot{E}_{OUT}$, is matched by the power being fed into the system by core-collapse supernovae (and stellar winds), $\dot{E}_{IN}$,
\begin{eqnarray}
\dot{E}_{OUT} & \equiv & \dot{E}_{IN} \nn \\
\dot{E}_{IN} & = & \dot{E}_{SN} + \dot{E}_{SW} \nn \\
\dot{E}_{OUT} & = & \dot{E}_{therm}^{hot} + \dot{E}_{kin}^{hot} + \dot{E}_{B} + \dot{E}_{CRp} \nn \\
&& + \dot{E}_{CRe} + \dot{E}_{kin}^{cold}+  \dot{E}_{rad}
\end{eqnarray}
Here $\dot{E}_{OUT}$ is composed of the following terms (cf. fig.~\ref{plotPwrFLAT}): 
the 
i)  thermal, $ \dot{E}_{therm}^{hot} \simeq 4.1 \ k_B \ T \ \dot{M}_{hot}/m_p $, and 
ii) kinetic power of the plasma in the outflow,  $\dot{E}_{kin}^{hot} = 0.5 \ \dot{M}_{hot} \ v_{hot}^2$; 
iii) the magnetic energy advected by the outflow per second due to field lines  frozen into the plasma, $ \dot{E}_{B} \simeq E_B^{GC} \ v_{hot}/h$ where $ E_B^{GC}$ is the total magnetic energy content of the GC region and $h \simeq 42$ pc its height;  
the power being lost into freshly-accelerated, non-thermal iv) protons and v) electrons;
vi) the kinetic power of the cold, entrained gas in the outflow, $\dot{E}_{kin}^{cold} = 0.5 \ \dot{M}_{cold} \ v_{cold}^2$; and
vii) the power radiated  by the plasma while it remains in the region,  $\dot{E}_{rad} = \Lambda[T,n_e,n_H] \ V_{GC}$ where we adopt the parameterization of \citet{Raymond1976} of the plasma volumetric cooling rate, $ \Lambda$, as a function of temperature, electron and proton density.

\subsection{Velocity of Outflow}

Analytical solutions exist for determining the  speed of a wind out of a starburst region as a function of distance {\it given certain simplifying assumptions} \citep{Chevalier1992,Zirakashvili2006,Strickland2009,Veilleux2005} like absence of halo drag, radiative losses, etc.
We take a numerical approach as it best suits our purposes, in particular, allowing for the self-consistent treatment of non-thermal ISM phases (magnetic field, cosmic rays) which may themselves represent a significant sink of injected mechanical energy.
Our approach 
allows the modelled system to distribute energy between different phases without theoretical prejudice as to thermalization efficiency, cosmic ray acceleration efficiency, etc.
Our modelling self-consistently 
determines both the {\it mass loading} and amount of cold gas {\it entrainment} that occurs in the outflow
(these appear as free parameters in the analytical wind solutions; see \S\ref{sctn_loading} for the distinction between these).
The outflow speed is given implicitly within our model by coupled equations, in particular, through the fact that the timescale over which the various ISM components are advected from the GC is one parameter controlling the power leaving the system (and the system is constrained to conserve energy globally) and also that this timescale is also a controlling parameter for the steady-state non-thermal particle spectra (and, therefore, their consequent radiation).
Note that, in general, we find below that -- as it leaves the boundary of the GC system -- the outflow is significantly sub-sonic (as a result of both mass loading and entrainment and mechanical power also being fed into non-thermal ISM phases).

\subsection{X-ray Luminosity of System}
\label{sctn_Xraychi2}

We model energy and mass flows through the system and our modelling therefore gives us a handle on the steady-state plasma conditions in the GC which we can compare against data.
We do not fix or constrain the temperature of the plasma beyond the empirically-motivated requirement that it be below $10^8$ K, but we do calculate its 2-10 keV X-ray luminosity via both continuum free-free  and line emission according to our parameterization of the results of \citet{Raymond1976}.
We add a term into our $\chi^2$ function accounting for the requirement that the predicted 2-10 keV luminosity reproduce the observed \citep{Belmont2005}
$4 \times 10^{37}$ erg/s (to within a factor 2 assumed to represent the 1$\sigma$ error in this luminosity).

\subsection{Number of Wolf-Rayet Stars in System}

The measured number of WR stars in the GC is 92 \citep{Mauerhan2010,Hussmann2011}. 
Allowing for the possibility that some such stars remain to be discovered, we conservatively assume that the steady-state number of WR stars predicted by our modelling
should fall between this number and twice this number.
The steady-state number of WR stars is given by
\begin{eqnarray}
\dot{E}_{OUT} &\equiv& \dot{M}_{SF} \times \left(\int^{M_{up}}_{M_{frag}} M \ \psi[M] \ dM\right)^{-1} \nn \\
&&
\times \int_{M^{min}_{WR}}^{M_{up}} \psi[M] \ \tau_{WR} [M] \ d M
%\frac{d N}{d M} [M]
\end{eqnarray}
where $ \tau_{WR}[M_{ZAMS}] $ is the duration of the WR phase experienced by a star of given $M_{ZAMS}$ and $M^{min}_{WR}$ is the minimum mass required in order that a star experience such a phase.
Here there are considerable theoretical uncertainties influenced by whether, at the population level, the most important channel for lifetime-integrated massive stellar mass loss is
via (single star) winds, eruptions, or binary mass transfer.
This is very much a field of active research and debate \citep{Dessart2011,Modjaz2011,Meynet2011,Smith2011,Boissier2009,Smartt2009,Eldridge2008,Meynet2003};
for definiteness we shall adopt a parameterization of the results of \citet{Meynet2003} for 
the massive, single-star evolution of 
non-rotating and stars rotating initially at 300 km/s.
Rotating stars lose mass more quickly and get to spend longer in the WR phase before core collapse, so evolution with rotation predicts a larger steady-state number of WR stars for a given SFR and IMF than the no rotation case
 are require that our modelling predict.
 Within the single star evolution paradigm $M^{min}_{WR} \simeq 25 \msun$.
To bracket the theoretical uncertainties we require that $N_{WR}(no \ rot) < 2 \times 92$ and $N_{WR}(rot) > 92$.

\subsection{Other constraints}

Other constraints we enforce are as follows:

\vspace{0.3 cm}
\noindent
{\bf Free-free emission from system:} Our detailed modelling accounts for thermal bremsstrahlung (free-free) emission from individual H{\sc ii} regions and/or dispersed plasma within the GC in fitting to the higher-frequency part of the radio continuum spectrum (and self-consistently accounts for free-free {\it absorption} of radiation at lower frequencies).
Free-free emission can be used to derive an empirically-calibrated lower limit  (because of dust-absorption) to the SFR \citep[see Eq.~11 of][]{Murphy2011} in the system and we demand that the SFR and free-free satisfy this implicit constraint.

\noindent
{\bf Total infrared emission from system:} $L_{TIR}$ offers a rather robust, empirically-calibrated (but IMF-dependent) handle on the SFR in the system. 
We do not directly model the system's total infrared output but we do conservatively require that the modelled SFR (of $> 5 \msun$ stars) predict \citep{Condon1992} a $L_{TIR}$
no more than a factor of two larger than the output observed  \citep[$1.6 \times 10^{42}$ erg/s][]{Launhardt2002} from the system.

\noindent
{\bf Ejecta masses:} 
We require that the masses ejected by the SN explosions ending stars' lives (accounting for modelled mass loss over their lifetimes and conservatively assuming a 1.4 $\msun$ compact remnant irrespective of $M_{ZAMS}$) be positive.

\noindent
{\bf Explaining radio continuum from GCL:}
As introduced above, there is good evidence that non-thermal electrons injected at the plane are carried out of the immediate GC region to synchrotron-illuminate the GCL.
In fact, as discussed, radio observations \citep{LaRosa2005,Crocker2010a} reveal a distinct, extended, non-thermal radio continuum source on even larger scales around the GC ($6^\circ$ in longitudinal extent) of which the GCL non-thermal emission forms only a part.
Our previous modelling has shown that the inner 200 pc region can supply enough power ($\sim 10^{38}$ erg/s) in hard-spectrum cosmic ray electrons to explain these observations (and this is borne-out by the modelling presented below); the only question is whether the outflow transports this population on timescales short enough with respect to the electrons' loss time in the dense and highly-magnetized GC ISM.
Given  the physical plausibility of the requirement that the GC supply the GCL electrons, we therefore demand that the modelled wind be able to transport the advected electron population to $\gtrsim 100$ pc over the electrons' loss time (at the  energy corresponding to synchrotron-emission -- for the modelled magnetic field -- into the highest frequency range where emission is still manifestly non-thermal in character, viz. 10 GHz).

\subsection{Collimation, mass loading, and entrainment}
\label{sctn_loading}

Following \citet{Strickland2009}, we distinguish here and in our modelling between (centralised) mass $loading $
and mass
$entrainment$.
In the former, 
additional ISM material is heated to plasma temperatures and co-mixed with the supernova and stellar wind material in the
energetically dominant `wind fluid' \citep{Strickland2000} thereby affecting the entire flow in a global, distributed process.
In the latter, local process, a cooler and denser gas phase, stripped from ambient cold gas by the ram pressure of the wind fluid,  is 
carried along with the flow but remains a distinct phase. 
Note the entrained gas  may actually come to dominate the mass efflux.
Physically, of course, a real wind or outflow is characterised by multiple co-evolving phases and entrainment, e.g., may lead to loading but we ignore these subtleties in our modelling.
The physical picture suggested to us by the data and which we advocate is that the dense and massive, star-forming gas ring \citep{Molinari2011} collimates the wind outflow (cf. \citealt{Zubovas2011,Strickland2000,Westmoquette2011}; the inferred large-scale, poloidal field structure may also have a role in collimating the outflow; \citealt{Morris2007}).
Indeed, the radio continuum data suggest that the outflow has a projected width matching the projected radius of this ring (see fig.\ref{HESSRadioCorres}).
Conversely, ram pressure stripping of $H_2$ from the inner edge of the ring by the escaping plasma outflow
supplies gas to be loaded on to the outflow.

We quantify gas entrainment following the prescription of \citet{Martin2005} who  shows that
the ram pressure of the hot outflow, of density $\rho_{wind}$ and speed $v_{wind}$,  accelerates entrained, cold gas clouds, of density $\rho_c$,  to a terminal speed given by 
\begin{equation}
v_{term} \simeq \left[\frac{3 \rho_{wind}}{2 \rho_c}\right]^{1/2} v_{wind} \, .
\end{equation}
We assume in our modelling that the density of the cold material is that 
corresponding to the number density  which the  non-thermal particle populations are primarily sampling and within which they generate the non-thermal radiation we detect.
The entrained gas can be expected to be shock heated by the plasma wind fluid, some fraction potentially to X-ray emitting temperatures \citep[e.g.,][]{Strickland1997} and potentially converting from `entrained' to `loaded';
we leave a detailed treatment of this for future work..

\section{Results}

Given the above model and constraints we delimit the parameter space providing a good description of the GC environment using  $\chi^2$ minimisation
to fit to the diffuse, broadband emission detected from the GC.
We are fit to 6 radio data points (at 74 MHz, 1.4 GHz, 2.4 GHz, 2.7 GHz, 8.4 GHz, and 10.3 GHz), 9 HESS $\gamma$-ray points (from $2.7 \times 10^{11}$ eV to $1.3 \times 10^{13}$ eV), and one diffuse X-ray flux datum (2 to 10 keV); there are 9 fitting parameters and one constraint (energy conservation) so 8 degrees of freedom overall.
We show an example fitted broadband spectrum in Appendix \ref{sctn_BBeg}.

\begin{figure}
 %\vspace{200pt}
 \epsfig{file=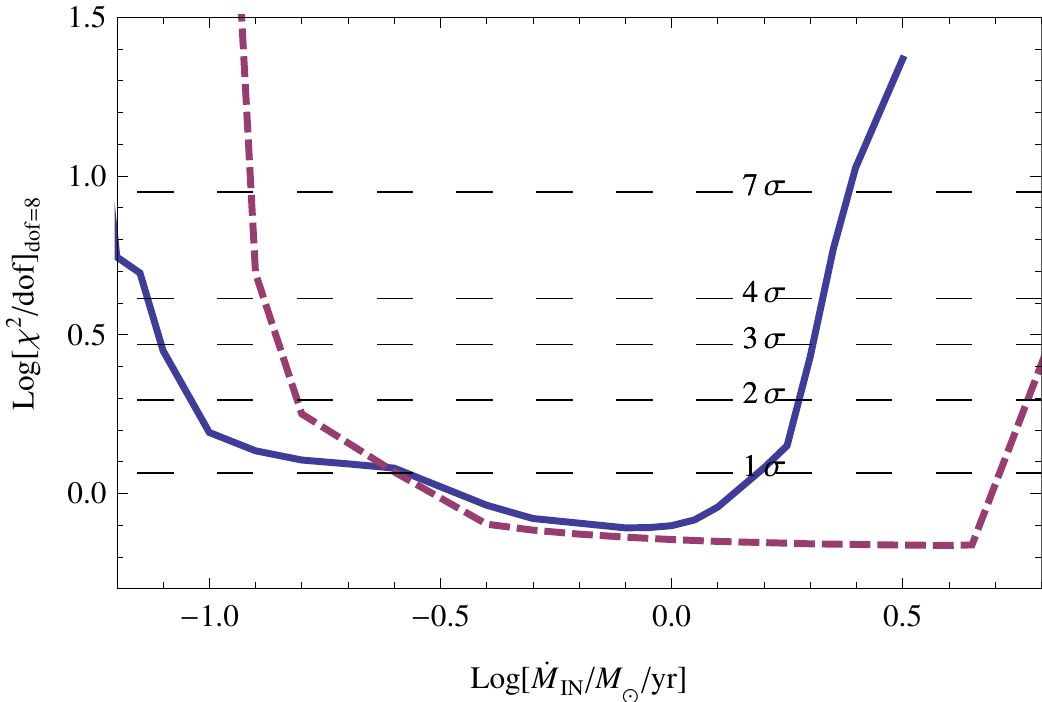,width=\columnwidth}
%{\includegraphics[width=\columnwidth]{plotRqrdWindSpeed.pdf}
\caption{$\chi^2$ as a function of the mass flux into the GC region, $\dot{M}_{IN}$, for the case of (solid,blue) $E_{SN} = 10^{51}$ erg (constant with respect to $M_{ZAMS}$) and
(dashed, purple) $E_{SN}$ an increasing function of $M_{ZAMS}$ as described in \S\ref{sctn_ESN}.
}
\label{plotChiSqrdCmprd}
%} 
\end{figure}

As discussed, we employ two prescriptions for the mechanical power delivered by core-collapse supernovae into the ISM to bracket the reasonable possibilities, viz.:
that this be invariant at $10^{51}$ erg per SN and that this be growing function of $M_{ZAMS}$ as described in \S \ref{sctn_ESN}.
In general, we find fits acceptable at 2$\sigma$ confidence for the control parameter $\dot{M}_{IN}$ in the ranges $0.1 - 2 \msun$/year and  $0.2 - 5 \msun$/year 
for these two cases
(fig.~\ref{plotChiSqrdCmprd}).

\subsection{Star formation rate}

\begin{figure}
 %\vspace{200pt}
 \epsfig{file=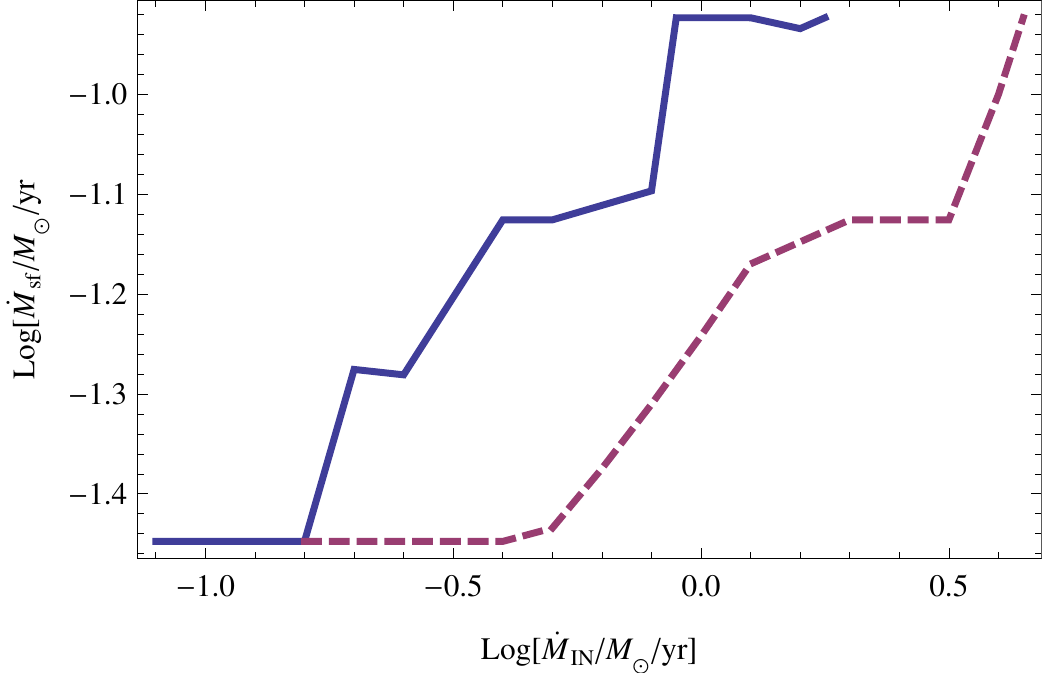,width=\columnwidth}
%{\includegraphics[width=\columnwidth]{plotRqrdWindSpeed.pdf}
\caption{Star formation rate as a function of the mass flux into the GC region, $\dot{M}_{IN}$, for the case of (solid,blue) $E_{SN} = 10^{51}$ erg (constant with respect to $M_{ZAMS}$) and
(dashed, purple) $E_{SN}$ an increasing function of $M_{ZAMS}$ as described in \S \ref{sctn_ESN}.
{\it In this figure and the following we show  modelled parameters only over  the range of  $\dot{M}_{IN}$ where they are acceptable at $2\sigma$-level.}
}
\label{plotSFR}
%} 
\end{figure}

On the basis of our constrained modelling, we determine a SFR in the system that, at $2\sigma$-level, lies in the range 0.04--0.12 $\msun/$year (fig.~\ref{plotSFR}) with best-fit values of 0.08 and 0.12 $\msun/$year for the cases, respectively,
 of $E_{SN} = 10^{51}$  and $E_{SN}[M_{ZAMS}]$.
 These values agree well with the previous, independent determinations set out in \S\ref{sctn_sfrGC}.

\begin{figure}
 %\vspace{200pt}
 \epsfig{file=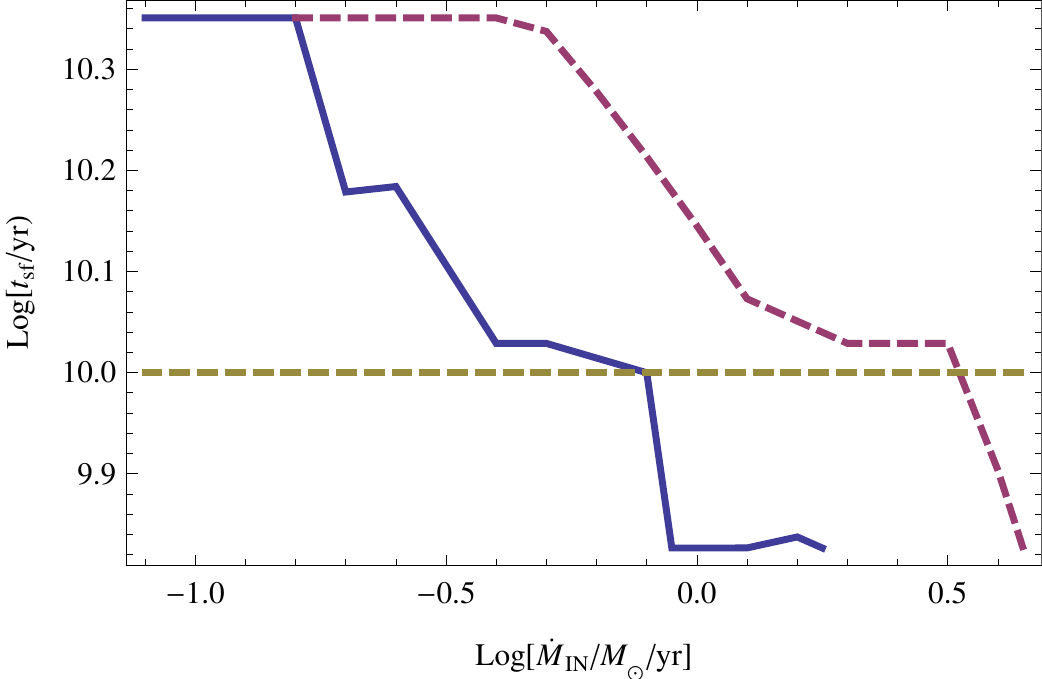,width=\columnwidth}
%{\includegraphics[width=\columnwidth]{plotRqrdWindSpeed.pdf}
\caption{The total time required for the GC to supply the $\sim8 \times 10^8 \msun$ stellar population \citep{Launhardt2002} of the $R < 120$ pc region of the Nuclear Bulge  for the case of (solid,blue) $E_{SN} = 10^{51}$ erg (constant with respect to $M_{ZAMS}$) and
(dashed, purple) $E_{SN}$ an increasing function of $M_{ZAMS}$ as described in \S\ref{sctn_ESN}.
The horizontal, dotted, yellow line shows the nominal, 10 Gyr age of the system.
}
\label{plotTimeToFormNB}
%} 
\end{figure}

\subsubsection{Inferred age of system}

Fig.~\ref{plotTimeToFormNB} displays the time required by the system to form the $\sim8 \times 10^8 \msun$ stellar population \citep{Launhardt2002} of the $R < 120$ pc region of the Nuclear Bulge being modelled.
Given an age to the system of $\lesssim$10 Gyr, this indicates minimum values for $\dot{M}_{IN}$ of $\sim 0.4$ and $2 \msun/$year for the cases, 
respectively, of $E_{SN} = 10^{51}$  and $E_{SN}[M_{ZAMS}]$.
Given we also find  upper limits (at the $2\sigma$ level) for the SFR
(corresponding to the largest plotted values of $\dot{M}_{IN}$ for both $\dot{M}_{sf}$ curves) corresponding to a formation time of $\sim$6 Gyr, this latter constitutes the {\it minimum} age of the system within our scenario.

\subsection{Modelled Power}

\begin{figure}
 %\vspace{200pt}
 \epsfig{file=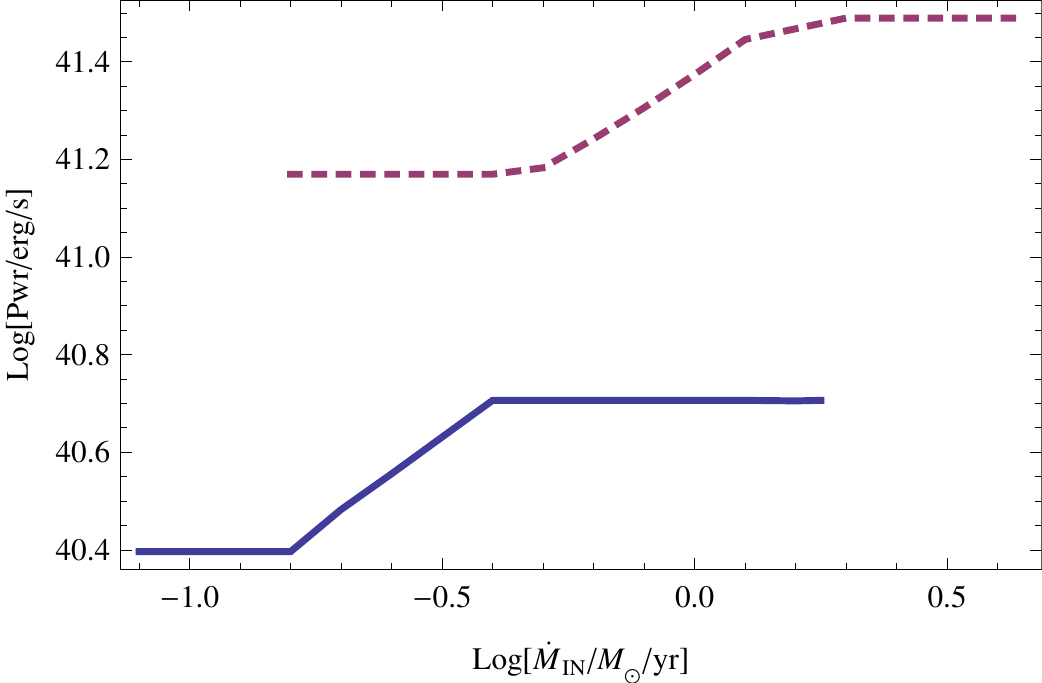,width=\columnwidth}
%{\includegraphics[width=\columnwidth]{plotRqrdWindSpeed.pdf}
\caption{$\dot{E}_{IN}$: Total power being delivered into the GC by supernovae and, sub-dominantly, stellar winds  for the case of (solid,blue) $E_{SN} = 10^{51}$ erg (constant with respect to $M_{ZAMS}$) and
(dashed, purple) $E_{SN}$ an increasing function of $M_{ZAMS}$ as described in \S\ref{sctn_ESN}.
}
\label{plotTotPwr}
%} 
\end{figure}

Our modelling (fig.~\ref{plotTotPwr}) reveals that the inner 200 pc region requires an input power of $\gtrsim 2 \times 10^{40}$ erg/s to sustain its non-thermal radiation and, more importantly energetically, the inferred outflows of non-thermal and thermal ISM phases.
One important implication of this  is that the finding by \citet{Yasuda2008} that the current level of star formation in the GC is low (with the total FIR being dominated by old K and M giants) seems difficult to sustain (i.e., our results are consistent with the SFR inferred were young, massive stars to dominate the system's radiative output as is the  case for star-burst-like environments and more generally; e.g. \citealt{Thompson2006}).

\begin{figure}
 %\vspace{200pt}
 \epsfig{file=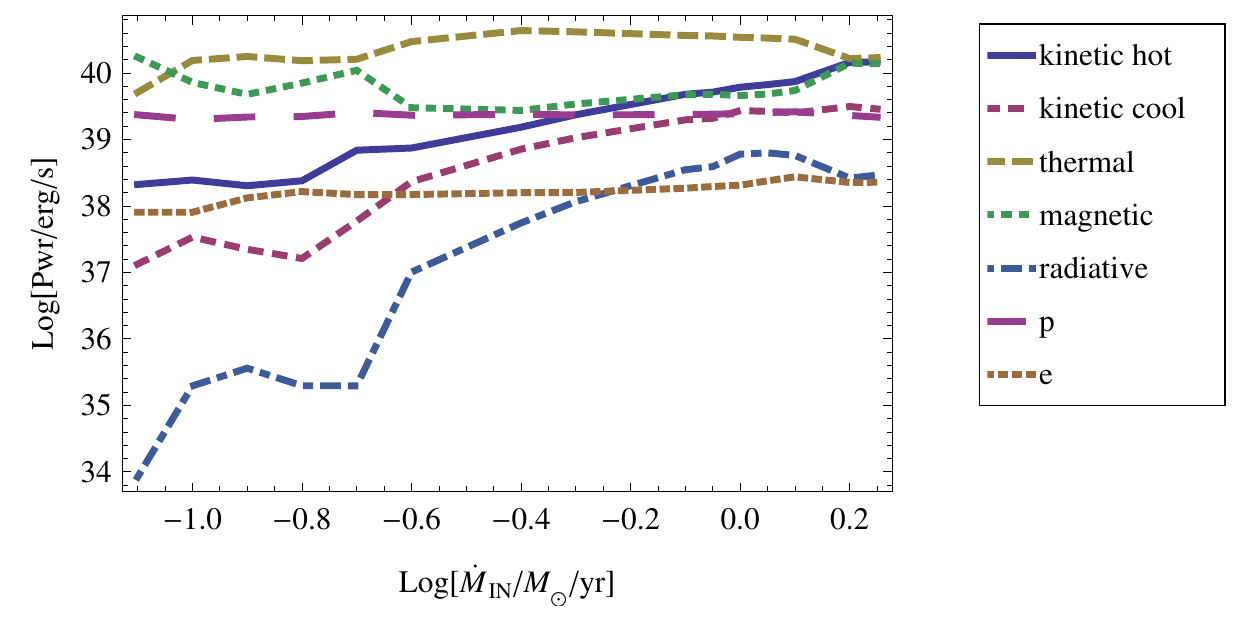,width=\columnwidth}
%{\includegraphics[width=\columnwidth]{plotRqrdWindSpeed.pdf}
\caption{$\dot{E}_{OUT}$: Power being delivered into various phases for the case of $E_{SN} = 10^{51}$ erg (constant with respect to $M_{ZAMS}$) as a function of the mass flux into the GC region, $\dot{M}_{IN}$. 
The curves, as denoted in the legend, are for 
i) the kinetic power of the hot plasma in the outflow, 
ii) the kinetic power of the cool gas entrained by the outflow, 
iii) the thermal power advected by the hot plasma in the outflow,
iv) the magnetic energy advected by the outflow per second due to field lines  frozen into the plasma,
v) plasma radiative  losses in the GC region, 
and power being delivered into the freshly-accelerated populations of non-thermal  vi) protons and vii) electrons
Note that the power going into the freshly-accelerated proton population is almost invariant at $\sim10^{39}$ erg/s over the well-fitting region of $\dot{M}_{IN}$; this is precisely the proton power required, in steady state, to sustain the $\gamma$-ray luminosity of the Fermi Bubbles.
Also note that the power going into the freshly-accelerated electron population is almost invariant at $\sim10^{38}$ erg/s over the well-fitting region of $\dot{M}_{IN}$; this is the electron power required \citep{Crocker2010a}, 
in steady state, to sustain the synchrotron luminosity of the extended, non-thermal emission detected \citep{LaRosa2005} around the GC including the GCL structure.
}
\label{plotPwrFLAT}
%} 
\end{figure}

\begin{figure}
 %\vspace{200pt}
 \epsfig{file=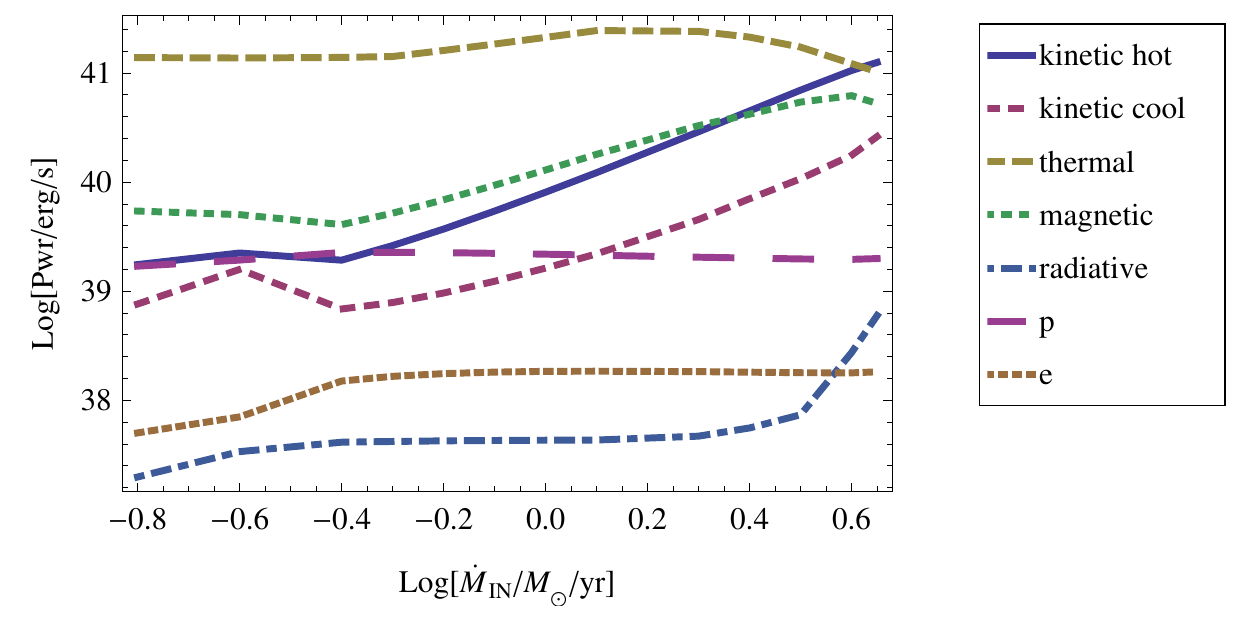,width=\columnwidth}
%{\includegraphics[width=\columnwidth]{plotRqrdWindSpeed.pdf}
\caption{$\dot{E}_{OUT}$: Power being delivered into various phases for the case of $E_{SN}[M_{ZAMS}]$.
The curves are as explained in the caption to fig.~\ref{plotPwrFLAT}
}
\label{plotPwrGROWING}
%} 
\end{figure}

\begin{figure}
 %\vspace{200pt}
 \epsfig{file=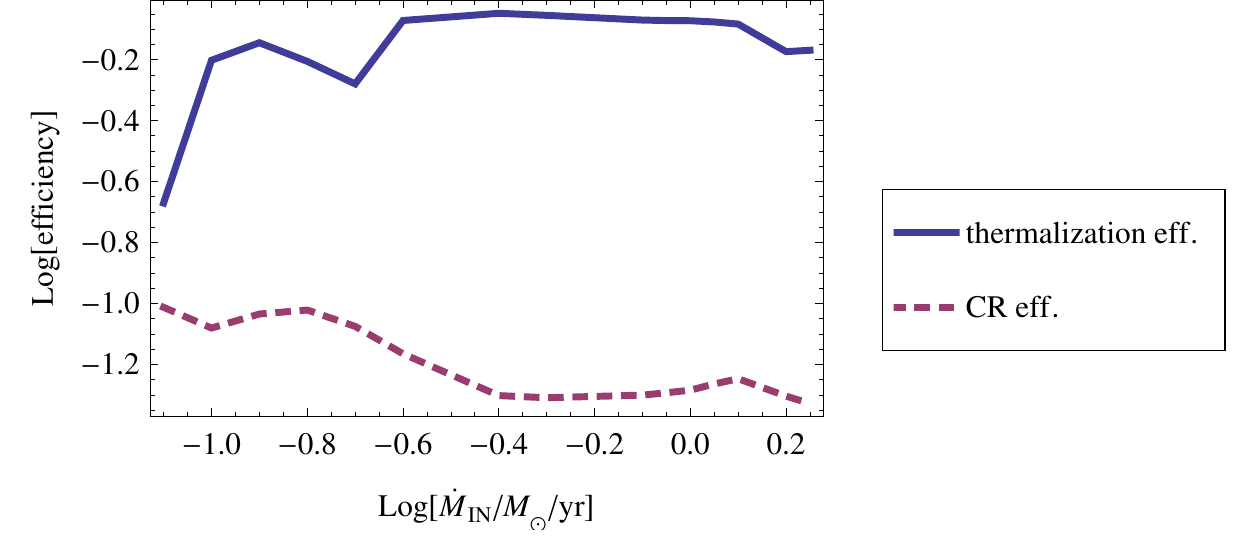,width=\columnwidth}
%{\includegraphics[width=\columnwidth]{plotRqrdWindSpeed.pdf}
\caption{Modelled thermalization efficiency and cosmic ray acceleration for the case $E_{SN} = 10^{51}$ erg.
}
\label{plotEfficiencyFLAT}
%} 
\end{figure}

\begin{figure}
 %\vspace{200pt}
 \epsfig{file=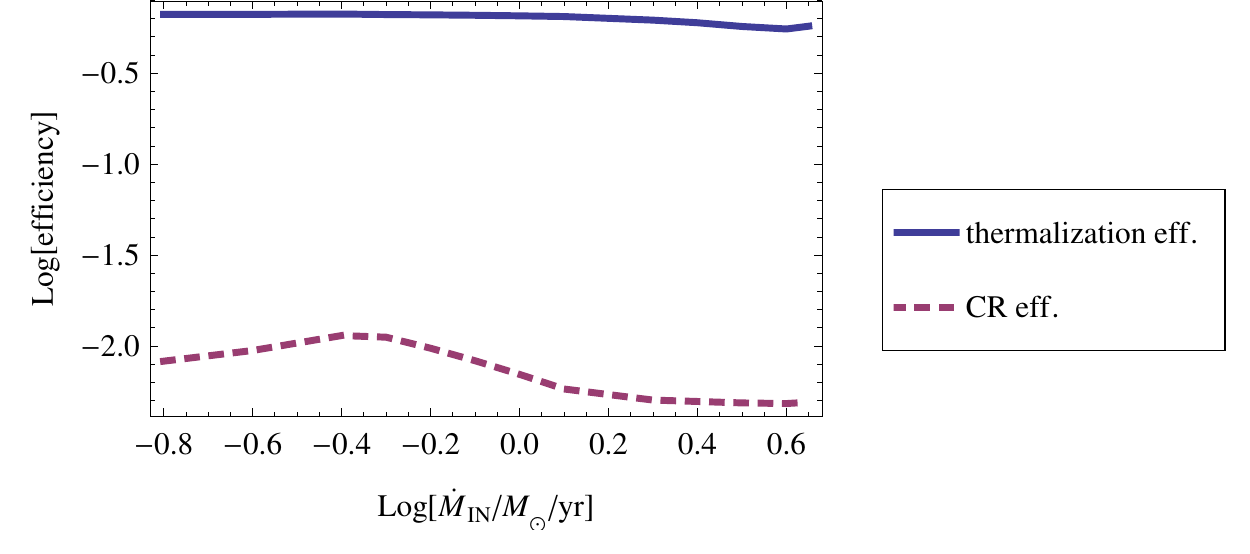,width=\columnwidth}
%{\includegraphics[width=\columnwidth]{plotRqrdWindSpeed.pdf}
\caption{Modelled thermalization efficiency and cosmic ray acceleration efficiency for the case $E_{SN}[M_{ZAMS}]$.
}
\label{plotEfficiencyGROWING}
%} 
\end{figure}

Our modelling also allows us to determine the efficiency with which the mechanical energy injected by supernovae is, on the one hand, converted into heating or moving the ISM (their `thermalization efficiency') and, on the other, converted into freshly-accelerated, non-thermal particles.
In common with some analyses of the dynamics of winds driven out of star-forming   nuclear regions of external galaxies \citep[e.g.,][]{Strickland2009} we find a high thermalization efficiency for the GC system for both the $E_{SN} = 10^{51}$ erg and $E_{SN}[M_{ZAMS}]$ cases.

With respect to the efficiency of the system as a cosmic ray accelerator, we can say that, were the  $E_{SN} = 10^{51}$ erg prescription correct, the system would be typically or close to typically \citep[e.g.,][]{Hillas2005} efficient, losing 5-10\% of power into these non-thermal particles.
Were the $E_{SN}[M_{ZAMS}]$ prescription correct, a rather low cosmic ray acceleration efficiency of $\lesssim 1 \%$ would be implied.

\subsubsection{Advected magnetic field}

Figs.~\ref{plotPwrFLAT} and \ref{plotPwrGROWING} make clear that  magnetic energy losses in the form of field-lines frozen into the advected plasma 
contribute substantially to the total energy budget of the system\footnote{In principle, the calculated energy loss into the magnetic field disregards the fact that the vertical outflow will {\it not} advect the poloidal magnetic field component and it therefore represents a {\it de facto} upper limit on the true losses. Giiven, however, that the outflow is entraining cooler gas and frozen-in field lines and -- for the dense gas -- the field lines have been sheared into a toroidal configuration consistent with polarimetry measurements \citep{Chuss2003} we expect the real magnetic energy loss rate to be close to that displayed. We thank Mark Morris for bringing up this point.}.
The magnetic energy losses are, in fact, larger than or comparable to the kinetic power of the outflow indicating that the magnetic field is an important determinant in the dynamics of the outflow \citep[e.g.][]{Beck1996}.
This is consistent with, on $\sim$ degree scales, the complex but ordered
phenomenology of the outflowing dust filaments
revealed by  MIR  observations (as briefly reviewed above). 
On tens of degree scales, magnetic field must equally play an important part in governing 
the evolution of the Fermi Bubbles as revealed by  the detection of polarised microwave radiation from the GCS and part of the edge of the Northern Bubble \citep{Jones2011}.
We will consider the evolution of the magnetic field -- governed by reconnection/relaxation and adiabatic losses -- between the scales of the GC region and the full size of the Bubbles  more fully below.

\subsection{Mass Outflows}

A first important point here is that, although the entrained, cooler gas is moving more slowly than the hot outflow, it represents a larger mass flux (fig.~\ref{plotMassEntrainment}). 
We deal with these different phases separately below.

\begin{figure}
 %\vspace{200pt}
 \epsfig{file=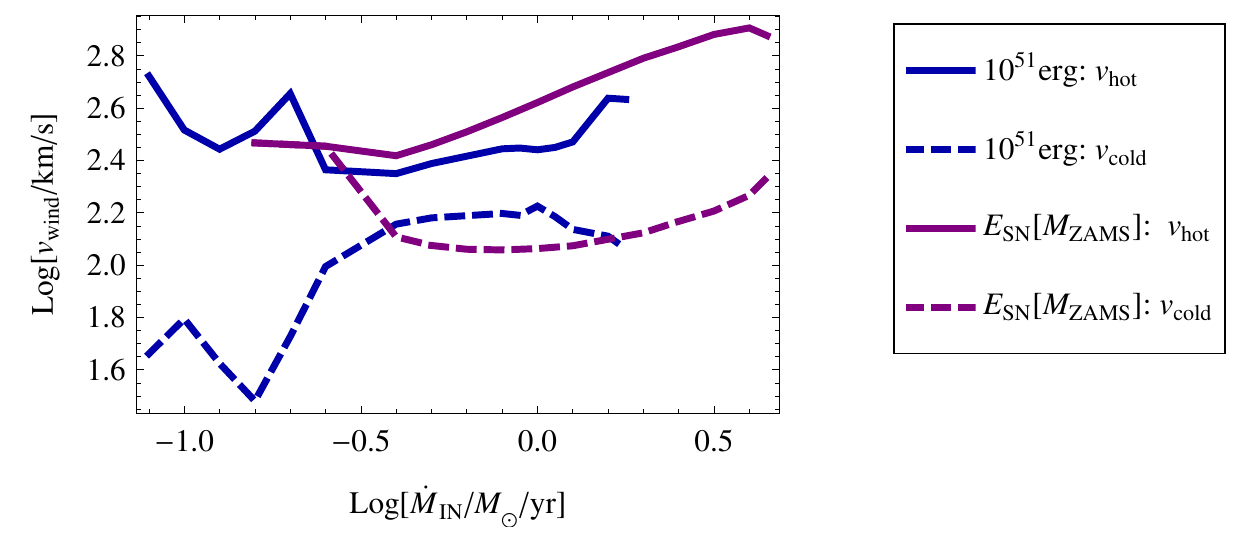,width=\columnwidth}
%{\includegraphics[width=\columnwidth]{plotRqrdWindSpeed.pdf}
\caption{Speed of hot (solid) and cold (dashed) outflows for (blue) the  case  of  $E_{SN} = 10^{51}$ erg; and (purple) 
$E_{SN}$ an increasing function of $M_{ZAMS}$ as described in \S\ref{sctn_ESN}
}
\label{plotOutflowSpeeds}
%} 
\end{figure}

\begin{figure}
 %\vspace{200pt}
 \epsfig{file=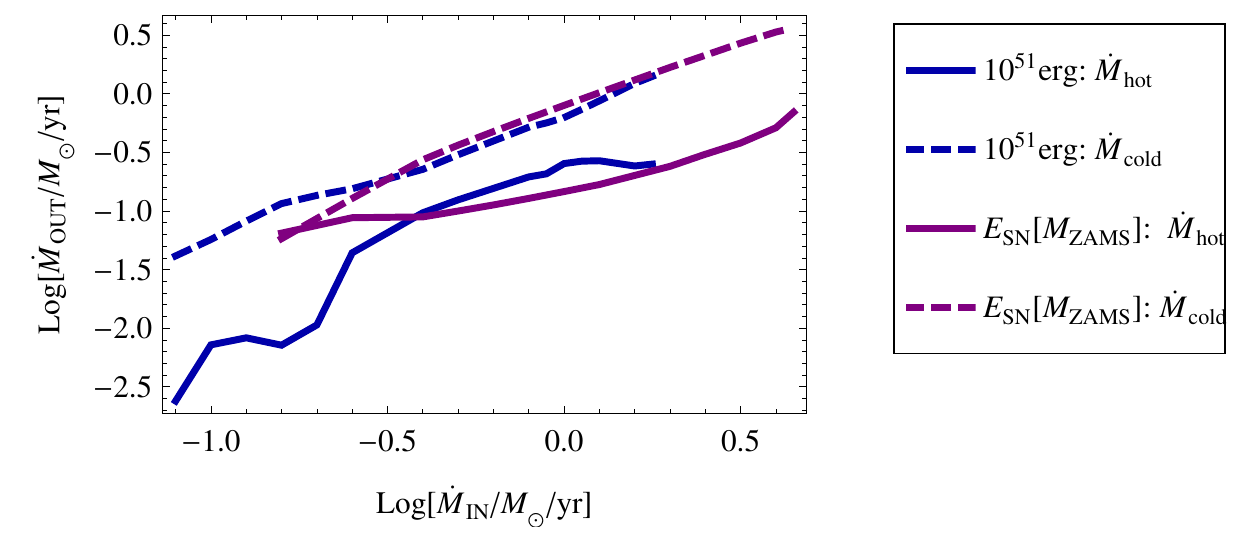,width=\columnwidth}
%{\includegraphics[width=\columnwidth]{plotRqrdWindSpeed.pdf}
\caption{Mass flux in hot (solid) and cold (dashed) outflows for (blue) the  case  of  $E_{SN} = 10^{51}$ erg; and (purple) 
$E_{SN}$ an increasing function of $M_{ZAMS}$ as described in \S\ref{sctn_ESN}
}
\label{plotMassFlows}
%} 
\end{figure}

\subsubsection{Hot Outflow}

\begin{figure}
 %\vspace{200pt}
 \epsfig{file=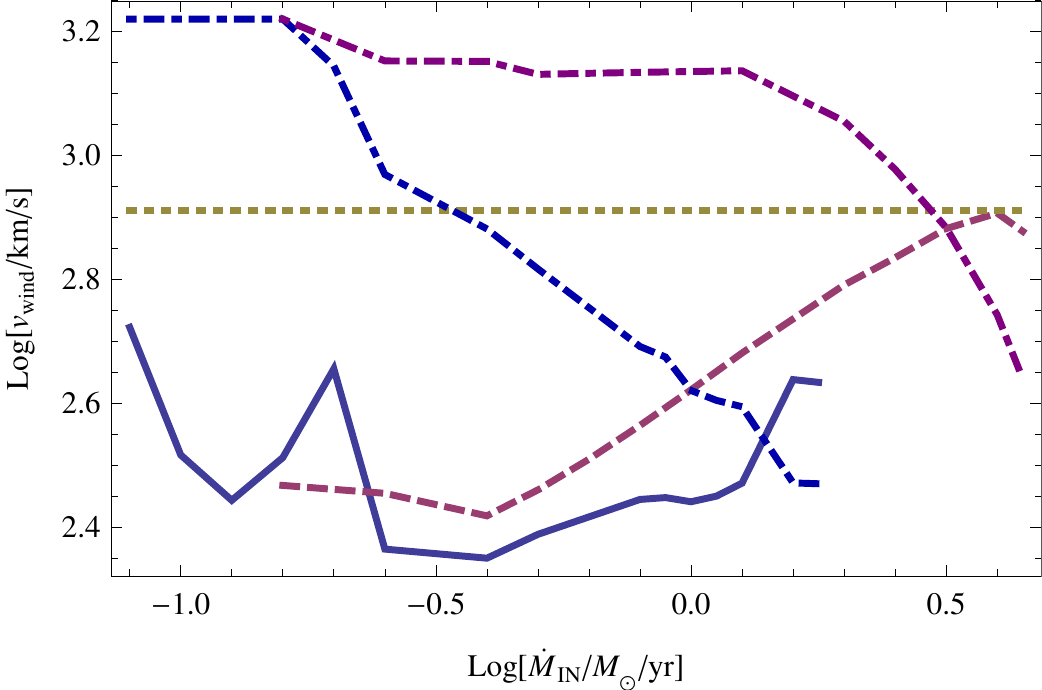,width=\columnwidth}
%{\includegraphics[width=\columnwidth]{plotRqrdWindSpeed.pdf}
\caption{The speed of the hot GC outflow for the case of (solid,blue) $E_{SN} = 10^{51}$ erg (constant with respect to $M_{ZAMS}$) and
(dashed, purple) $E_{SN}$ an increasing function of $M_{ZAMS}$ as described in \S\ref{sctn_ESN}.
The dotted, yellow horizontal line shows the escape velocity for material launched from a height of 42 pc given our assumed gravitational potential.
The blue and purple dot-dashed lines show the sound speeds in the plasma given its modelled temperature for, respectively, the $E_{SN} = 10^{51}$ erg and $E_{SN}[M_{ZAMS}]$ cases.
The long dashed curves likewise show the Aflv{\' e}n speed for the same two cases.
Note that the outflow has only just reached its Alv{\' e}nic point on the boundary of the modelled region for favoured values of $\dot{M}_{IN}$; inside this region, the magnetic field will rotate rigidly, consistent with the phenomenology of the region's non-thermal filaments \citep{Yusef-Zadeh1987,Morris1989}.
}
\label{plotWindSpeed}
%} 
\end{figure}

\begin{figure}
 %\vspace{200pt}
 \epsfig{file=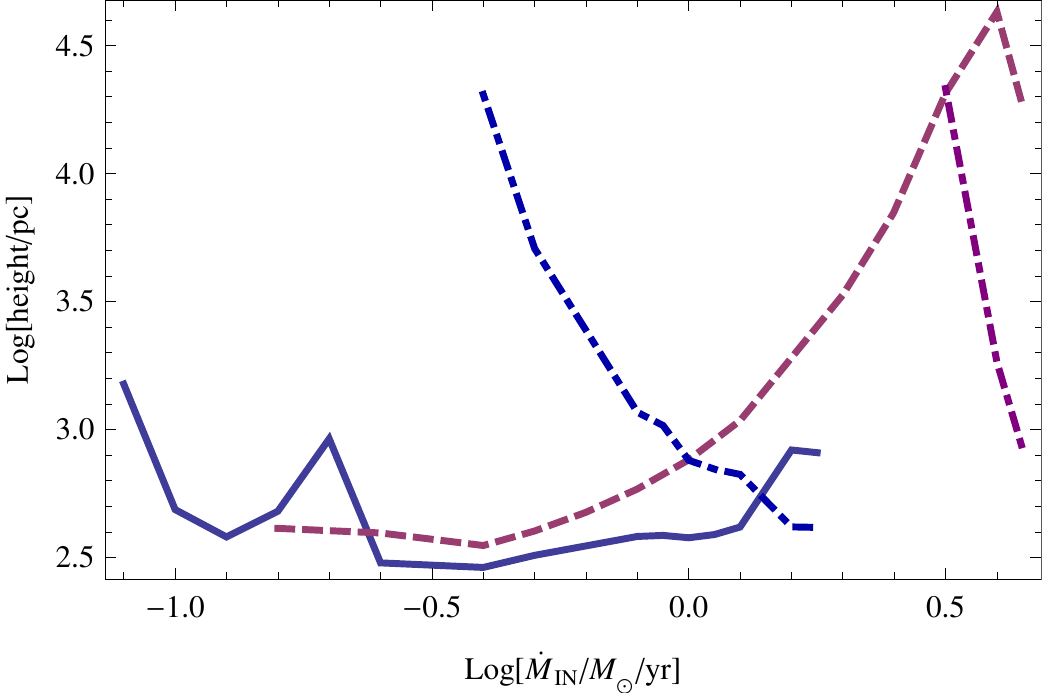,width=\columnwidth}
%{\includegraphics[width=\columnwidth]{plotRqrdWindSpeed.pdf}
\caption{The height attained by the hot material in the outflow assuming it is moving ballistically after crossing the boundary of the region.
This is unlikely to be true: the pressure represented by thermal and non-thermal components will likely further accelerate the outflow and  it will therefore reach greater heights than plotted, which therefore represent lower limits.
A rough upper limit to the height of the outflow is presented for the case that 
the outflowing material is assumed to cross the boundary at the sound speed (in our assumed gravitational potential the escape speed is $\sim 810$ km/s at 42 pc height; if the upper limit on the outflow height is not plotted the sound speed is above $ 810$ km/s).
The case of  $E_{SN} = 10^{51}$ erg (constant with respect to $M_{ZAMS}$) is shown in blue for (solid) the minimum height and (dot-dashed) the maximum height;
the case of  $E_{SN}$ an increasing function of $M_{ZAMS}$ as described in \S\ref{sctn_ESN} is shown in purple for (dashed) the minimum height and (dot-dashed) the maximum height.
}
\label{plotOutflowHeight}
%} 
\end{figure}

\begin{figure}
 %\vspace{200pt}
 \epsfig{file=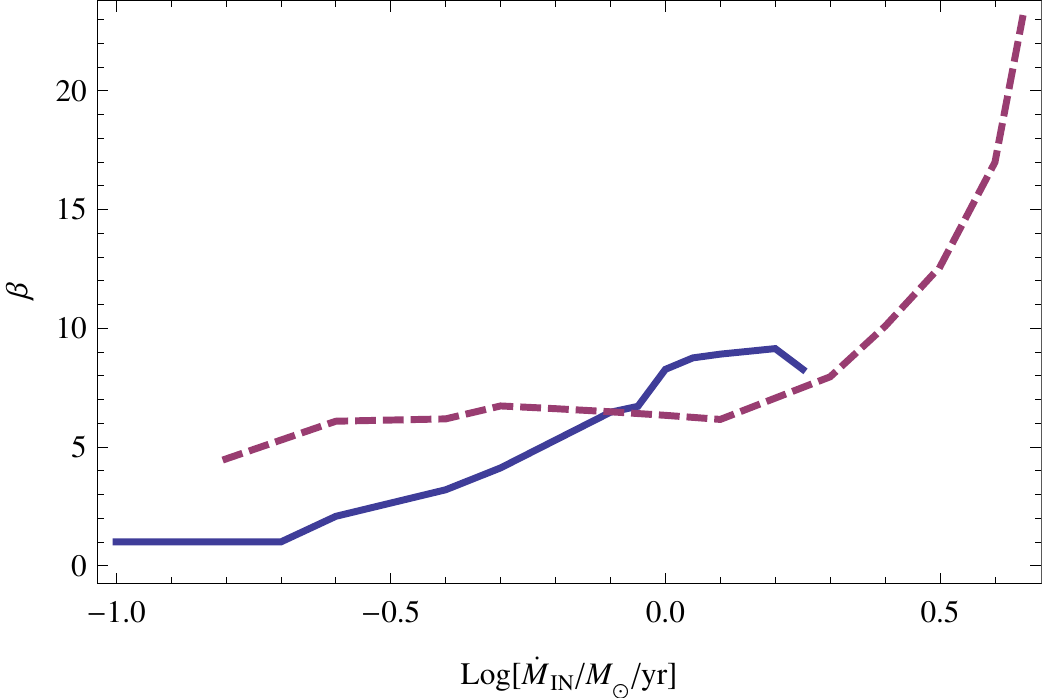,width=\columnwidth}
%{\includegraphics[width=\columnwidth]{plotRqrdWindSpeed.pdf}
\caption{Inferred amount of centralised 
mass loading  for the case of (solid,blue) $E_{SN} = 10^{51}$ erg (constant with respect to $M_{ZAMS}$) and
(dashed, purple) $E_{SN}$ an increasing function of $M_{ZAMS}$ as described in \S\ref{sctn_ESN}. 
}
\label{plotMassLoading}
%} 
\end{figure}

We calculate the height that the escaping hot material might reach under the assumption that it is moving ballistically after crossing the boundary of the region.
We assume that the strong gravitational potential of the region is as given by the parameterization of \citet{Breitschwerdt1991} for the case of vanishing Galactocentric radius.
This gives an escape speed of $\sim 810$ km/s for material launched from $z = 42$ pc  (the half-height of our region).

Given the strong gravitational deceleration, ballistically-moving hot material only reaches heights of 300 -- 1000 pc for the case of 
$E_{SN}$ = const, with considerably higher ranges predicted for the $E_{SN}[M_{ZAMS}]$ case.
This, however, is likely to be an underestimate of the true height the material reaches: the pressure gradient represented by thermal and non-thermal components will likely further accelerate the outflow after it passes the region's boundary.
Here a rough upper limit -- accounting for this effect -- is given by the assumption that the material moves at the sound speed: this would imply that the outflow is not gravitationally bound for $\dot{M}_{IN} <  0.4 \msun$/year and $ 3 \msun$/year for the cases, respectively, of 
$E_{SN}$ = const and $E_{SN}[M_{ZAMS}]$; however, these low ranges of $\dot{M}_{IN}$ are disfavoured by the consideration that they result in SFRs too small to fill-out the Nuclear Bulge stellar population over the age of the Galaxy as already discussed (cf. fig.~\ref{plotTimeToFormNB}). 
Note that for the favoured $\dot{M}_{IN}$ range and the $E_{SN} = 10^{51}$ erg case the outflow velocity at the region's boundary and the sound speed differ by less than a factor of 3.

A rather firm conclusion of our modelling, then, is that the material leaving the GC does not escape to infinity, i.e., the outflow is not a true wind but rather a fountain \citep{Bregman1980}.
Qualitatively, this matches the UV-absorption phenomenology introduced above (cf. \S\ref{sectn_UV}) which demonstrates the existence of highly-ionized 
 material fountaining up to heights of $12 \pm 1$ kpc \citep{Keeney2006}, a scale interesting close to the $\sim 10$ kpc heights of the Fermi Bubbles.
The launching of material to such heights requires a mechanism distinct from -- and significantly more powerful than -- the `standard' disk-halo gas connection mediated by blow-out of super bubbles around disk supernova associations \citep{deGouveiaDalPino2010} and even seems difficult to explain within our model for the $E_{SN} = 10^{51}$ erg case (as fig.~\ref{plotOutflowHeight} demonstrates).
If we {\it demand} that the global outflow shoot material directly to $\sim 10$ kpc then it would seem $E_{SN}[M_{ZAMS}]$ is preferred.

However, even if the global `super'-outflow we model is incapable of reaching these sort of distances directly,  there are two effects that may 
mean GC material still reaches these heights:
\begin{enumerate}

\item The activity of individual mini star-bursting events -- leading to the creation of the GC's super stellar clusters (of which the GC, Arches, and Quintuplet are merely the most recent examples) -- might achieve this \citep{Rodriguez-Gonzalez2009}.
The GC Spur may represent evidence for just such an outflow-within-an-outflow as previously discussed (\S\ref{sctn_GCS}).

\item The ejected low-density thermal and CR  plasma may  rise buoyantly once ejected into the bulge (\citealt{Su2010}, also cf. \citealt{Rodriguez-Gonzalez2009}).
X-ray and radio observations \citep[e.g.,][]{McNamara2005} reveal that such a mechanism certainly operates on the scale of galaxy clusters; the rather slow velocities involved would, moreover, be entirely consistent with the general expectation afforded by our previous work \citep{Crocker2011a} that the formation timescale for the Fermi Bubbles is rather long.
In this regard, we note the work of \citet{Raley2007} who modelled the buoyant ascent of bubbles formed by individual supernovae in the Galactic halo and found 
 surprising slow rise speeds of only 5-15 km/s (not too much larger than 1--2 km/s Bubble growth speed implicit in our scenario) taking into account a very high, effective drag coefficient for the bubbles in the halo plasma.

\end{enumerate}

\subsubsection{Centralised mass loading}

We can calculate the amount of centralised mass-loading, $\beta$, within our model (fig.~\ref{plotMassLoading}); this is the ratio between the total hot mass efflux (which includes  swept-up, heated ISM gas) and the directly injected hot gas originating as supernova ejecta and stellar winds.
We find $\beta \sim$ 3--10 for the $E_{SN} = 10^{51}$ erg case over the favoured region of parameter space ($\log[\dot{M}_{IN}/M_\odot/\textrm{year}] \gtrsim -0.4$) and 
$\beta \gtrsim$10 for the case of $E_{SN}[M_{ZAMS}]$.
In comparison, the recent study of \citet{Strickland2009}  finds that $\beta$   lies in the range 1-3 for the star-burst conditions in M82, with a practical upper limit, for this system at least, at $\sim 10$.
Other studies have countenanced or suggested somewhat higher values for star-burst environments (e.g., $\sim 5$ according to \citealt{Suchkov1996},  $\sim 10$ according to \citealt{Martin2005}).
Taking these independent estimates at face value, the $E_{SN}[M_{ZAMS}]$ scenario does appear to be disfavoured by the large amount of mass loading it requires.

\subsubsection{Cold Outflow}

\begin{figure}
 %\vspace{200pt}
 \epsfig{file=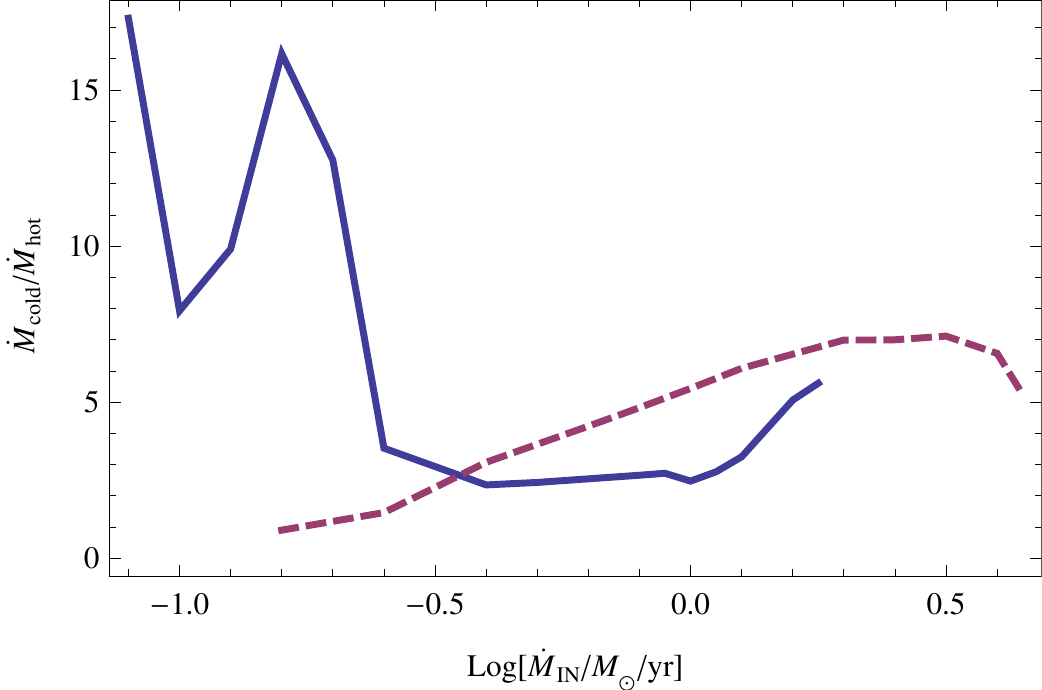,width=\columnwidth}
%{\includegraphics[width=\columnwidth]{plotRqrdWindSpeed.pdf}
\caption{The mass flux in  the cold (entrained) outflow relative to the hot outflow.
}
\label{plotMassEntrainment}
%} 
\end{figure}

We find a mass flux in the cold outflow 2--6 times larger than in the hot outflow over the favoured $\dot{M}_{IN}$ parameter space (fig~\ref{plotMassEntrainment}).
The entrained cold gas  is (relatively) slow moving and only 
reaches heights of a few $\times 100$ pc from the plane (at most).
It will, therefore,  fall back on the GC.
Thus, while the fountaining cold gas seems 
to dominate the mass flux out of the system, it also makes a large contribution to the total mass flux {\it into} the system,  ($\dot{M}_{IN}$) and the {\it net} mass flux represented by the cold gas  is small.
Still, this cold gas circulation is likely to have important dynamical effects.
Firstly, the fall of this material back on to the plane likely represent a significant source of turbulent stirring of the gas there
\citep[e.g.][]{Klessen2010}, additional to the in-situ effect of expanding H{\sc ii} bubbles, stellar winds, and supernovae consequent to the region's star 
formation\footnote{The kinetic power in the entrained gas approaches $10^{39}$ erg/s for the favoured $\dot{M}_{IN}$ range -- cf. fig~\ref{plotPwrFLAT}. 
We previously estimated the power lost into turbulence dissipation in the region at $4 \times 10^{39}$ erg/s: see Appendix B 2.5 of \citet{Crocker2011b}.}.
Secondly -- as we describe below -- the ejection of cold, high-metallicity gas into the halo and its subsequent mixing with the in-situ plasma should `catalyse' the further condensation and accretion of plasma out of the halo \citep{Marinacci2010,Marinacci2011,Binney2011}.

\begin{figure}
 %\vspace{200pt}
 \epsfig{file=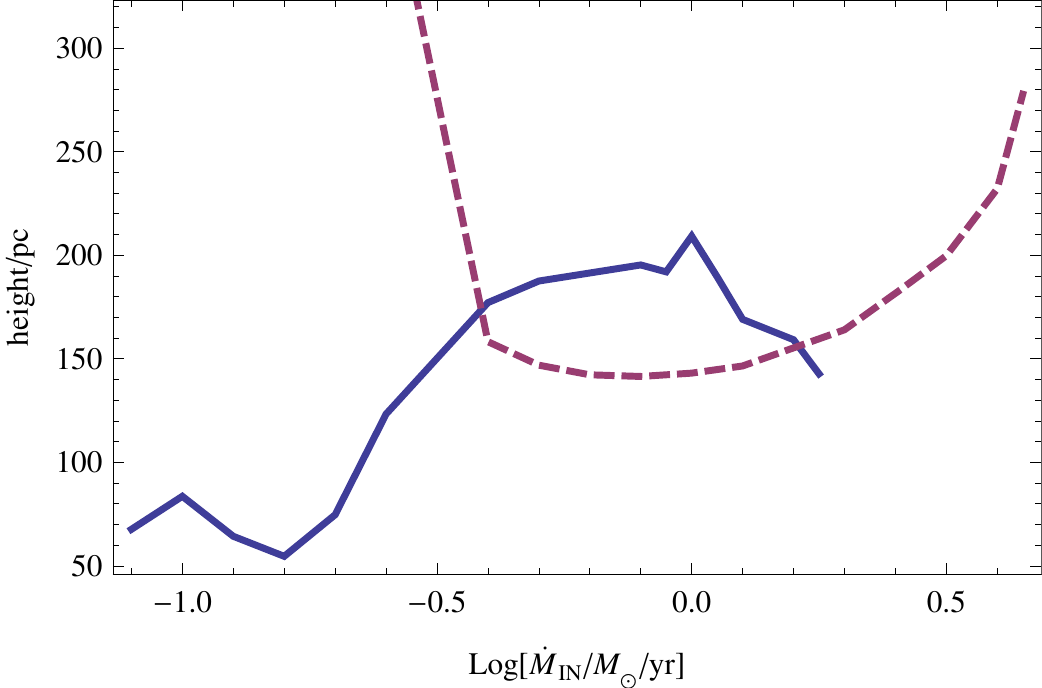,width=\columnwidth}
%{\includegraphics[width=\columnwidth]{plotRqrdWindSpeed.pdf}
\caption{The height attained by entrained, cold gas after being launched from 42 pc height were it to move a ballistic trajectory for the case of (solid,blue) $E_{SN} = 10^{51}$ erg (constant with respect to $M_{ZAMS}$) and
(dashed, purple) $E_{SN}$ an increasing function of $M_{ZAMS}$ as described in \S\ref{sctn_ESN}..
Recent Nanten-II measurements (see \S\ref{sctn_extndH2}) indicate a total mass of few $\times 10^6 \msun$ of molecular gas in an extended (100-200 pc) halo around the GC region.
}
\label{plotHeightEntrained}
%} 
\end{figure}

\begin{figure}
 %\vspace{200pt}
 \epsfig{file=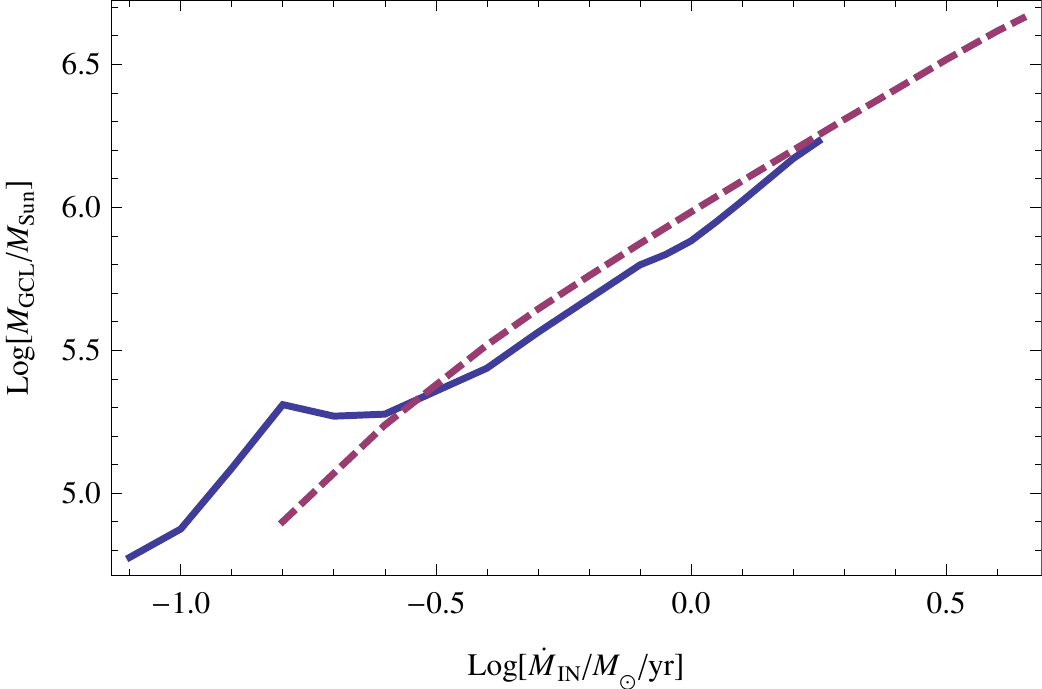,width=\columnwidth}
%{\includegraphics[width=\columnwidth]{plotRqrdWindSpeed.pdf}
\caption{The steady state mass of  cold gas which is either entrained by the outflow or falling back to the plane after reaching its maximum height  for the case of (solid,blue) $E_{SN} = 10^{51}$ erg (constant with respect to $M_{ZAMS}$) and
(dashed, purple) $E_{SN}$ an increasing function of $M_{ZAMS}$ as described in \S\ref{sctn_ESN}.
Recent Nanten-II measurements indicate a total mass of few $\times 10^6 \msun$ of molecular gas in an extended (100-200 pc) halo around the GC region 
(see \S\ref{sctn_extndH2}); this represents  independent support to the notion that $\dot{M}_{IN} \gtrsim 1 \msun$/year.
}
\label{plotGCLMass}
%} 
\end{figure}

\begin{figure}
 %\vspace{200pt}
 \epsfig{file=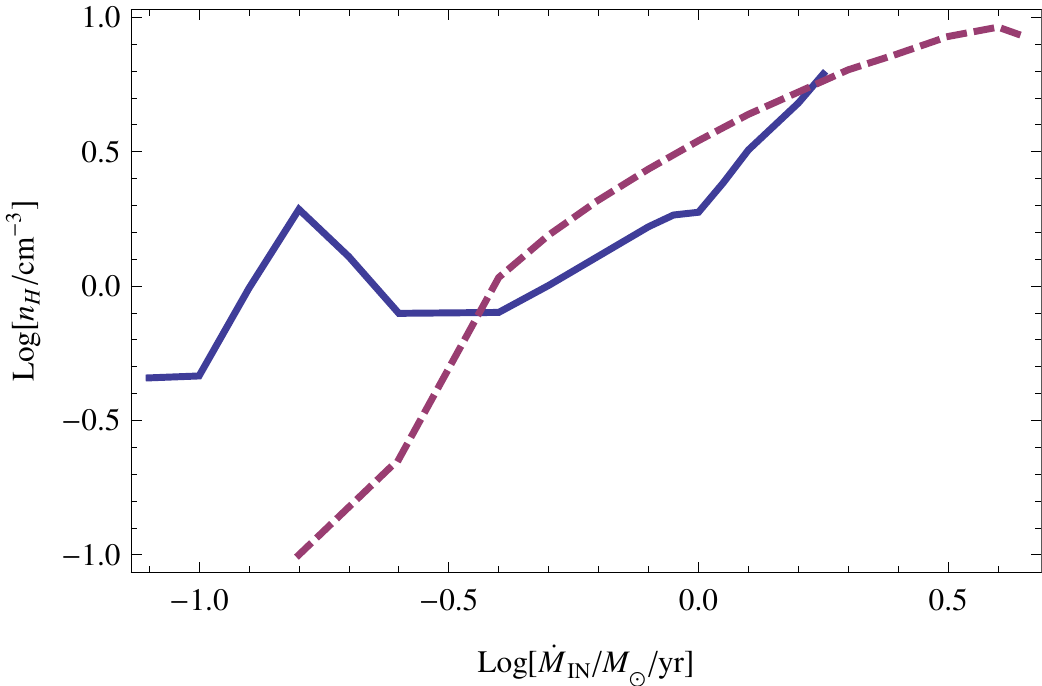,width=\columnwidth}
%{\includegraphics[width=\columnwidth]{plotRqrdWindSpeed.pdf}
\caption{Modelled density of the phase wherein the observed non-thermal radiation is generated, assumed to be identical to the  cold gas entrained by the outflow.
}
\label{plotDensityEntrained}
%} 
\end{figure}

\subsubsection{Cold gas filling factor}

\begin{figure}
 %\vspace{200pt}
 \epsfig{file=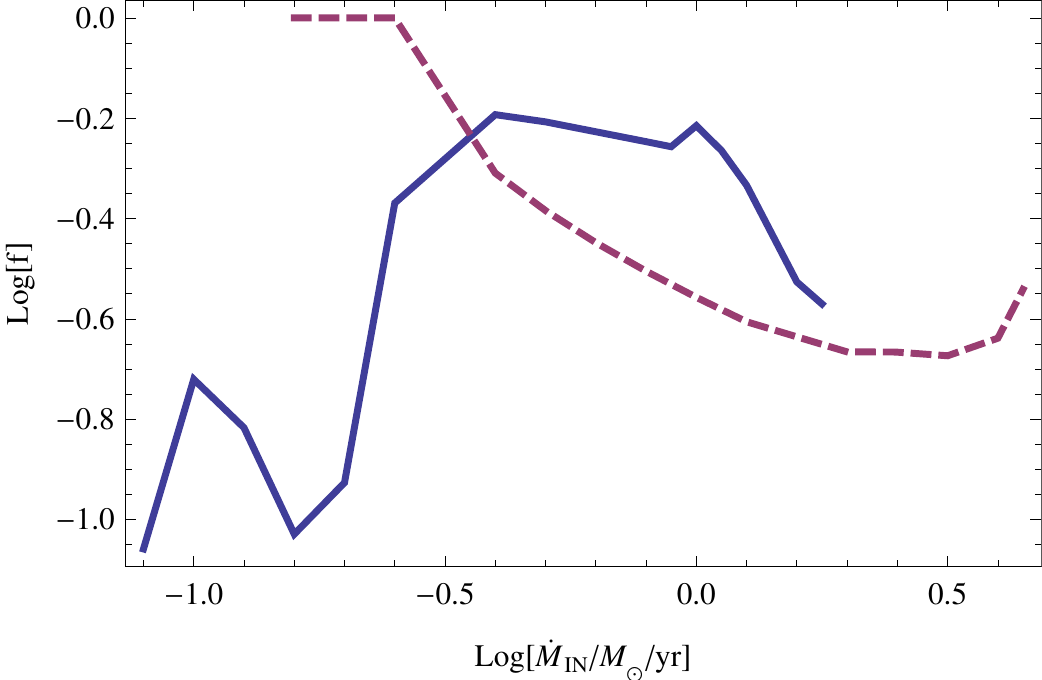,width=\columnwidth}
%{\includegraphics[width=\columnwidth]{plotRqrdWindSpeed.pdf}
\caption{Inferred filling factor of the cold gas entrained by the outflow (also assumed to be  the phase wherein the observed non-thermal radiation is generated).
}
\label{plotFillingFactor}
%} 
\end{figure}

The modelled filling factor of the entrained cold gas is shown in fig.~\ref{plotFillingFactor}.
For the case of both $E_{SN} = const$ and $E_{SN}[M_{ZAMS}]$ the  filling factor 
is rather large over some of the favoured range of the control parameter $\dot{M}_{IN}$ with respect to the expectation from star-burst winds \citep[which would suggest a range 0.1 -- 0.01; e.g.,][]{Strickland1997}.
This may indicate a break-down of the model assumption that the non-thermal emission dominantly arise in the (relatively) cold, entrained material.
Alternatively, the filling factor result may be correct and indicate a point of difference between the GC situation and true star-burst environments.

One piece of evidence tending one to the latter view is that analysis of the large  $H_3^+$ columns observed in absorption towards the region \citep{Goto2008,Yusef-Zadeh2007,Oka2005} and  different emission lines of $^{12}$CO \citep{Oka1998} and other molecules  \citep{Rodriguez-Fernandez2001}
indicates the presence of a  highly-ionized ($\gtrsim 10^{-5}$), comparatively diffuse ($\sim 100$ cm$^{-3}$) and hot ($\sim$ 250 K) molecular phase.
This `envelope' $H_2$, which represents $\sim$30\% of the total molecular gas by mass \citep{Ferriere2007}, appears to be unique within the Galaxy and has been claimed to have a high filling factor, perhaps approaching 100 \%  \citep{Goto2008}.

\subsection{Inferred Mass Accretion Rate}

As revealed above, a number of indicators come together to suggest that our control parameter, $\dot{M}_{IN}$ -- the total mass being fed into the system -- has a lower limit at around 0.4 $\msun$/year: 
i) given the star-formation rate determined by our modelling as function of $\dot{M}_{IN}$, we find that to fill out the stellar population of the inner regions of the nuclear bulge in $\sim$ 10 Gyr at least this $\dot{M}_{IN}$ is required;
ii) to explain the amount of mass in the GC lobe 
at least this $\dot{M}_{IN}$ is required,
more generally,  the recently-identified $\gtrsim 10^6 \msun$ molecular halo around the CMZ,  seems to require $\dot{M}_{IN} \gtrsim 0.4 \msun$/year;
iii) if we require that the GC sit on the \citet{Kennicutt1998}-Schmidt relation we seem to require $\dot{M}_{IN} \gtrsim 0.4 \msun$/year;
and iv) the requirement that the Bubbles be slightly positively (or even neutrally) buoyant in our scenario implies $\eta \gtrsim 1$ which is again satisfied for $\dot{M}_{IN} \sim (0.6 - 1) \msun/$year for the $E_{SN} = 10^{51}$ erg case.

We find a 2$\sigma$ upper limit on $\dot{M}_{IN}$  (for the $E_{SN} = 10^{51}$ case) at 1.8 $\msun$/year.
It is  intriguing that over the favoured $\dot{M}_{IN}$ range for the $E_{SN} = 10^{51}$ case  there are very similar energy loss rates into the kinetic power of the hot outflow, cosmic ray acceleration, and advected magnetic field (cf. fig.~\ref{plotPwrFLAT} and \ref{plotPwrGROWING}).

\begin{figure}
 %\vspace{200pt}
 \epsfig{file=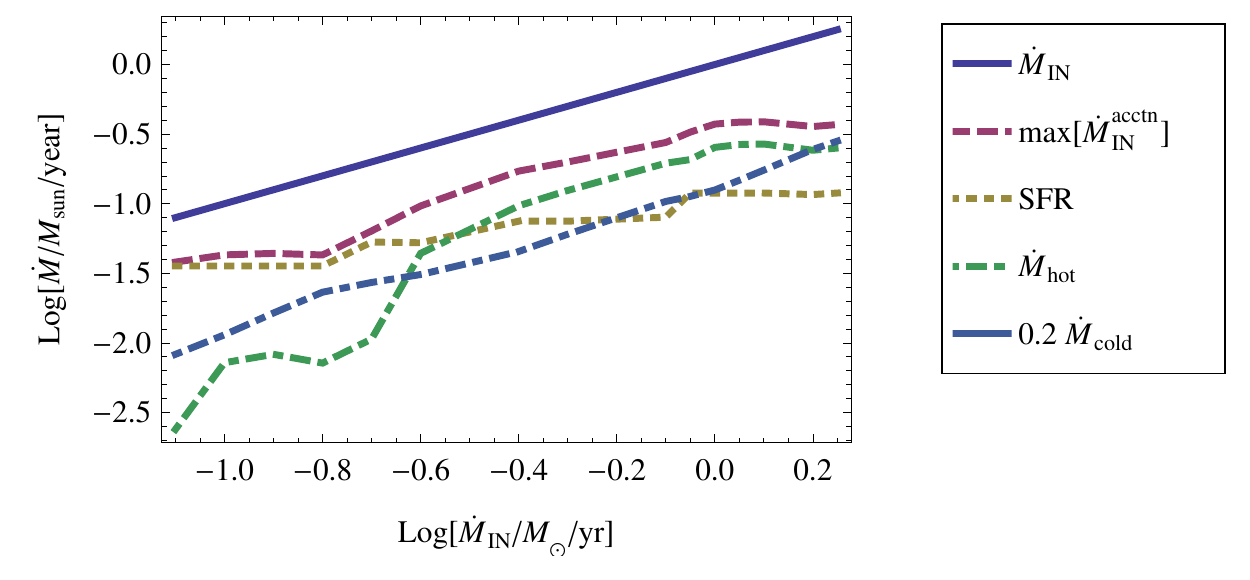,width=\columnwidth}
%{\includegraphics[width=\columnwidth]{plotRqrdWindSpeed.pdf}
\caption{Mass flows for the case of  $E_{SN} = 10^{51}$ erg (constant with respect to $M_{ZAMS}$).
Given that the cold gas is likely to fountain back on to the system, we can determine an upper limit on the true, net inflow into the system $\dot{M}_{IN}^{acctn} < SFR + \dot{M}_{hot}$.
Following the work of \citet{Marinacci2011} we also  display 20\% of the cold mass flow as a rough estimator of the scale of the net mass growth (with every cycling of the cold gas) due to the phenomenon of 
high-metalicity, outflowing H{\sc i} mixing with and subsequently `catalysing' the condensation and accretion of halo plasma.
}
\label{plotMassFlowsFLAT}
%} 
\end{figure}

\begin{figure}
 %\vspace{200pt}
 \epsfig{file=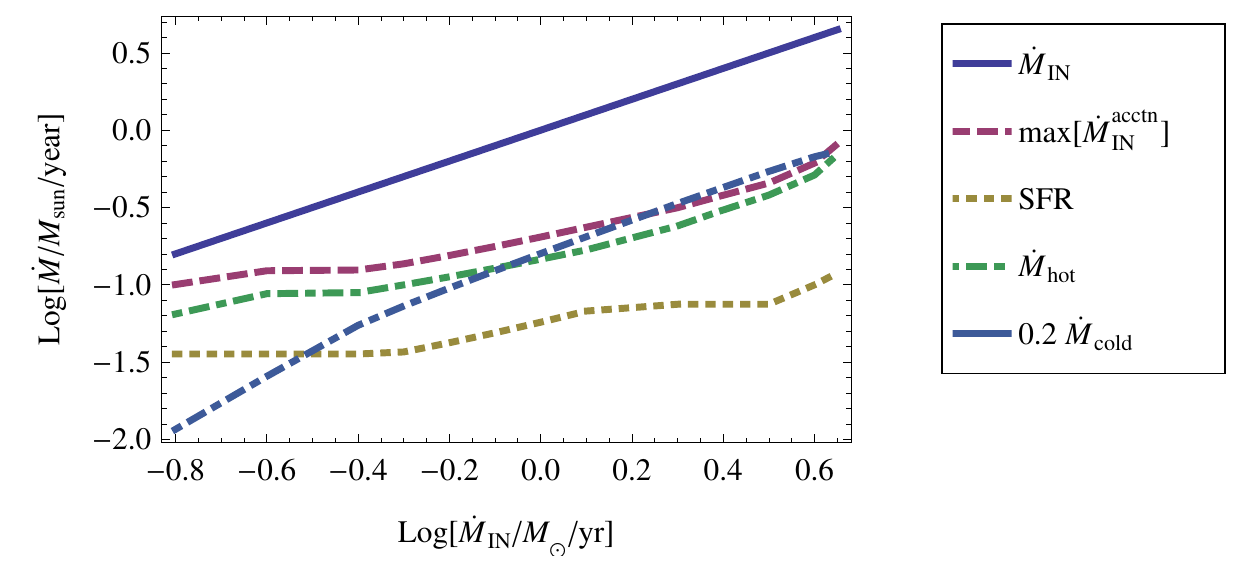,width=\columnwidth}
%{\includegraphics[width=\columnwidth]{plotRqrdWindSpeed.pdf}
\caption{Mass flows for the case of  $E_{SN}$ an increasing function of $M_{ZAMS}$ as described in \S\ref{sctn_ESN}.
}
\label{plotMassFlowsGROWING}
%} 
\end{figure}

\begin{figure}
 %\vspace{200pt}
 \epsfig{file=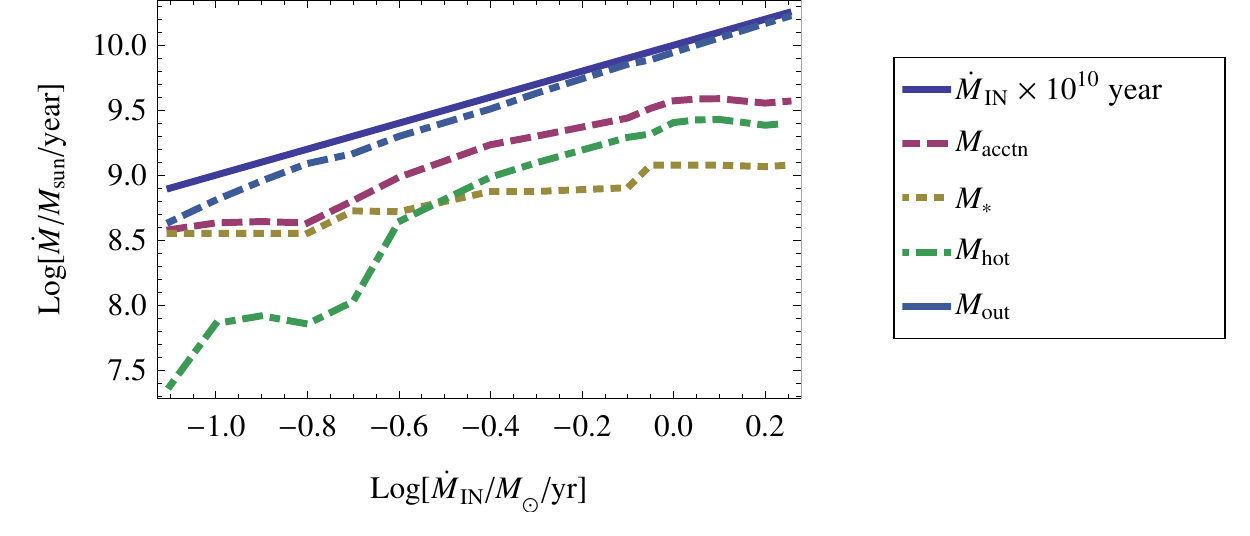,width=\columnwidth}
%{\includegraphics[width=\columnwidth]{plotRqrdWindSpeed.pdf}
\caption{Total masses assembled/processed over an assumed $10^{10}$ years and for the  $E_{SN} = 10^{51}$ erg case.
}
\label{plotTotalMassesFLAT}
%} 
\end{figure}

Our modelling plus the requirement that the GC star-formation fill out the stellar population of the Nuclear Bulge over 10 Gyr or less point to a total gas mass that has been processed through the GC over this timescale of $\gtrsim (3-10) \times 10^9 \msun$.
The mass flux out of the system is dominated by the cold, entrained material but we have already seen that this material 
fountains back on to the GC.
As this material is cycling -- rather than being truly ejected -- $\dot{M}_{IN} \equiv SFR + \dot{M}_{OUT} = SFR + \dot{M}_{hot} + \dot{M}_{cold}$  would represent an over-estimate of the mass accretion rate required to sustain the system in steady state.
This is, instead, given approximately by $\dot{M}_{IN}^{acctn} \leq SFR + \dot{M}_{hot}$.
Note this is  an still upper limit because -- given that even the hot outflow does not reach the escape speed -- 
some proportion of this plasma (given by the details of the interaction between the hot outflow and halo)  will also fountain back on to the region.
We plot the upper limit on $\dot{M}_{IN}^{acctn}$ in figs.~\ref{plotMassFlowsFLAT} and \ref{plotMassFlowsGROWING} for the two cases of $E_{SN}$ investigated: it is $\sim 0.5 \msun$/year over the favoured parameter space for both these cases.
From fig.~\ref{plotTotalMassesFLAT} we can infer that  $\gtrsim 3 \times 10^9 \msun$ of gas must be accreted on to the GC over 10 Gyr.
We discuss below the implications of this.

\subsection{GC ISM conditions}

\subsubsection{Magnetic field}

\begin{figure}
 %\vspace{200pt}
 \epsfig{file=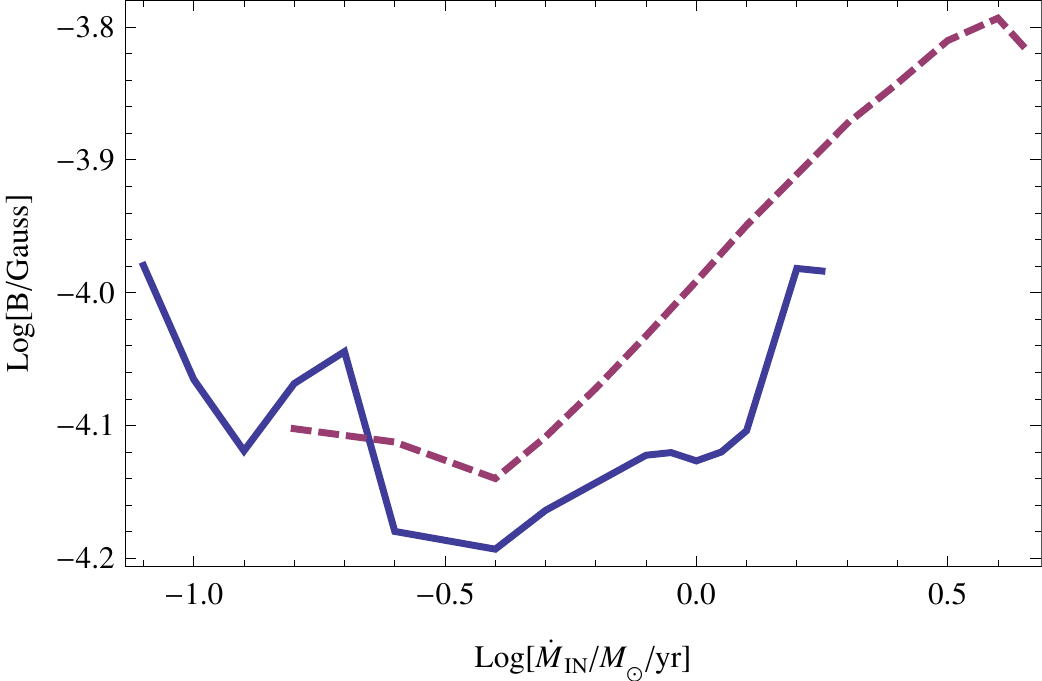,width=\columnwidth}
%{\includegraphics[width=\columnwidth]{plotRqrdWindSpeed.pdf}
\caption{Inferred magnetic field amplitude in the GC ISM for the cases of (solid,blue) $E_{SN} = 10^{51}$ erg (constant with respect to $M_{ZAMS}$) and
(dashed, purple) $E_{SN}$ an increasing function of $M_{ZAMS}$ as described in \S\ref{sctn_ESN}.
}
\label{plotMagneticField}
%} 
\end{figure}

Consistent with previous work (\citealt{Crocker2010a,Crocker2010b,Crocker2011a}; also see \citealt{Contini2011,Spergel1992})
we find our modelling favours a very high magnetic field amplitude in the GC, of order 100 $\mu$G (i.e., $\sim$400 times the energy density of the local ISM field).
As we discuss above and below, such a strong field can have important gas dynamical effects.
Another very important aspect of GC magnetic phenomenology is the existence of the non-thermal filaments (NTFs).
These are thin, synchrotron-illuminated structures running mostly perpendicular to the Galactic disk   and characterised by very high magnetic field amplitudes, $\sim$ mG  \citep{Yusef-Zadeh1987,Morris1989}.
Given that we have claimed that it is enhanced star formation in the GC region
that drives all the non-thermal phenomenology dealt with here and a large-scale outflow, it is intriguing that the NTFs are unique within the Galaxy to
the longitude range of this region.
In fact,  magnetic field advection by the wind may well  have a crucial role in forming the NTFs as analogues to cometary plasma tails, formed from the interaction of the large-scale, magnetized plasma outflow draping the region's dense molecular clouds. \citep{Shore1999,Boldyrev2006}.

\subsubsection{Plasma conditions}

Our modelling indicates that it is {\it possible} for the system to heat the outflowing plasma to the `very-hot' temperatures $\sim 7 \times 10^7$ K pointed to by X-ray observations.
However, that the plasma be this hot is clearly disfavoured for the $E_{SN} = 10^{51}$ erg case at the sort of $\dot{M}_{IN}$ values suggested by the
 considerations given above; even the $E_{SN}[M_{ZAMS}]$ case favours somewhat cooler temperatures over some of its favoured $\dot{M}_{IN}$  range.

\begin{figure}
 %\vspace{200pt}
 \epsfig{file=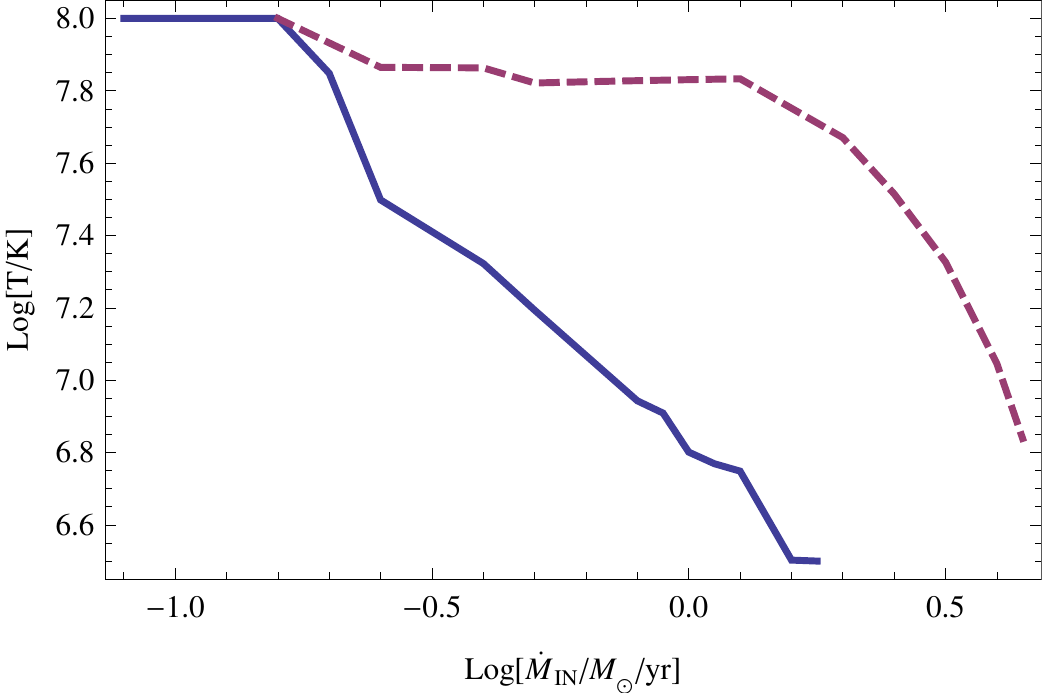,width=\columnwidth}
%{\includegraphics[width=\columnwidth]{plotRqrdWindSpeed.pdf}
\caption{Temperature of the diffuse plasma in the GC for the case of (solid,blue) $E_{SN} = 10^{51}$ erg (constant with respect to $M_{ZAMS}$) and
(dashed, purple) $E_{SN}$ an increasing function of $M_{ZAMS}$ as described in \S\ref{sctn_ESN}.
}
\label{plotTemperature}
%} 
\end{figure}

\section{Implications}

\subsection{Modes of Gas Accretion on to Inner Galaxy}

\citet{Chandran2000} estimate -- to order of magnitude precision -- that $3 \times 10^9 \msun$ of gas has fallen on the central $\sim200$ pc over the Galaxy's lifetime; this would imply an average accretion rate of $\sim0.3 \msun$/year.
\citet{Figer2004} estimate a mass accretion rate of $\sim0.4 \msun$/year from dividing the mass in molecular ring circumscribing the CMZ, $8 \times 10^6 \msun$, by  its orbital period, $2 \times 10^7$ year.
A conservative upper limit on the accretion rate can be derived from the assumption
that all the material falling on the Galactic disk, $< 10 \msun$/year  \citep{Combes2004} -- and more probably $0.5-5 \msun$/year \citep[e.g.][and references therein]{Wakker1999,Klessen2010}--  finds its way to the GC.

\noindent
{\bf Accretion through the disk:}
The overall position and size of the CMZ  is  presumably controlled by the gravitational dynamics dictated by the Galactic bar \citep{Binney1991,Morris1996,Stark2004}.
The non-axisymmetric gravitational potential of the  bar induces torques on disk gas that cause it to fall inwards at a rate estimated \citep{Morris1996} at 0.1-1 $\msun/$year.

Other mechanisms could, however, be acting to supplement accretion of gas on to the GC and may be particularly important in achieving  further  transfer of gas inward of the characteristic $\sim$100 pc radius of the X2 orbit family.
Such additional mechanisms include the action of a secondary, nested bar \citep{Namekata2009}, dynamical friction, and shocks associated with the X1 $\to$ X2 orbit transition \citep[see][\S 3.2 for a more inclusive list]{Morris1996}.
Also of interest are:

\noindent
{\bf Magnetic torquing:}
If sufficiently strong, the GC magnetic field might have appreciable effects on gas dynamics in the region \citep{Morris1996}.
In particular, magnetic viscosity may provide a channel for accretion of gas to radii smaller than that of the molecular ring established by the larger-scale gravitational dynamics.
Scaling the results of \citet{Balbus1998} and \citet{Beck1999} \citep[also see][]{Beck2005b}, we find that magnetic stress should provide a mass inflow rate inside 100 pc of $\dot{M_{IN}}^{mag} \simeq B^2 \ h \ \Omega^{-1} \simeq 0.2 \msun$/year, where the total magnetic field amplitude is assumed to be $B = 100 \ \mu$G, the gas scale height $\sim$13 pc \citep{Ferriere2007} and $\Omega \sim 100$ km/s/100 pc is the angular velocity.
This is interestingly comparable to other accretion channels.

\noindent
{\bf Bulge stellar mass loss:}
A minimum level of $\sim0.2 \msun$/year \citep{Jungwiert2001} of gas is supplied to the outer bar \citep{Stark2004} by mass loss from the evolved bulge stars.
Accretion of material out of the slowly-rotating Galactic bulge represents a dilution of disk specific angular momentum \citep[and references therein]{Morris1996}.

We have already seen (\S\ref{sctn_diskHalo1})  that there is interesting evidence for a small amount of
relatively pristine gas in the GC region \citep{Lubowich2000}, especially in the outer X1 orbit family or in the process of being transferred from these outer orbits to the inner X2 orbits \citep{Riquelme2010,JonesBurton2011}.
This represents rather compelling evidence for the accretion of material out of the halo on to the GC which would require a mechanism or mechanisms distinct from those listed above.

Three, non-mutually-exclusive mechanisms might operate to collect this gas, all potentially related to the outflow we have identified here and, therefore,
potentially self-catalysing.

\noindent
{\bf Dust cooling:} Ambient dust grains provide for the collisional cooling of plasmas
\citep[e.g.,][]{Montier2004,Natale2010}.
The operation of this mechanism in the halo above the GC requires the injection of dust into the region.
Mid-infrared  maps trace outflows of dust coincident with radio continuum spurs running north and south from the GC region \citep{Bland-Hawthorn2003,Stolovy2006,Morris2006} as already noted (\S\ref{sctn_extndH2}), thus there is  good evidence for entrainment of dust in GC outflows and, presumably, its injection into the halo above the GC where  it may serve to catalyse the cooling of the halo plasma; we leave a quantitive treatment of this potential mechanism to further work\footnote{We thank Richard Tuffs for raising this possibility.}.

\noindent
{\bf Shocks ahead of Giant Molecular Loops:} Also as already noted,  observations with the Nanten-II instrument \citep{Fukui2006} suggest the presence of giant, rising loops of  molecular material in the GC region.
The somewhat surprising presence of large amounts of molecular material arching over the entire length of each of these loops (rather than being strongly concentrated at the `foot-points' where the gas might be expected to fall subsequent to each loop's rise under magnetic flotation) may simply be a question of timing but it does seem, in general, to place the magnetic floatation via the Parker instability interpretation under some strain \citep{Morris2006}.
\citet{Riquelme2010}, moreover, have also found relatively nuclear-unprocessed molecular gas coincident with the GML footpoints, 
which again is somewhat surprising if the loops represent  molecular gas magnetically-levitated out of the disk close to the GC 
(for, were this true, the footprint isotopologue ratios would be consistent with the highly-processed gas found in the rest of the inner CMZ).
Thus, a somewhat different scenario explaining the arrival of fresh gas at the footprints of the GMLs may be required \citep{Riquelme2010}.
Such may be the idea
 \citep{Morris2006b,Torii2010} that 
the loops are magnetically-floated but  accrete gas by driving shocks into the halo above the GC in which H{\sc i} reaches sufficient densities to efficiently cool and condense into $H_2$.
Of course, this general idea might be extended to cover any sort of outflow that drives a sufficiently strong shock.

\noindent
{\bf H{\sc i} self-catalysed:} Finally, we note that the Kelvin-Helmholtz instability will strip gas off the (relatively) cool and metal-rich H{\sc i} clouds ejected into the halo.
Subsequent mixing of this stripped gas in the turbulent cloud wakes with in situ plasma allows for the cooling and condensation of the latter (allowing a phase transition from plasma to H{\sc i}) and the subsequent accumulation of some amount of this less astrated material \citep{Marinacci2010,Marinacci2011,Binney2011}.
Thus, not only does the fountaining of the cold, entrained gas present a mechanism for the {\it net} accretion of H{\sc i} in each cycle, it also potentially explains how new, relatively nuclear unprocessed matter finds its way into the GC system.
Following the modelling of \citet{Marinacci2011}  of this process in the disk of the Galaxy, net accumulation of gas of $\sim 20$\% per fountain cycle is suggested; we thus plot 
$0.2 \times \dot{M}_{cold}$ in fig.~\ref{plotMassFlowsFLAT} as a rough estimator of the mass flow in this accretion channel.
At the back-of-the-envelope level, this channel can supply a significant fraction -- in fact saturate for high $\dot{M}_{IN}$ in the $E_{SN}[M_{ZAMS}]$ scenario -- the net accretion rate demanded by star-formation and the hot outflow, $\dot{M}_{IN}^{acctn}$; clearly modelling particularised to conditions typical for the GC/inner Galaxy are required here to render this conclusion confident, however.

Any of the three  channels proposed above for accretion of matter out of the halo is of potential interest from a number of points of view: 
i) as stressed, these mechanisms allow for the admixture of relatively pristine gas as demanded by observations -- in general, this gas acts to counter-balance the effect of the high levels of astration in the GC region, thereby keeping overall metallicity somewhat in check; 
ii) such mechanisms show, in general,  how the star-formation processes might be self-sustaining in the sense that the star-formation-driven outflow actually catalyses the further accretion of gas;
and 
iii) following from this point, such a mechanism may explain how the long-term stability of the system -- in particular, a star-formation rate apparently held rather constant over many dynamical times -- is achieved \citep[cf.][]{Binney2011,Marinacci2011}.

\subsection{Strong, in situ magnetic field}

As remarked, our modelling shows that the in situ magnetic field is large $\sim 100 \ \mu$G, potentially large enough to have important dynamical effects in the region.
How is such a field established and -- in light of the losses due to advection and turbulent diffusion \citep{Beck1999} -- maintained?
Here a compelling mechanism is the long timescale accumulation of magnetic field lines frozen into gas accreted on to the region along the disk of the Galaxy \citep{Chandran2000}.
This process produces a large-scale magnetic field orientation that, in agreement with observations, is predominantly vertical: compression amplifies the disk-perpendicular field component and advection of frozen-in field in the outflow (such as we have modelled) and/or ambipolar diffusion removes plane-parallel field components.
 \citet{Chandran2000} estimate that the GC should accrete a total mass of $\sim 3 \times 10^9 \msun$ over its lifetime 
 (dependent on the amplitude of the assumed pre-Galactic field) in order to  explain its present-day magnetic phenomenology;
fig.~\ref{plotTotalMassesFLAT}  shows that this requirement is nicely matched for the regions of parameter space we favour \citep[though we note that][favour a somewhat bigger, large-scale field amplitude than pointed to by our analysis]{Chandran2000}.

\subsection{Implications of expelled, frozen-in magnetic field}

Figs.~\ref{plotPwrFLAT} and \ref{plotPwrGROWING}
point to magnetic energy losses at a rate of $\gtrsim 3 \times 10^{39}$ erg/s.
If sustained over 10 Gyr, these would imply the injection of $\sim 10^{57}$ erg by the outflow into the Fermi Bubbles, corresponding to a magnetic field amplitude of 40 $\mu$G over the entire volume of the Bubbles; this is, however, a severe overestimate of the likely final field amplitude once the system has relaxed via reconnection and adiabatic losses are accounted for \citep[e.g.,][]{Braithwaite2010}.
Magnetic field reconnection generically provides for the local injection of energy in the Bubbles, delivering heat and perhaps even providing for the acceleration of a non-thermal particle population in a distributed fashion in the Bubbles\footnote{Though  note that the maximum rate of reconnection energy injection is a factor of a few short of being able to maintain the temperature of the  plasma injected into the Bubbles: these radiative losses are sustained in steady state by the thermal power of the plasma  itself after it is injected hot at the GC}.
Following \citet{Braithwaite2010}, we may roughly calculate the  amplitude of the final, relaxed field, $B_f$, as 
$B_f \sim B_{out} \sqrt{(r_{out}/r_{Bub}) \ (\rho_{Bub}/\rho_{out})} \sim 100 \  \mu$G $ \times \sqrt{(110 \ \textrm{pc}/4.5 \ \textrm{kpc}) \ (0.01 \ \textrm{cm}^{-3}/ 0.1 \ \textrm{cm}^{-3})} \sim 5 \  \mu$G (here `$out$' denotes outflow and `$Bub$' is for Bubble).
This is characteristic amplitude over the entire volume of the Bubbles: we expect the field closer to the plane to be stronger in general 
(consistent with the fact that the non-thermal WMAP haze has now been shown to fairly sharply cut-off for $b \gtrsim 35^\circ$: \citealt{Dobler2011})
and there to be local regions of enhanced field as demanded by the microwave polarisation observations \citep{Jones2011}.
We note that  within our model \citep{Crocker2011a} that the Fermi Bubbles' $\gamma$-ray emission is supplied by hadronic processes, 
we have an expectation for the luminosity of the Bubbles due to synchrotron emission from secondary electrons and positrons (created in the same $pp$ collisions delivering the $\gamma$-rays); 
to supply the observed non-thermal, microwave emission detected as the WMAP haze a $\sim 10 \  \mu$G field is required.
A further important effect of the injected magnetic field will be to stabilise the Bubble surfaces against fluid instabilities \citep[e.g.][]{Gourgouliatos2012}.

\subsection{Total masses ejected}
\label{sctn_TotMass}

Fig.~\ref{plotTotalMassesFLAT} also demonstrates something rather interesting about the total mass of hot gas processed through the system: this is $(1-3) \times 10^9 \msun$ over the assumed $10^{10}$ years for the favoured range of $\dot{M}_{IN}$.
In principle, there are three possible fates for this gas:
i) it is ejected to infinity; 
ii) it rains back on to plane away from the GC; or
iii) it falls back on the GC itself.
Here ejection to infinity (in a true wind) can be firmly rejected as already noted.
The exact proportion of material that falls back on to the GC will depend on the details of the interaction of the outflowing plasma with the differentially-rotating bulge \citep{Marinacci2011}; we expect a non-negligible fraction of this low-angular momentum material to do so (see below).

\subsection{Relation to Fermi Bubbles}

\begin{figure}
 %\vspace{200pt}
 \epsfig{file=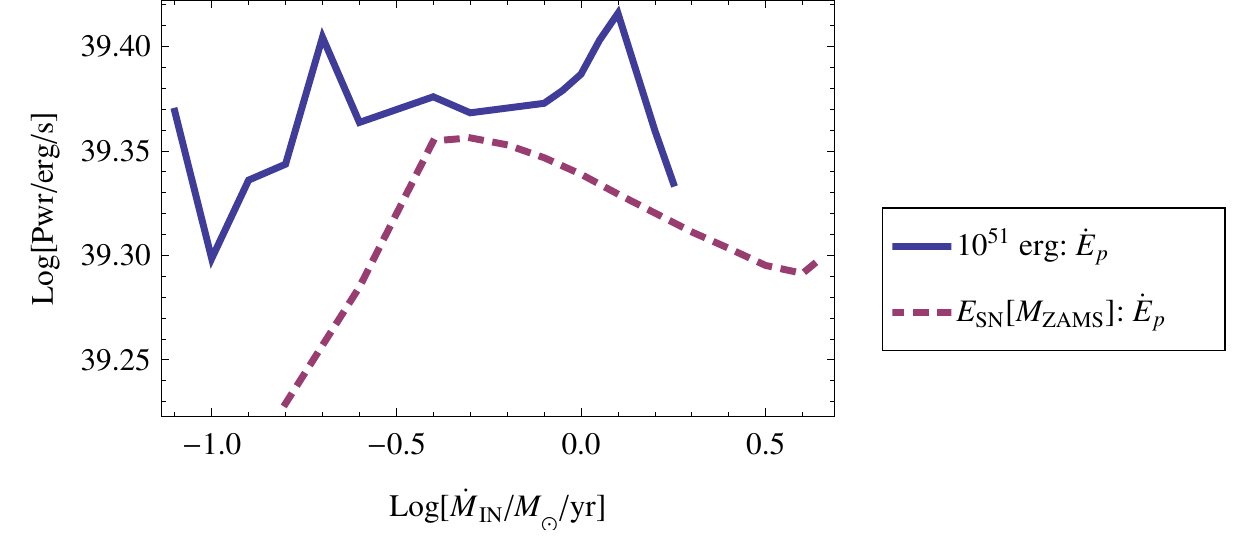,width=\columnwidth}
%{\includegraphics[width=\columnwidth]{plotRqrdWindSpeed.pdf}
\caption{Model predictions for the power being fed into freshly-accelerated protons for the
 case  (blue, solid)  of  $E_{SN} = 10^{51}$ erg; and (purple, dashed) 
$E_{SN}$ an increasing function of $M_{ZAMS}$ as described in \S\ref{sctn_ESN}.
}
\label{plotCRPower}
%} 
\end{figure}

Evident from fig.~\ref{plotCRPower} is that the power going into the locally-accelerated cosmic ray proton particle population is almost invariant at $\sim 10^{39}$ erg/s across the region of parameter space that fits the non-thermal data well.
We nowhere constrain our modelling to reproduce this result -- it is a prediction.
This power is precisely enough to sustain the $\sim$GeV $\gamma$-ray emission from the Fermi Bubbles assuming a saturation\footnote{{\bf Here `saturation' implies that i) the system has reached steady state and that ii) the loss timescale of $pp$ collisions is the smallest relevant timescale in the problem, i.e., the Bubbles constitute thick targets to the injected cosmic ray protons because the $pp$ loss time is, in particular, shorter than the escape time from the Bubbles for protons of the relevant energy to generate $\sim$GeV $\gamma$-rays. 
If these conditions apply and cosmic rays remain co-entrained with the parcels of gas and magnetic field in which they are injected into the base of the Bubbles by the outflow, the volumetric emission from the Bubbles becomes independent of the gas density in each such parcel (unless this is so low that the $pp$ loss time becomes longer than the age of the Bubbles).
As detailed by \citet{Crocker2011a} this effect implies that the overall $\gamma$-ray emission from the Bubbles is rather uniform as required by the observations.}} situation \citep{Crocker2011a}.
It is also sufficient to supply the total few $\times 10^{56}$ erg enthalpy of the Bubbles were this power delivered for the multi-Gyr timescale  independently suggested by the long $pp$ loss time (and the requirement that the steady-state be reached, achieving saturation; note that contributions from the outflowing plasma and magnetic field to the total enthalpy of the Bubbles would further decrease this timescale).
This is surely  not a coincidence.
Also significant is the fact that we explicitly confirm (see fig.~\ref{plotPwrFLAT}) that the outflow injects thermal plasma power into the base of the Bubbles  easily sufficient to explain their thermal emission  of $\sim 2 \times 10^{39}$ erg/s into the 0.5-2 keV energy band \citep{Almy2000,Snowden1997} or an inferred (temperature-dependent) bolometric luminosity of $\sim 10^{40}$ erg/s   \citep{Crocker2011a}.

\begin{figure}
 %\vspace{200pt}
 \epsfig{file=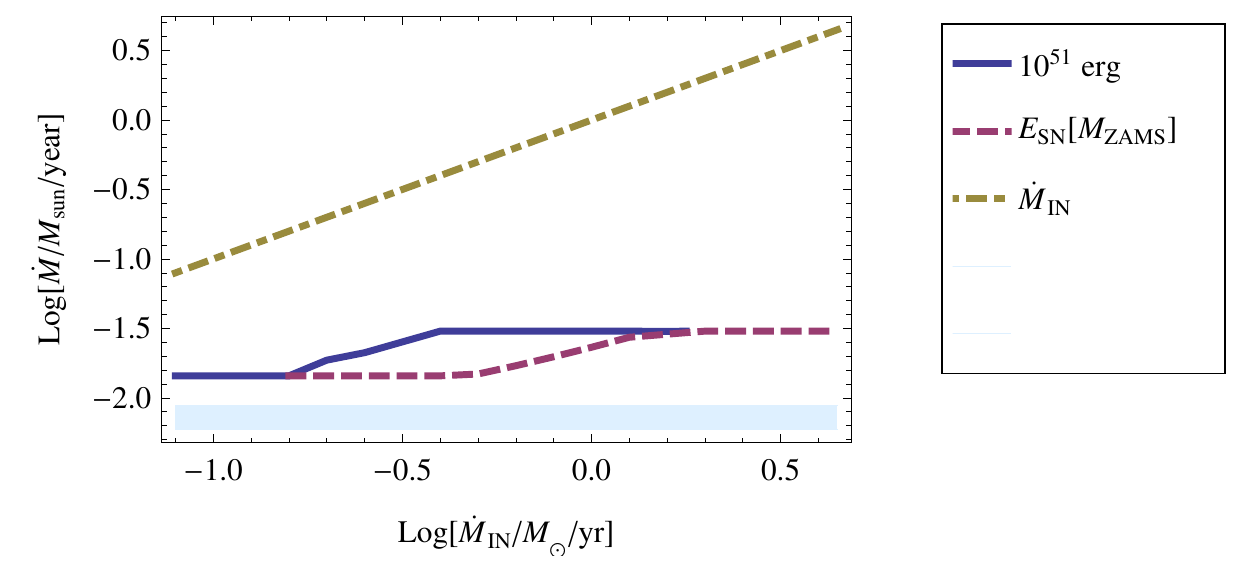,width=\columnwidth}
%{\includegraphics[width=\columnwidth]{plotRqrdWindSpeed.pdf}
\caption{Hot mass flows due to `directly' injected material (i.e., supernova ejecta and stellar winds)
for the case  (blue, solid)  the case of  $E_{SN} = 10^{51}$ erg; and (purple, dashed) 
$E_{SN}$ an increasing function of $M_{ZAMS}$ as described in \S\ref{sctn_ESN}.
The blue band shows the mass growth rate inferred \citep{Crocker2011a} for the Fermi Bubbles.
}
\label{plotDetailedMassLosses}
%} 
\end{figure}

The hot mass outflow suggested by our modelling (\S\ref{sctn_TotMass}) is easily sufficient to supply the $\lesssim 10^8 \msun$ of plasma in the Fermi Bubbles over 10 Gyr but there is some sensitivity to the particularities of our modelling in this conclusion because the hot outflow is dominated by the centrally-loaded and heated ISM material.
We can render the conclusion that GC star-formation can supply the plasma required to fill-out the Bubbles robust, however, by considering $only$ the 
mass flux in the `directly'-injected plasma (composed of supernova ejecta and stellar wind matter and neglecting the entrained ISM gas: there is little play in this number).
Fig.~\ref{plotDetailedMassLosses} shows directly-injected plasma mass flux; even it alone is sufficient to sustain the mass growth rate of the Fermi Bubbles we have previously inferred \citep{Crocker2011a} within our hadronic model (conclusion is robust with respect to both $\dot{M}_{IN}$  and choice of $E_{SN}$ and really just depends on the average SFR and the time over which it has been sustained), i.e., to supply $\sim 10^8 \msun$ of gas over $\sim$10 Gyr.

Let us now proceed under the assumption that the GC outflow is feeding the Bubbles and that our model's treatment of mass loading of the hot outflow is correct.
Then, if $M_9^{hot} 10^9 \msun$ of gas is ejected in the GC's hot outflow over $t^{life}_{10 \ \textrm{\tiny{Gyr}}}$ 10 Gyr, and  $f_{ret} M_9^{hot} 10^9 \msun$ of gas
is returned to the GC
(after each cycle of the hot material from GC, into the halo and back again)
 and given the mass of the Fermi Bubbles if $M_8^{FB}  10^8 \msun$ \citep{Su2010,Crocker2011a},  the (initially) hot gas must  cycle through the GC 
10 $(f_{ret}  M_9^{hot}/M_8^{FB})$ times over $t^{life}$, implying a cycle time of 1 $\times \ [t^{life}_{10 \ \textrm{\tiny{Gyr}}} M_8^{FB}/(f_{ret}  M_9^{hot})]$ Gyr. 
Self-consistently, this timescale is very similar to the cooling time of the $\sim 10^7$ K, $\sim$0.01 cm$^{-3}$ plasma in the Bubbles.
The implication of this reasoning is that the gas in the Bubbles is slowly
churning and, while the structures themselves may persist for $\sim 10$ Gyr,  individual parcels of gas injected into Bubbles  remain aloft for only a fraction of this time with, presumably, a trajectory involving relatively high-speed injection in the outflow, a period of neutral or positive buoyancy relatively high in the halo \citep{Rodriguez-Gonzalez2009}, and finally, condensation into H{\sc i} and fall-back to the GC once sufficient cooling has taken place.
\citet{Zech2008} have previously interpreted data on high velocity clouds at inner Galactic longitudes as representing exactly such a circulation.

\begin{figure}
 %\vspace{200pt}
 \epsfig{file=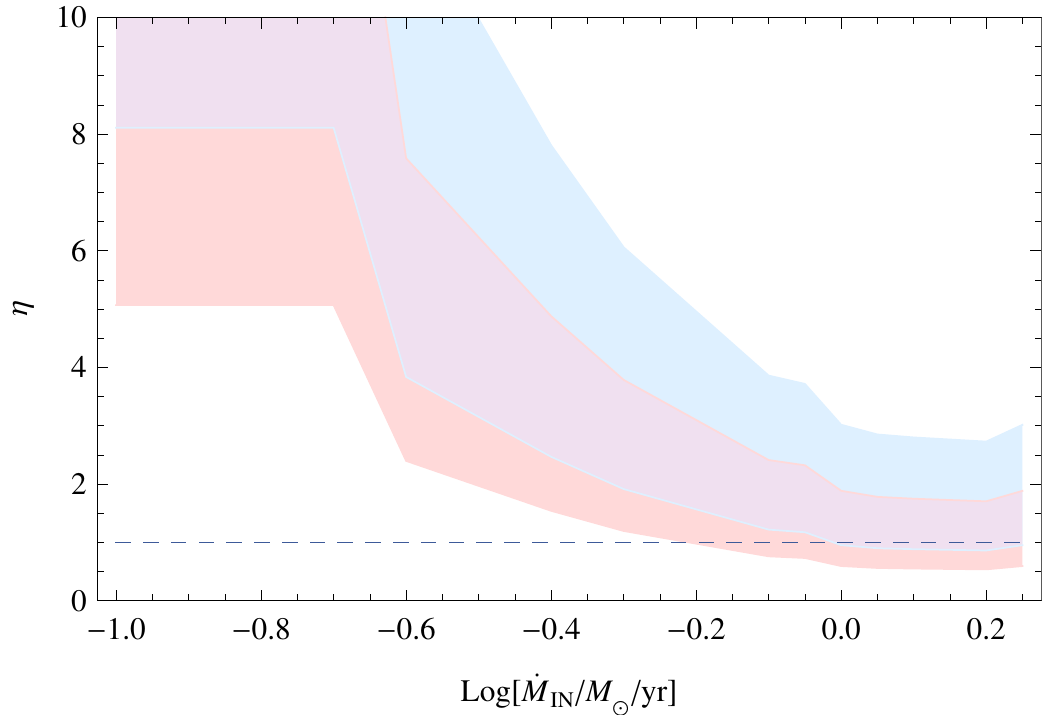,width=\columnwidth}
%{\includegraphics[width=\columnwidth]{plotRqrdWindSpeed.pdf}
\caption{Density contrast between external, halo plasma and Bubble interiors as filled by the outflow we model. 
The blue band is for an assumed halo plasma temperature of $10^{6.5}$ K and the red for $10^7$ K with the upper and lower range to each band given by taking the outflow fluid to be, respectively, a non-relativistic fluid ($\gamma = 5/3$) and a relativistic fluid ($\gamma = 4/3$).
In this calculation adiabatic evolution is assumed and the mass flux into the Bubbles is just as given by $\dot{M}_{HOT}$ ignoring mass drop-out due to cooling and condensation.
}
\label{plotDensityContrastFLAT}
%} 
\end{figure}

If the Bubbles  really are held aloft by buoyancy they must have an interior matter density less than the external halo plasma, implying a density contrast satisfying $\eta \equiv \rho_{ext}/\rho_{Bub} > 1$.
On the other hand, the Bubbles' asymptotic ascent speed is controlled by $\eta$ with a larger density contrast implying a faster ascent \citep[at least until additional, dynamical effects entrain significant amounts of material into the Bubbles; e.g.,][]{Pope2010}.
Following \citet{Hinton2007}, our modelling allows us to calculate the density contrast  assuming adiabaticity 
as
\be
\eta = \frac{\gamma-1}{\gamma} \ \frac{\mu \ L_{out} m_p}{\dot{M} \ k_B \ T_{ext}}
\ee
where $\mu \simeq 0.61$ (assuming for simplicity a plasma of solar composition), the temperature of the halo plasma is $T_{ext} \sim (3 - 10) \times 10^6$ K \citep{Almy2000}, and  $ L_{out}$  and $\dot{M}$ are the mechanical power and mass injection rate of the outflow, respectively.
In fig.~\ref{plotDensityContrastFLAT} we calculate $\eta$ for the case of $E_{SN} = 10^{51}$ erg, taking the mass growth rate of the Bubbles to be given by $\dot{M}_{HOT}$ (therefore ignoring mass drop-out due to cooling and condensation).
Note that the star-formation-driven outflow we model naturally predicts an initial  $\eta \sim (1-10)$; this is in contrast to AGN-driven outflows 
where initial density contrasts $\eta \gg 1$ are generically expected \citep[e.g.,][]{Hinton2007}.
We further find that $\eta \to 1$ (from above) in the range of $\dot{M}_{IN}$ preferred by other indicators predicting that -- without fine-tuning -- we expect the buoyant ascent of the Bubbles to be slow.

\subsection{Model-derived star formation and supernovae parameters}

\begin{figure}
 %\vspace{200pt}
 \epsfig{file=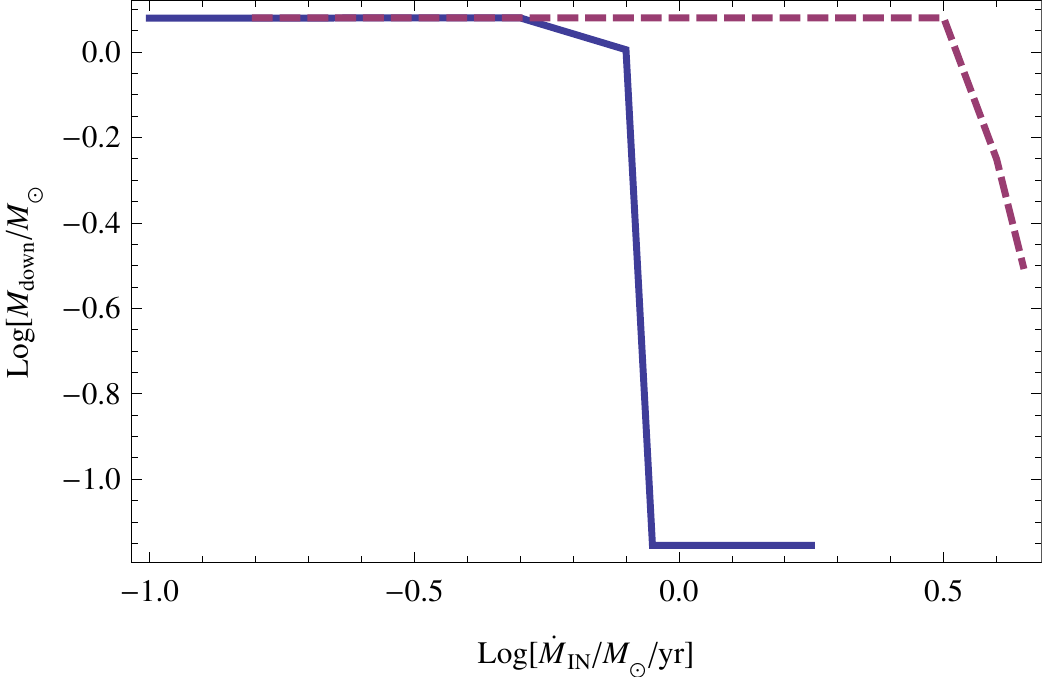,width=\columnwidth}
%{\includegraphics[width=\columnwidth]{plotRqrdWindSpeed.pdf}
\caption{The lower $M_{ZAMS}$ cut-off -- i.e., the fragmentation mass -- as a function of $\dot{M}_{IN}$.
Note that at the very top of the allowed $\dot{M}_{IN}$ ranges the modelled fragmentation mass is butting-up against the (purely) numerical constraint that this parameter be larger than $\sim 0.3 \msun$.
}
\label{plotFrgmntnMass}
%} 
\end{figure}

On the basis of our modelling we can calculate -- at least at the population level -- various parameters characterising the region's star-formation and resultant supernovae.
Fig.~\ref{plotFrgmntnMass} shows that our modelling suggests 
that as the favoured $\dot{M}_{IN}$ parameter range (for either $E_{SN}$ case) is approached from below we see a transition from the modelling preferring high (1.2$\msun$) to low (i.e., conventional: 0.07$\msun$) values of the
 fragmentation mass (the lower integration limit of the $M_{ZAMS}$ parameter).
This is an interesting coincidence but numerical experiments we have performed indicate that
our modelling does not seem to confer particularly strong sensitivity to this parameter.

\begin{figure}
 %\vspace{200pt}
 \epsfig{file=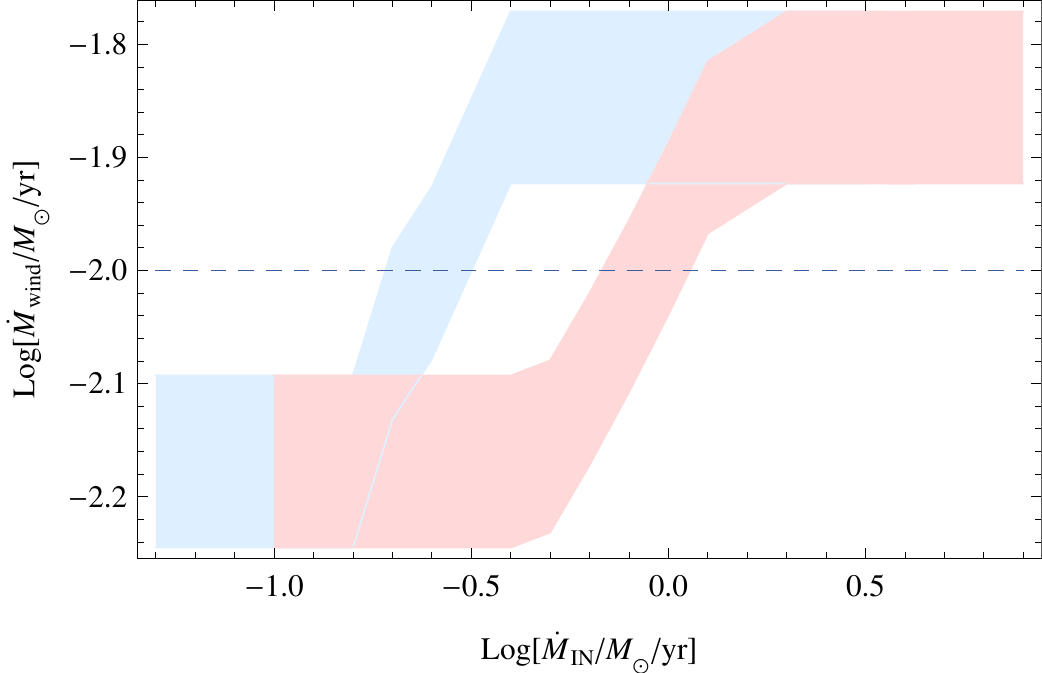,width=\columnwidth}
%{\includegraphics[width=\columnwidth]{plotRqrdWindSpeed.pdf}
\caption{Stellar mass loss predicted by our modelling employing the prescription for single-star stellar winds presented by \citet{Meynet2003} for the cases of (blue) $E_{SN} = 10^{51}$ erg (constant with respect to $M_{ZAMS}$) and
(red) $E_{SN}$ an increasing function of $M_{ZAMS}$ as described in \S\ref{sctn_ESN}. 
The upper and lower bounds of each band are for the case, respectively, of stars rotating at 300 km/s and not rotating.
The dashed line shows our previous estimate \citep{Crocker2011b} for all GC stellar wind losses, $0.01 \msun$/year.
}
\label{plotWindLosses}
%} 
\end{figure}

Fig.~\ref{plotWindLosses} demonstrates that, coincidentally or not, the modelled stellar wind losses within our scenario  -- as informed by a parameterization of the results presented by  \citet{Meynet2003} for {\it single star evolution} -- 
match rather well with the total stellar wind losses of $\sim 0.01 \msun$ from the region determined by observations \citep[see][and references therein]{Crocker2011b}.

\begin{figure}
 %\vspace{200pt}
 \epsfig{file=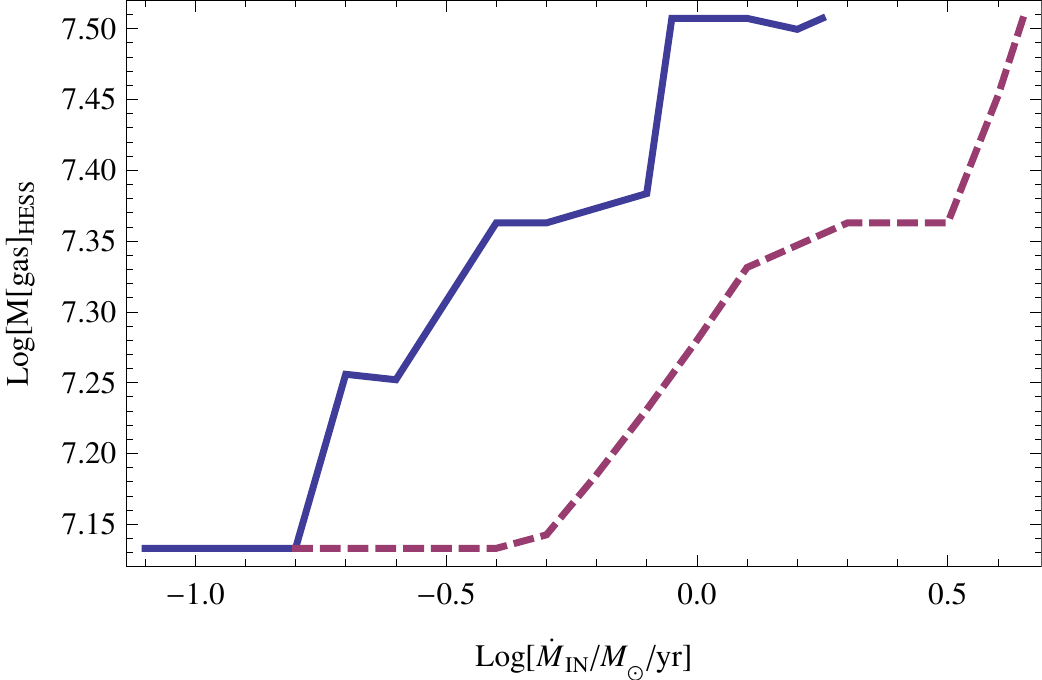,width=\columnwidth}
%{\includegraphics[width=\columnwidth]{plotRqrdWindSpeed.pdf}
\caption{The total molecular gas mass in the GC region
inferred from the \citet{Kennicutt1998}-Schmidt relation and the modelled SFR
 for the case of (solid,blue) $E_{SN} = 10^{51}$ erg (constant with respect to $M_{ZAMS}$) and
(dashed, purple) $E_{SN}$ an increasing function of $M_{ZAMS}$ as described in \S\ref{sctn_ESN}.
}
\label{plotInferredCMZMass}
%} 
\end{figure}

Given the calculated amount of star-formation going on in the region, we can calculate an expectation for the region's total gas mass on the basis of the  \citet{Kennicutt1998}-Schmidt relation (see fig.~\ref{plotInferredCMZMass}); this expectation matches reasonably well to the estimated gas mass of $3 \times 10^7 \msun$ of the region \citep[e.g.,][]{Molinari2011}.
In fact, if we demand that  the GC be typically-efficient at turning gas into stars, $\dot{M}_{IN} \gtrsim 1 \msun$/year and $\dot{M}_{IN} \gtrsim 3 \msun$/year are implied for the cases of  $E_{SN} = 10^{51}$ and $E_{SN}[M_{ZAMS}]$, respectively.
This is another (though perhaps rather weak) piece of evidence supporting $\dot{M}_{IN} \gtrsim 1 \msun$/year.

\subsubsection{Supernovae}

\begin{figure}
 %\vspace{200pt}
 \epsfig{file=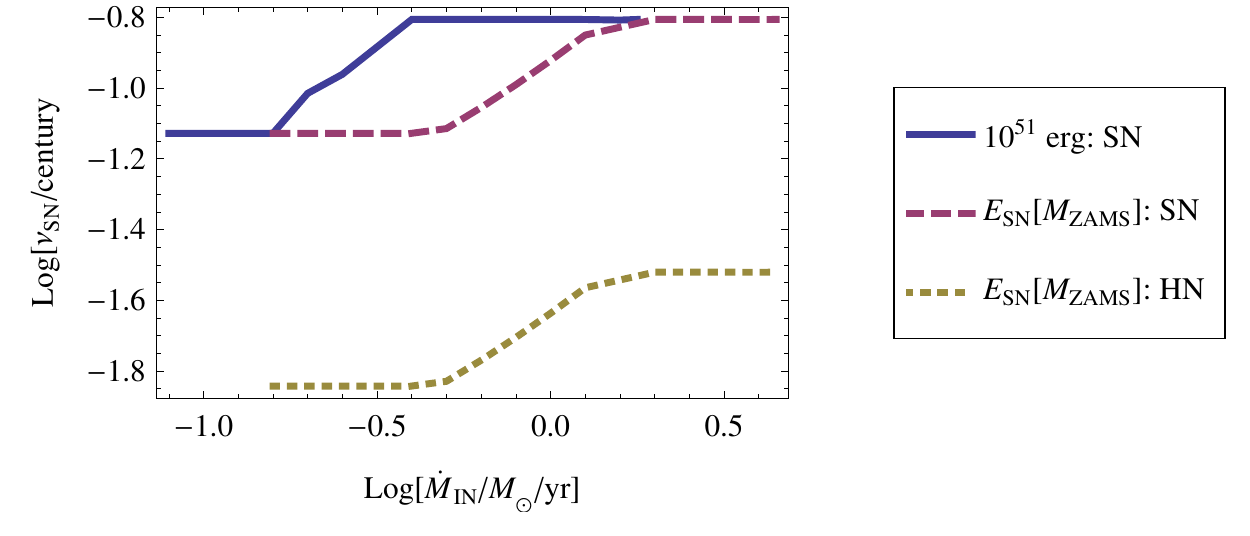,width=\columnwidth}
%{\includegraphics[width=\columnwidth]{plotRqrdWindSpeed.pdf}
\caption{Supernova rate for  the case of (solid,blue) $E_{SN} = 10^{51}$ erg (constant with respect to $M_{ZAMS}$) and
(dashed, purple) $E_{SN}$ an increasing function of $M_{ZAMS}$ as described in \S\ref{sctn_ESN} and (dotted, yellow) the
`hypernova' rate, i.e., the  number rate at which stars of $M_{ZAMS} > 26.5 \msun$ -- for which  our parameterization satisfies $E_{SN}[M_{ZAMS}] \simeq  10^{52}$ erg -- are formed.
}
\label{plotSNRate}
%} 
\end{figure}

Fig.~\ref{plotSNRate} shows that the modelled supernova rate in the region is 0.08--0.16 per century (corresponding to an expected time between supernovae of 630--1300 years);
this is at the upper end of the range we previously determined \citep[viz. 0.02--0.08 per century; see Appendix B2.2 of][]{Crocker2011b}, 
but certainly consistent with gross estimates (e.g., that the GC is responsible for a $\sim$few--10 \% of the Galaxy's massive star formation) and not in significant conflict with
an upper limit of $\sim$0.1/century obtained from counting point-like radio source pulsar candidates (\citealt{Lazio2008,Deneva2009}; also see \citealt{Wharton2011}).
Much of the difference between our current supernova rate determination and our previous estimates is due to our adoption 
of a floating, lower  cut-off to the fragmentation mass which, as shown above, 
the modelling prefers larger than the conventionally-adopted $\sim 0.07 \msun$ over much of the $\dot{M}_{IN}$ parameter space.

Also shown in this figure is the rate at which stars more massive than 26.5 $\msun$ are produced which -- in our adopted parameterizaton of $E_{SN}[M_{ZAMS}]$ -- is equivalent to the rate of SN explosions releasing more than $10^{52}$ erg mechanical energy, i.e., hypernovae.
This is similar to the total rate of broad-lined SNIbc expected in the Galaxy (adopting 2 core collapse supernovae per century, \citealt{Diehl2006}, of which 1-2 \% are SNIbc BL; e.g., \citealt{Smith2011}).

\begin{figure}
 %\vspace{200pt}
 \epsfig{file=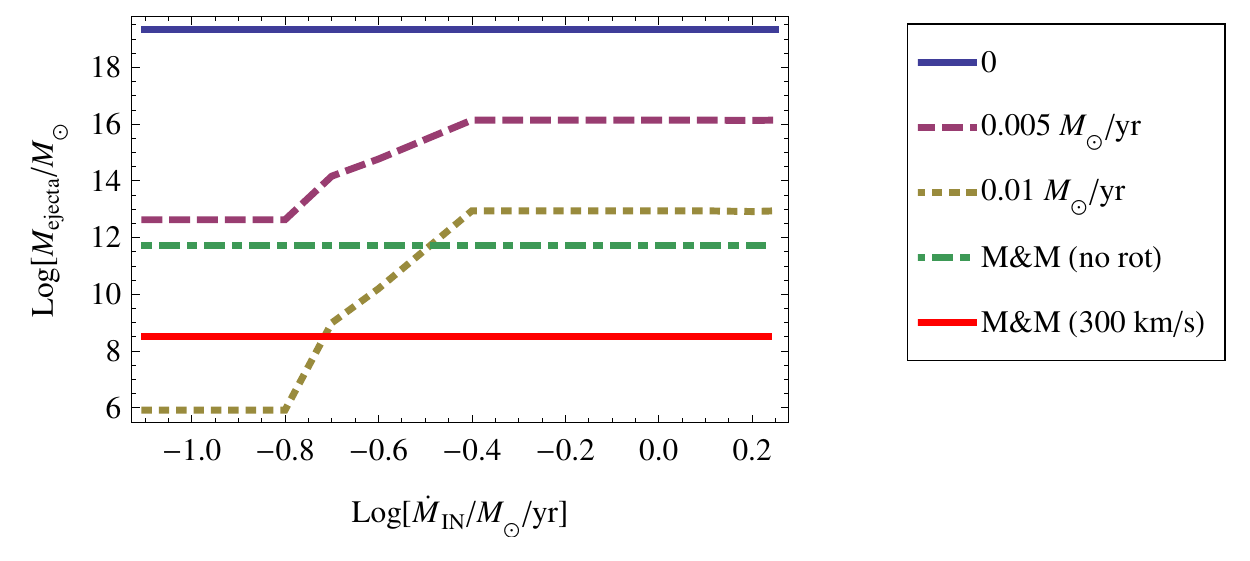,width=\columnwidth}
%{\includegraphics[width=\columnwidth]{plotRqrdWindSpeed.pdf}
\caption{Mean ejecta mass from GC core-collapse supernovae assuming an invariant 1.4 $\msun$ remnant produced by each SN and for different assumptions concerning annual mass loss rate for GC stars  (which could, in principle be any or all of stellar winds, binary mass transfer, and eruptive mass loss)  as indicated by the key with 
`M\&M' denoting the IMF-integrated effect of the single star stellar winds inferred from the prescription presented by \citet{Meynet2003} for non-rotating and stars initially rotating at 300 km/s, respectively.
}
\label{plotEjectaMass}
%} 
\end{figure}

Our modelling also allows us to calculate the mass ejected by each supernovae into the ISM (given certain assumptions: see fig.~\ref{plotEjectaMass}).
We can compare these modelled masses with  supernova ejecta masses inferred from observations of supernova light curves:

\noindent
{\bf SNIbc:}
With reasonable assumptions of average photospheric velocities \citet{Drout2011} have estimated  the mean ejecta masses for SN Ib, Ic, and  Ic (broad line) types  from photometric observations of a total of 25 
 Ibc supernovae detected within 150 Mpc with  the robotic Palomar 60-inch telescope.
Consistent with the picture that either stellar winds, massive eruptions, or binary mass transfer (or some combination of these; e.g., \citealt{Smith2011}) removes most of the envelopes of the massive progenitors of such explosions, \citet{Drout2011} determine mean ejecta masses for SN Ib and Ic supernovae of $\sim 2 \msun$ (and $\sim 5 \msun$ for the more energetic, broad-line explosions); this result should be relatively unbiased by selection effects towards unusually bright SN Ibc\footnote{Maria Drout, private communication}.

\noindent
{\bf SNII:}
Unfortunately, we are not aware of such an unbiased mean ejecta mass determination for SNII supernovae being presented in the literature.
With this in mind, from fig.~S3 from the Supplementary Information to the work by \citet{Perets2010},   we find a rough mean SN II ejecta mass of 16 $\msun$; 
 from the data presented by \citet{Utrobin2011} on SN IIP, we find a mean 19.7 $\msun$ ejecta mass.
These latter determinations are quite consistent with SN II pregenitor's experiencing little mass loss at all during their lifetimes: with our assumed Kroupa IMF, the mean mass of stars formed between 8 and 150 $\msun$ is  20.7 $\msun$, neglecting the mass of compact remnant, or 17.4  $\msun$, if a 1.4 $\msun$ neutron stars is formed by all progenitors with $M_{ZAMS}$ between 8 and 20 $\msun$ and a $\sim 8 \msun$ black hole is formed for more massive progenitors \citep{Belczynski2011}.
Of course, these averages are presumably affected by selection biases as noted.

\begin{figure}
 %\vspace{200pt}
 \epsfig{file=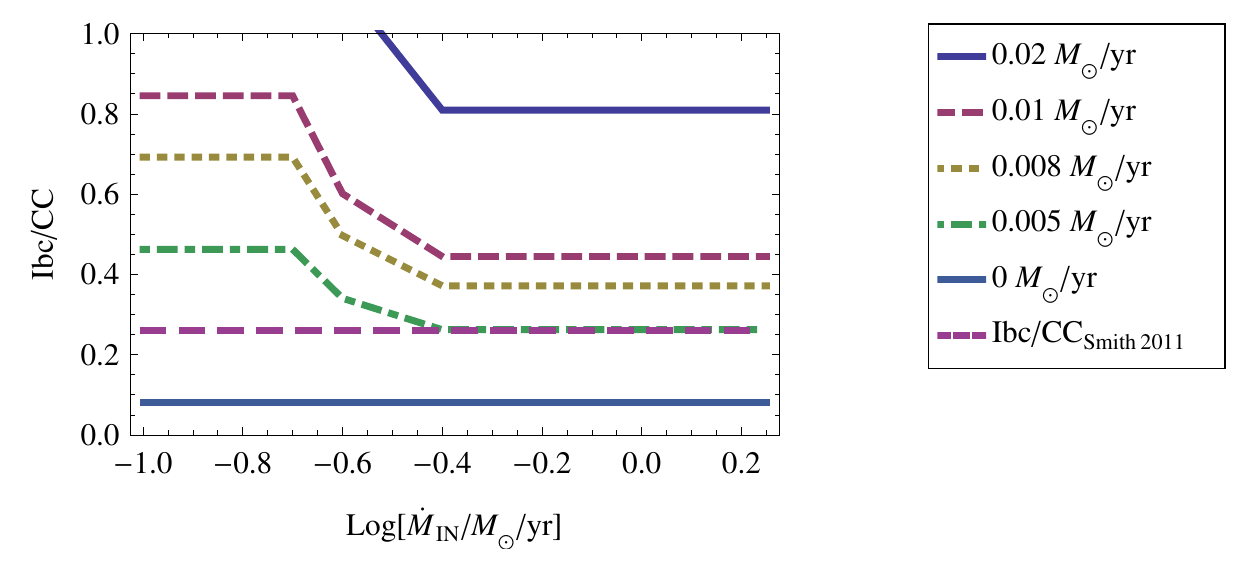,width=\columnwidth}
%{\includegraphics[width=\columnwidth]{plotRqrdWindSpeed.pdf}
\caption{Inferred rate of SNIb and SNIc explosions as a fraction of all core-collapse supernovae in the GC with different assumptions for the total mass-loss rate
experienced by all GC stars (i.e., through all channels, potentially including stellar winds, binary mass transfer, and eruptive mass losses).
The Ibc/CC$_{\tiny\textrm{Smith  2011}}$ horizontal line shows the cosmological average  26\% contribution the sum of all SNIb and Ic make to the total core-collapse rate as inferred by \citet{Smith2011}.
}
\label{plotSNIbcFrctn}
%} 
\end{figure}

Broadly, we find ejecta masses consistent with some admixture of different SN types (fig.~\ref{plotEjectaMass}).
In principle, we can actually infer the relative fractions of SNII and SNIbc types in the GC environment on the basis of our modelling and adopting the observational ejecta mass averages given above: see fig.~\ref{plotSNIbcFrctn} which shows the minimum fraction of  SNIb and SNIc supernovae with respect to all core-collapse supernovae in the GC environment given the ejecta mass determined in our modelling and assuming that the progenitors of SNII experience negligible mass loss and form 1.4 $\msun$ compact remnants. 
Adopting our fiducial number for the mass loss due to all stellar winds in the region, a SNIbc fraction $\gtrsim 50$\% is indicated, substantially larger than the cosmological average of 26\% for all SNIb and Ic's arrived at by \citet{Smith2011}.
Allowing for other sources of mass loss (binary mass transfer, eruptive mass loss) would indicate an even higher fraction.
These determinations are susceptible to many uncertainties, however, and we leave detailed treatment of this aspect of the problem to future work.
For the moment we remark it is interesting that our modelling implies that
the fraction of SNIb and SNIc events in the GC is higher than the cosmological average.
This is consistent with the finding that this fraction
tends to  increase in higher metallicity environments \citep[e.g.,][]{Boissier2009,Modjaz2011} and/or towards the brightest star-forming regions of galaxies \citep[e.g.,][]{Leloudas2010}.

\subsubsection{Stellar remnants}

\begin{figure}
 %\vspace{200pt}
 \epsfig{file=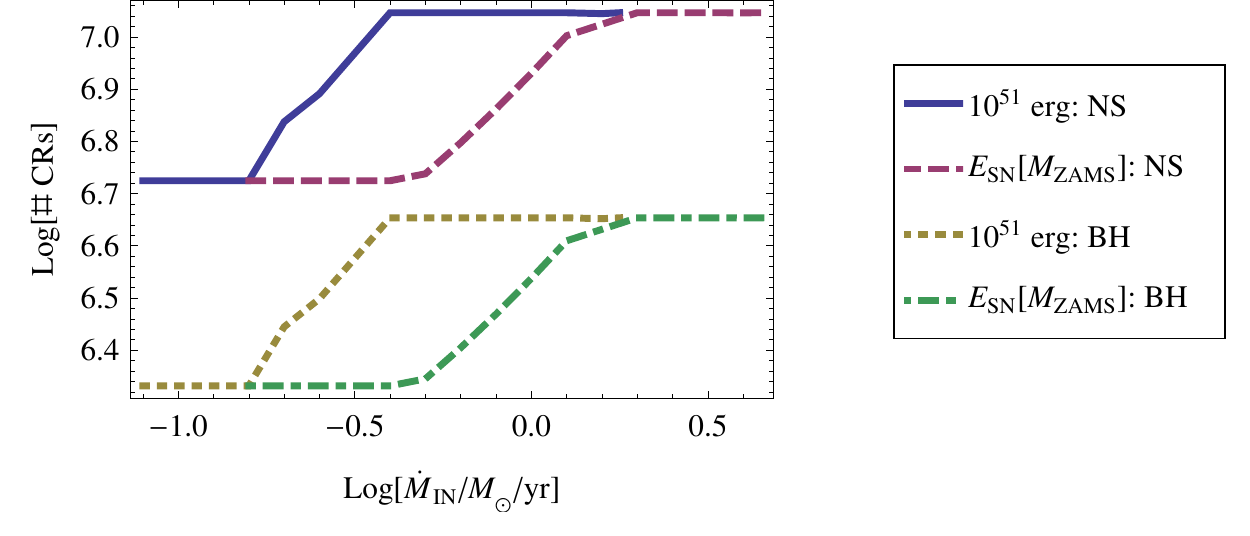,width=\columnwidth}
%{\includegraphics[width=\columnwidth]{plotRqrdWindSpeed.pdf}
\caption{Total numbers of compact stellar remnants formed over an assumed 10 Gyrs in the GC.
Cases are given in the key; `NS' denotes neutron stars  and `BH' denotes black holes.
We  assume that all stars $8 < M_{ZAMS}/\msun < 20$ form neutron stars and all stars $M_{ZAMS} > 20 \msun$ form black holes.
}
\label{plotCompactRemnants}
%} 
\end{figure}

Our modelling gives us a rough handle on the total number of compact remnants produced by GC star formation.
Inspired by the findings of \citet{Belczynski2011}, we assume that stars with $8 < M_{ZAMS}/\msun < 20$ form neutron stars and all stars $M_{ZAMS} > 20 \msun$ form  black holes (with population-mean mass $\sim 8 \msun$). 
These assumptions would indicate (fig.~\ref{plotCompactRemnants}) 
the formation of $\sim 10^7$  neutron stars and  $\sim 4 \times 10^6$  stellar mass black holes
over the assumed 10 Gyr age of the system (with more total mass in black
holes)\footnote{Some fraction of the formed neutron star population will be both retained in the region and `recycled' into the $\mu$s period range by spin-up due to accretion from a binary companion.
In turn, some fraction -- depending on beaming and other effects -- of this milli-second pulsar (MSP) population can be expected to radiate at $\sim$GeV energies \citep{Abdo2009,Malyshev2010}.
As we have previously remarked \citep{Crocker2011b}, the  $\sim$ GeV flux data covering the region has a spectrum morphologically consistent with being dominated by such MSPs (which would have to amount to a few hundred in number).
Given the many large uncertainties (particularly surrounding the recycling fraction), it is currently difficult to convert the neutron star formation rate (and assumed timescale) arrived at in our modelling to an accurate prediction of the number of extant and detectable MSPs.
The analysis of \citet{Wharton2011} seems to broadly show that our numbers are reasonable given the play in the relevant parameters.}.

\subsection{Implifications for picture of GC as starburst-like environment}

In disk environments -- in both our own Galaxy and in external, `normal' spiral galaxies -- it is generally held that $\sim10\%$ of the initial kinetic energy delivered by a SN explosion is thermalized \citep[and references therein]{Strickland2009}.
In the GC environment we find, in contrast, a situation much more akin  -- from this perspective -- to starburst environments \citep{Heckman1990}.
For instance, the recent detailed analysis of the starburst nucleus of M82 by \citet{Strickland2009} points to a fraction approaching unity of each SN's energy being lost into heating or moving the ISM.
Very hot plasma temperatures approaching $10^8$ K also seem to be prevalent in such systems.

As we have seen, for the GC, something similar seems to be happening.
{\it Prima facie}, given the large amount of ambient molecular gas and the consequently large volumetric average gas densities in these systems, this appears surprising: why isn't more energy lost radiatively (cf. fig.~\ref{plotPwrFLAT})? 

Certainly, the volume filling factor of the densest cores of this gas (where most of the mass is tied up), is very small \citep[e.g.,][]{Oka2005}.
Moreover, the localisation of GC star-formation and its temporally-extended, continuous nature imply that the conditions in the GC resemble a sort of steady-state super bubble -- or, indeed, starburst -- where previous generations of stars and/or supernovae have presumably
riven the ambient dense gas with cavities, blow-outs and channels \citep[cf.][]{Westmoquette2009,Contini2011}.
These conditions mean shocks sent out into the ISM have a reduced opportunity to interact with dense ISM phases (that would then efficiently radiate) and, in fact, the energy injected by supernovae is efficiently thermalized as demanded \citep[cf.][]{Higdon2005}.

The same conditions mean that GC supernova remnants apparently remain good cosmic ray accelerators (assuming the $E_{SN} = 10^{51}$ erg prescription is correct). 
The cosmic rays, moreover, only ever `see' a small fraction of the total ambient gas mass
\citep{Crocker2011b}; both plasma and cosmic rays are apparently blown out of  the system so quickly that they they do not penetrate into the dense gas cores where stars are being formed.

\section{Discussion}

\subsection{Is the `very hot' plasma real or not?}

We have already seen (fig.~\ref{plotTemperature}) that the $E_{SN} = 10^{51}$ erg case indicates a plasma temperature $\lesssim 10^7$ K over the favoured $\dot{M}_{IN}$ parameter space; on the other hand, the $E_{SN}[M_{ZAMS}]$ case provides for a temperature  approaching  $\sim 7 \times 10^7$ K at the lower end of its favoured $\dot{M}_{IN}$
range. 
Moreover, if we constrain our modelling so that the modelled plasma temperature is $\geq 5 \times 10^7$ K we find good fits to the data for even the  $E_{SN} = 10^{51}$ erg  case, though, of course,  the plasma temperature is no longer then a $prediction$ of the modelling.
It would seem yet premature, therefore, to rule the idea of the very hot plasma out on the basis of our modelling.

Two other facts should also be kept in mind here. 
Firstly,  we are modelling the plasma's temperature as it leaves the region of interest; 
this is {\it after} the centralised mass loading (i.e., the heating of swept-up, ambient ISM gas to plasma temperatures), also pointed to by our analysis, has taken place.
This mass loading process has to  take place over some physical scale (and may, indeed, continue once the outflowing gas has left the region).
The natural temperature scale for the plasmas in the centres of star-forming regions {\it before} any mass loading has taken place (and assuming themalization efficiency  $\beta \to 1$, not too much at variance with what we find) is $\sim 3 \times 10^8$ K \citep[e.g.,][]{Veilleux2005}.
So, in principle, the X-ray observations may be revealing the very hot, central temperature of the plasma before too much mass loading takes place (as the gas is driven out of the system).
Indeed, emission from the putative very hot component is concentrated on {\it somewhat} smaller size scales than the full 200 $\times$ 80 pc$^2$ region modelled here \citep{Yamauchi1990,Belmont2005}.

{\bf Previous work on the question of the existence of the putative very hot plasma \citep[e.g.,][]{Muno2004} has focused on the energetic difficulties  apparently encountered were the plasma real and escaping at approximately the sound speed because of the inadequacy of known power sources to sustain the thermal energy carried off by the plasma.
Setting aside the fact that our modelling apparently favours a lower temperature for the escaping diffuse plasma anyway,
it is worth emphasising here that i) the mass loading and entrainment processes we have modelled here slow down the outflow significantly with respect to the sound speed and ii) the supernova rate we find -- close to 1 per 1000 years -- is {\it by construction} sufficient to sustain the energy losses represented by the outflow.}

Note that other pieces of evidence continue to suggest the plasma is real: the suggestive pressure equilibrium that would exist between such a plasma and the other GC ISM phases \citep{Spergel1992}; 
the fact that such very hot, diffuse plasmas are seen in the centres of active star-forming systems like M82 that are driving `super' outflows \citep{Strickland2009} and, 
strengthening this connection, the fact that the 6.7 keV Fe He$\alpha$ line flux measured from the GC  \citep[$\sim 4 \times 10^{34}$ erg/s as inferred from  table 2 of][]{Koyama2007} is exactly as expected \citep[see \S 5.5 of][]{Strickland2009} given the GC SFR and overall size of the  star-forming region.

On the other hand, as explained above, we find evidence for (relatively) cool, entrained gas in our outflow, apparently with rather large filling factor, consistent with the envelope $H_2$ phase identified by $H_3^+$   and other measurements; it has been claimed \citep{Goto2008} that it is difficult to reconcile the existence of this phase with the existence of a (necessarily) large filling factor very hot plasma.
In summary, it seems that for the moment the jury remains out on the important question of whether the very hot, diffuse plasma exists or not.

\subsection{Is the system really in steady state? What about the contribution of individual star-bursts?}

The steady state modelling presented here indicates   
that $\sim10^{7.2}$ core-collapse supernovae have occurred in the GC in the post-infant Galaxy.
This number is ultimately determined by the secular accretion of gas on to the region which has led to the formation of $\sim 10^9 \msun$ of stars in the nuclear bulge.
Of course there are stochastic variations in the GC star-formation rate.
According to the analysis of \citet{Sjouwerman1998} of their observations of oxygen-rich, cool giant stars, the most recent, significant burst of SF in the GC occurred approximately 1 Gyr ago and should have resulted \citep{Higdon2009} in $\sim10^6$ core-collapse supernovae.
Thus even this relatively dramatic event in the GC's history only contributes at the $\lesssim$10\% to the total star-formation (and resultant energy-deposition) history of the region.

Observations of today's GC super stellar clusters of stochastic variation in the star-formation rate on much more recent timescales, few $\times 10^6$ Myr, but the total mass of stars formed in these systems/events, $\lesssim 10^5 \msun$, is {\it much} less than the $\sim10^9 \msun$ of stars  represented by the nuclear bulge.
Such events will only be associated with the ejection of significant amounts of energy if they are accompanied by substantial jumps in the accretion rate on to Sgr A$^*$ \citep[cf.][]{Zubovas2011}.

\subsection{How is the long-term stability of the SFR maintained?}

The long-term star formation scenario requires that the GC continues to accumulate gas over a similar timeframe; {\it prima facie}
this  requires that a driving stellar bar remains extant, too, over the requisite multi Gyr timescale \citep{Serabyn1996}.
The fact that bars are common in spiral galaxies \citep{Eskridge2000} is consistent with this requirement.
On the other hand, 
we have listed a number of mechanisms that may
 allow the GC to pull gas directly out of the halo.
 The operation of some such  mechanism seems to be required by the detection of
 gas that has undergone relatively little astration in the GC region and  may imply a `guaranteed' accretion rate that means that GC  is not completely at the mercy of conditions prevailing on much wider scales in the disk to sustain its star-formation.

The scenario investigated here of a mass-loaded, sub-sonic outflow is interestingly reminiscent of the situation identified by \citet{Tang2009}
who model long-lasting feedback due to outflows from galactic nuclei (driven by Type Ia supernovae) wherein there is little active star-formation.
These authors find that, by dropping out cool, entrained gas at large radii, such outflows are stabilised to the extent that they remain essentially stationary for a few Gyr.
As we have previously argued \citep{Crocker2011a} and continue to find in the research described here, the evidence for the GC system and the outflow it drives is that  a similar situation of quasi-stationarity has been reached.

Finally, \citet{Kim2011} in their numerical treatment of GC star-formation note that a negative feedback mechanism may be implied by the fact that GC star-formation peaks sharply in a region consistent in extent with the expected size of the X2 orbits (i.e., a smaller region than the whole CMZ) but seems to be relatively suppressed beyond this region (in the X1 orbit region) despite the presence of gas.
They postulate that  heating of the relatively more diffuse gas in a UV photon field of $\sim 30$ eV cm$^{-3}$ 
-- not unreasonable for the inner GC environment -- may provide for this suppression.
Remembering our finding above and previously \citep{Crocker2010b,Crocker2011b} that cosmic rays do not seem to penetrate into the dense gas in the time
they remain in the system,
we note here that cosmic ray heating may equally-well affect (or at least supplement) 
the requisite negative feedback  \cite[cf.][and references therein]{Suchkov1996,Yusef-Zadeh2007,Crocker2011b}.
Consistent with this, we find steady-state cosmic ray energy densities in our modelling of $\geq 20-40$ eV cm$^{-3}$.

\section{Conclusions}
\label{sctn_Conclusions}

We have modelled the mass and energy flows through the Galactic centre in a one-zone model and shown that  star-formation -- and resultant supernovae -- in the GC are well-sufficient to drive the gross dynamics of the system and to explain its non-thermal phenomenology.
None of this is to deny the importance of, e.g., stellar-radiation driving of gas dynamics in particular regions of the GC, but  radiation pressure is not required to explain the large-scale mass movements we infer here.

Current evidence does not seem strong enough for us to promote the empirically-inspired scaling of supernova mechanical energy with zero-age, main sequence progenitor mass that we have tested above; if anything, the `standard' $E_{SN} = 10^{51}$ erg assumption seems to work better in a number of cases (e.g., in suggesting a mass loading factor closer to expectation for star-burst-like environments; a SFR closer to other, independent estimates; supplying the total stellar mass of the nuclear stellar bulge for less extreme values of $\dot{M}_{IN}$).
In any case, as emphasised above, the $E_{SN}$ scaling we adopted  certainly constitutes an upper limit to the true, population-averaged evolution of $E_{SN}$ with stellar mass.
Likely, further modelling, probably involving other constraints, will be needed before a better handle on  $E_{SN}[M_{ZAMS}]$ can be arrived at.
Given these considerations, we now particularise our discussion to the case of  $E_{SN} = 10^{51}$ erg.

A number of indicators come together to suggest that our control parameter, $\dot{M}_{IN}$ -- the total mass being fed into the system -- has a lower limit at around 0.4 $\msun$/year.%
We find a 2$\sigma$ upper limit on $\dot{M}_{IN}$  (for the $E_{SN} = 10^{51}$ case) at 1.8 $\msun$/year.
We emphasise, however, that  $\dot{M}_{IN}$ is an overestimate of the total mass flux accreting out of the plane of the Galaxy on to the GC.
Our modelling suggests that there is an  outflow of plasma and cosmic rays from the system that entrains cool gas.
This entrained mass constitutes most of the mass flux but will fountain back on the GC.
This naturally accounts for the recently-observed halo of warm molecular hydrogen found to be surrounding the CMZ and may self-catalyse the accretion of relatively pristine corona gas into the system.

We find that our modelling robustly predicts an almost invariant $10^{39}$ erg/s for the power going into the freshly accelerated cosmic ray proton population in the GC region.
We nowhere constrain our model to produce this result: it emerges from the  numerical minimisation procedure.
As we have previously emphasised \citep{Crocker2011a,Crocker2011b} this power is {\it precisely} enough to sustain the $\gamma$-ray emission from the Fermi Bubbles in a hadronic saturation scenario and allow for the inflation of the Bubbles against the pressure of the external medium in a few Gyr (probably assisted by the injected magnetic field and injected plasma).
Equally, we find that the modelling robustly predicts that the GC system injects $\sim 10^{38}$ erg/s into hard-spectrum cosmic ray electrons; this is sufficient \citep{Crocker2010a,Crocker2011b} to explain the non-thermal synchrotron radiation detected from the GC lobe and wider diffuse, non-thermal source region detected around the GC \citep{LaRosa2005,Crocker2010a}.

Together with the evidence that the GC SFR has been quasi-stationary for Gyr timescales and that the outflow from the GC advects most non-thermal particles out of the acceleration region before they lose much energy, we consider our finding that the GC accelerates $10^{39}$ erg/s in cosmic rays as a very strong indication that GC star-formation essentially explains the Fermi Bubbles.
This argument is strengthened by the facts that the same star-formation processes can inject the plasma mass and thermal power required to fill-out the Bubbles and sustain their X-ray emission {\it and} to inject the magnetic fields that can stabilise the Bubble surfaces against fluid instabilities, trap their cosmic ray and plasma contents for long timescales and explain their microwave synchrotron emission.

\vspace{1 cm}

The Galactic centre is not particularly distinguished in the night sky -- but this belies its true activity: the many orders of magnitude of visual extinction arising from the column of dust we view it through is the reason for this $extrinsic$ dimness.
Equally, the GC is not particularly impressive as a non-thermal radiation source;  its $\sim$ GeV $\gamma$-ray luminosity, at few $\times 10^{36}$ erg/s, is an order of magnitude short of the 5-10\% of the Galaxy's output one might guess on the basis of the amount of massive star-formation happening in the system \citep[the Galactic $\sim$ GeV $\gamma$-ray luminosity is about $3 \times 10^{38}$ erg/s as inferred from fig.~1 of][]{Strong2010}.
As we have emphasised previously \citep{Crocker2010b,Crocker2011a,Crocker2011b} {\it this is because the cosmic rays accelerated in the region are mostly leaving before they can radiate}.
The radiation that these particles finally do emit is writ large in the Fermi Bubbles: it is from these structures that we detect the $\sim$ 10\% of Galactic $\gamma$-ray luminosity (i.e., $4 \times 10^{37}$ erg/s) that we expect on the basis of the GC's share of Galactic star-formation. 

The fact that we can detect the Fermi Bubbles at all is testament to the long-term stability of the GC as  star-forming system.
We have seen hints above as to how this stability can be established, in particular, how GC star-formation activity can be insulated from the vicissitudes of conditions in the Galactic disk: it seems that a minimal level of accretion in the GC system is self-catalysed.
This is consistent with the presence of relatively pristine gas accreted out of the halo via a mechanism or mechanisms directly related to the star-formation-driven outflow 
(i.e., injection of cool, high metallicity gas and/or dust and/or shocks into the halo plasma).

We find complete consistency between the long-timescale-averaged power required to drive the GC system and the power  injected by the star-formation we can  infer is currently taking place in the system.
We need not invoke periods of very bright AGN-type activity of the supermassive black hole to explain the dynamics of the GC.
In fact, as a final speculation: these studies hint that the importance of sustained, GC star-formation 
is that it effectively erects a curtain wall around the SMBH, either using-up gas directly by creating new stars or blowing it away before much of it can reach Sgr A$^*$.
This prevents mass accretion at rates that would allow the SMBH to undergo phases of  activity sufficient to heat the Galaxy's coronal gas
to such high temperatures that further accretion on the Galactic disk would become impossible\footnote{We emphasise that this certainly does not imply there is {\it no} accretion on to the SMBH or that it must be in a state of absolute quiescence.}.
This, in turn, enables  the long-term sustenance of {\it disk} star formation   \citep[cf.][]{Binney2011}.
Such a mechanism would explain the emerging finding \citep{Erwin2011} -- to which the Milky Way adheres -- that the mass of a nuclear star cluster correlates with the {\it total} stellar mass of its host galaxy rather than the galaxy's bulge.

\section{Acknowledgements}

RMC gratefully acknowledges useful conversation or correspondence with   Rene Belmont, Geoff Bicknell, Joss Bland-Hawthorn,  Valenti Bosch-Ramon, 
Dieter Breitschwerdt,
Michael Burton,  Doug Finkbeiner,  Ilya Gurwich, David Jones, Cornelia Lang, Casey Law, Andrea Maccio, Karl Menten, Mark Morris, Giovanni Natale, Masayoshi Nobukawa, Tomo Oka, J{\"u}rgen Ott, Christoph Pfrommer, Wolfgang Reich, Frank Rieger, Rahul Shetty, Tracy Slatyer, Meng Su,   Heinz V{\"o}lk, 	Benjamin Winkel, and Daniel Wang, particularly thanks Felix Aharonian, Maxim Barkov, Rainer Beck, Maria Drout, Werner Hofmann, Richard Tuffs,  and Andrea Stolte and is very grateful to Brian Keeney, Fulvio Melia, and Mark Morris for detailed comments on the manuscript and to the referee, Vladimir Dogiel, for an expeditious and helpful report.
RMC  thanks Nicole Bell and Ray Volkas for hospitality at the School of Physics, University of Melbourne where some of this research was carried out.
For most of the duration of this project RMC was the grateful recipient of IIF-Marie Curie fellowship awarded by the European Research Council.
RMC thanks the Max-Planck-Institut f{\" u}r Kernphsik for supporting this research.

%\begin{thebibliography}{99}
%\begin{thebibliography}{} <---- correct one

\appendix

\section{Metallicity of GC environment}

There remain considerable uncertainties surrounding the metallicity of the GC environment with different observational probes suggesting different results.
Broadly, however, these different approaches suggest that metallicity of the environment is between solar and twice solar: $Z_\odot \lesssim Z_{GC} \lesssim 2 Z_\odot$.

In more detail, from measurements of the nitrogen surface abundance of WN stars in the Arches cluster \citet{Najarro2004} report a solar metallicity.
\citet{Grafener2011} have, however, claimed that these measurements actually represent a lower limit to the initial C + N + O abundance with a best-fit value around $2 
Z_\odot$.
Recent  spectral analysis of luminous cool stars \citep{Cunha2007} and LBVs \citep{Najarro2009} indicates an approximately solar Fe
abundance and enhanced abundances of $\alpha$-elements.
\citet{Liermann2010}, citing the \citet{Martins2008} survey of the Arches cluster, determine a slight metallicity enrichment: $Z_{GC} = (1.3-1.4) Z_\odot$.
\citet{Mauerhan2010} in their Paschen-$\alpha$ survey of the inner $\sim$90 pc 
determine a dominance of WNL and WCL Wolf-Rayet subtypes over WNE and WCE subtypes.
This is consistent with a metallicity for the region that is {\it at least} solar.

From X-ray observations \citet{Wang2006} report a plasma Fe abundance $1.8^{+0.8}_{-0.2}$ solar; \citet{Koyama2007} and \citet{Nobukawa2010} report an upper limit at twice solar for the inner $\sim$ 200 pc region.
\citet{Borkowski2010} report super-solar metal abundances in the GC supernova remnant G1.9+0.3 on the basis of X-ray line observations.
On the basis of high-latitude UV spectroscopy \citet{Zech2008} report the existence of super solar metallicity gas likely associated with a GC wind or fountain.

Finally, studies of radio recombination line emission from H{\sc ii} regions reveal a clear gradient in metallicity with Galactic radius \citep{Shaver1983,Afflerbach1996}
and predict that the warm plasma phase of the GC ISM should be rather cool;
radio recombination line studies \citep{Law2009} of the GC reveal exactly this. 

\section{Example broadband fit}
\label{sctn_BBeg}

We show here an example broad-band spectral fit obtained by our $\chi^2$-minimization procedure.

\begin{figure}
 %\vspace{200pt}
 \epsfig{file=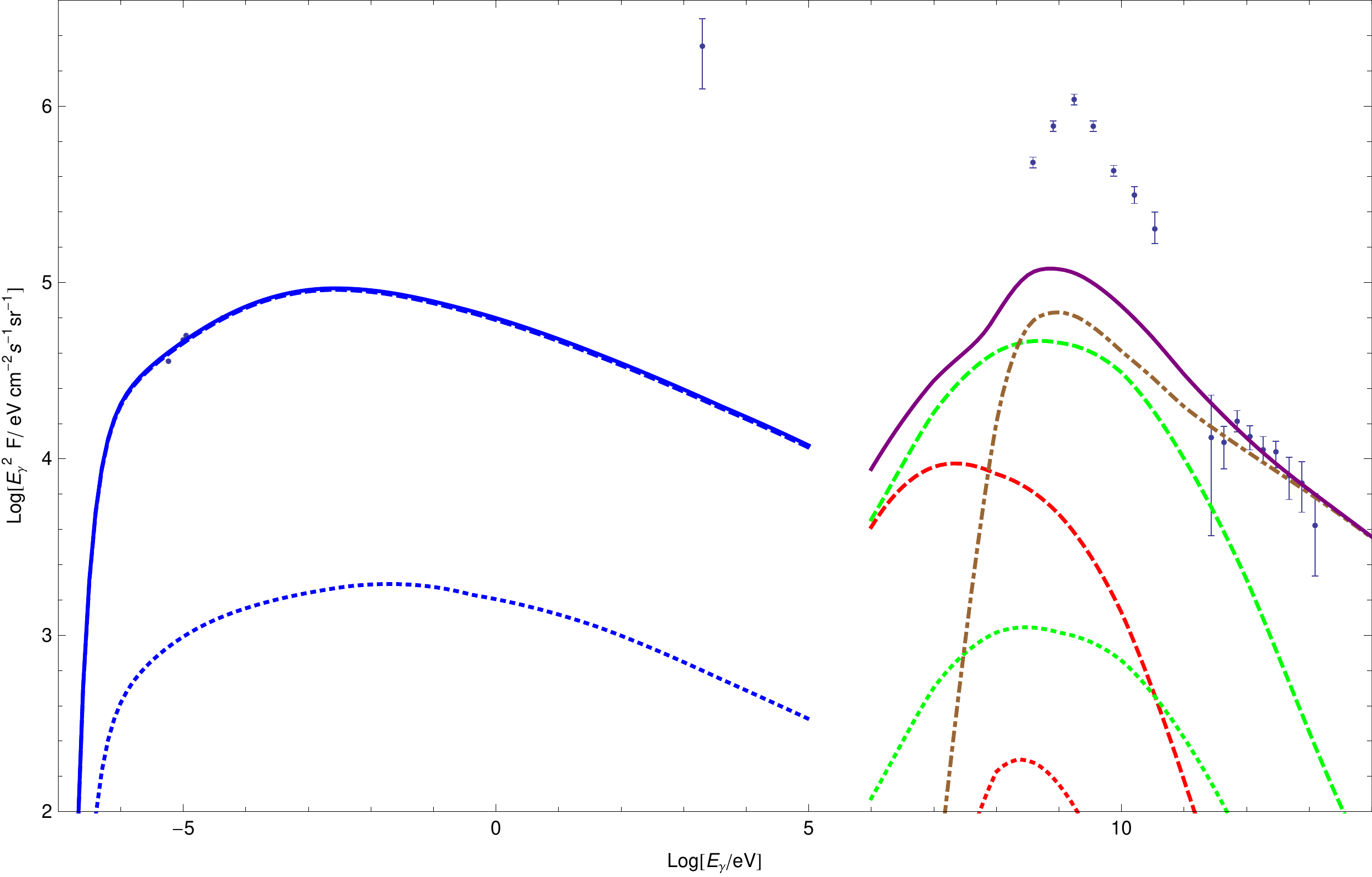,width=\columnwidth}
%{\includegraphics[width=\columnwidth]{plotRqrdWindSpeed.pdf}
\caption{An example modelled broadband spectrum for the region (showing non-thermal components only), in this case for the control parameter $\dot{M}_{IN} = 1 \msun$/year.
The curves are as follows: 
i) solid blue: total synchrotron;
ii) dashed blue: synchrotron from primary electrons (almost indistinguishable  from total synchrotron in this case);
iii) dotted blue: synchrotron from secondary electrons;
iv) solid purple: sum of  $\gamma$-ray fluxes from all processes;
v) dot-dash brown: $pp \to$ neutral mesons $\to \ \gamma$-rays;
vi) dashed green: inverse Compton by primary electrons;
vii) dotted green: : inverse Compton by secondary electrons;
viii) dashed red: bremsstrahlung by primary electrons; and
ix) dotted red: bremsstrahlung by secondary electrons.
Note the $\sim$GeV data points and the X-ray flux datum only constitute weak upper limits to the diffuse, non-thermal emission (the X-ray datum is explained via thermal bremsstrahlung emission by the plasma, an emission process we also model).
}
\label{plotBBeg}
%} 
\end{figure}

\end{document}